\documentclass[12pt]{article}
\makeatletter \@addtoreset{equation}{section} \makeatother
\renewcommand{\theequation}{\thesection.\arabic{equation}}
\newcommand{\ba}{\begin{array}}
\newcommand{\ea}{\end{array}}
\newcommand{\beq}{\begin{equation}}
\newcommand{\eeq}{\end{equation}}
\newcommand{\bea}{\begin{eqnarray}}
\newcommand{\eea}{\end{eqnarray}}

\def\bce{\begin{center}}
\def\ece{\end{center}}

\def\nonu{\nonumber}

\def\pa{\partial}

\def\be{\beta}

\def\la{\lambda}

\def\eps6{{\displaystyle \mathop{\epsilon}^{6}}{}}
\def\g6{{\displaystyle \mathop{g}^{6}}{}}

\def\nab6{{\displaystyle \mathop{\nabla}^{6}}{}}

\def\0{{\sst{(0)}}}
\def\1{{\sst{(1)}}}
\def\2{{\sst{(2)}}}
\def\3{{\sst{(3)}}}
\def\4{{\sst{(4)}}}
\def\5{{\sst{(5)}}}
\def\6{{\sst{(6)}}}
\def\7{{\sst{(7)}}}
\def\8{{\sst{(8)}}}

\def\ba{\begin{array}}
\def\ea{\end{array}}
\def\beq{\begin{equation}}
\def\eeq{\end{equation}}
\def\be{\begin{equation}}
\def\ee{\end{equation}}

\def\la{\lambda}
\def\eps{\epsilon}

\def\ba{\begin{array}}
\def\ea{\end{array}}
\def\beq{\begin{equation}}
\def\eeq{\end{equation}}
\def\be{\begin{equation}}
\def\ee{\end{equation}}

\def\la{\lambda}
\def\eps{\epsilon}

\def\eps6{{\displaystyle \mathop{\epsilon}^{6}}{}}

\def\nab6{{\displaystyle \mathop{\nabla}^{6}}{}}

\newcommand{\bean}{\begin{eqnarray*}}
\newcommand{\eean}{\end{eqnarray*}}

\begin{document}
\thispagestyle{empty} \addtocounter{page}{-1}
   \begin{flushright}
\end{flushright}

\vspace*{1.3cm}
  
\centerline{ \Large \bf   
Higher Spin Currents 
} 
\vspace*{0.3cm}
\centerline{ \Large \bf 
in the Orthogonal Coset Theory } 
\vspace*{1.5cm}
\centerline{{\bf Changhyun Ahn } 
} 
\vspace*{1.0cm} 
\centerline{\it 
Department of Physics, Kyungpook National University, Taegu
41566, Korea} 
\vspace*{0.8cm} 
\centerline{\tt ahn@knu.ac.kr
} 
\vskip2cm

\centerline{\bf Abstract}
\vspace*{0.5cm}

In the coset model $(D_N^{(1)} \oplus D_N^{(1)},D_N^{(1)})$
at levels $(k_1,k_2)$, the higher spin $4$ current that
contains the quartic WZW currents contracted with completely symmetric
$SO(2N)$ invariant $d$ tensor of rank $4$   is obtained.
The three-point functions with two scalars are obtained for any
finite $N$ and $k_2$ with $k_1=1$.  They are
determined also in the large $N$ 't
Hooft limit. When one of the levels is the dual Coxeter number of
$SO(2N)$, $k_1=2N-2$, the higher spin $\frac{7}{2}$
current, which contains the septic adjoint fermions contracted with
the above $d$ tensor and the triple product of structure constants,
is obtained from the operator product expansion (OPE)
between the spin $\frac{3}{2}$ current living in the ${\cal N}=1$
superconformal algebra and the above higher spin $4$ current.
The OPEs between the higher spin $\frac{7}{2}, 4$ currents
are described.
For $k_1=k_2=2N-2$ where both levels are equal to the dual Coxeter number of
$SO(2N)$, the higher spin $3$ current of $U(1)$ charge
$\frac{4}{3}$, which contains the six product of
spin $\frac{1}{2}$ (two) adjoint fermions contracted with
the product of $d$ tensor
and two structure constants,
is obtained. The corresponding ${\cal N}=2$
higher spin multiplet is determined by calculating the remaining higher
spin $\frac{7}{2}, \frac{7}{2}, 4$ currents with the help of
two spin $\frac{3}{2}$ currents in the ${\cal N}=2$ superconformal
algebra. The other
${\cal N}=2$ higher spin multiplet, whose $U(1)$ charge is opposite to
the one of above ${\cal N}=2$ higher spin multiplet, is obtained.
The OPE between these two ${\cal N}=2$ higher spin mutiplets
is also discussed.

\baselineskip=18pt
\newpage
\renewcommand{\theequation}
{\arabic{section}\mbox{.}\arabic{equation}}

\tableofcontents

\section{Introduction}

The Gaberdiel and Gopakumar proposal \cite{GG1011},
the duality between the
higher spin gauge theory on $AdS_3$ space \cite{PV1998}
and the large $N$
't Hooft limit of a family of $W_N (\equiv W A_{N-1})$
minimal models is the
natural analogue of the Klebanov and Polyakov duality \cite{KP}
relating the $O(N)$ vector model in three-dimensions to
a higher spin theory on $AdS_4$ space.
Then the obvious generalization of \cite{GG1011}
is to consider the Klebanov and Polyakov duality
in one dimension lower. 
By replacing the $SU(N)$ group by $SO(2N)$,  
the relevant  most general coset  model is described 
as \cite{LF1990,GKOplb}
\bea
\frac{\hat{SO}(2N)_{k_1} \oplus \hat{SO}(2N)_{k_2}}{\hat{SO}(2N)_{k_1+k_2}}.
\label{generalcoset}
\eea
One can also consider the case where
the $SU(N)$ group by $SO(2N+1)$ but
this is not described in this paper.
It is known that
the conformal weight (or spin) of the primary state  is equal to 
the quadratic Casimir 
eigenvalue divided by the sum of the level and
the dual Coxeter number of the
finite Lie algebra \cite{BS,DMS}.
For example, 
for $SO(2N)$, the quadratic Casimir
eigenvalue for the adjoint representation
is given by $2N-2$ while
the dual Coxeter number is $2N-2$. Then we are 
left with the adjoint fermion of spin $\frac{1}{2}$
at the critical level 
which is equal to the dual Coxeter number.
One can apply this critical behavior to the two numerator  
factors in (\ref{generalcoset}) simultaneously.
In the description of these adjoint free fermions,
the central charge grows like $N^2$ in the large
$N$ 't Hooft limit: so-called  stringy coset model \cite{GG1207}. 
See also the relevant works in
\cite{GG1205,GG1305,GG1406,GG1501,GG1512}.

Although some constructions on the higher spin currents in
\cite{Ahn1202} have been done,
there are two unknown coefficients in the expression of
higher spin $4$ current.
Moreover, the spin $1$ currents in the numerators of (\ref{generalcoset})
are described with the double index notation.
Each index is a vector representation of $SO(2N)$
and because of antisymmetric property of these
spin $1$ currents, the number of independent fields
is given by $\frac{1}{2} \left[ (2N)^2 -2N \right] = N(2N-1)$.
In order to obtain the description of above free adjoint fermions,
one should write down the spin $1$ currents with a single
adjoint index.
It is known that the real free fermions transforming
in the adjoint representation of $SO(2N)$
realize the affine Kac Moody algebra for the critical level.
It is equivalent to the theory of $\frac{1}{2} 2N (2N-1) =
N(2N-1)$ free fermions \cite{DMS}.

Before one considers the adjoint free fermion description,
one should obtain the higher spin $4$ current 
from the spin $1$ currents living in the numerator factors of
(\ref{generalcoset}) and having a single adjoint index.
The higher spin $4$ current is $SO(2N)$ singlet field \cite{BS}.
Then one should have a quantity contracted with the quartic
terms in the above spin $1$ currents. This is known as
$d$ symbol which is completely symmetric $SO(2N)$ invariant tensor
of rank $4$.
In calculation of any OPE between the higher spin currents,
one should use various contraction identities between the above
$d$ symbol and the structure constant $f$.
Recall that in the defining OPE between the spin $1$ currents,
the structure constant $f$ symbol appears.
As far as I know, there are no known identities between
$f$ symbol and $d$ symbol except of $f f$ contraction
in the literature.
This is one of the reasons why 
the double index notation in \cite{Ahn1202} is used.

In this paper,
one starts with the definition of $d$ symbol which is given by
one half times
the trace over six quartic terms in the $SO(2N)$ generators. 
When one meets the relevant contraction identities in the
calculation of any OPE, one can try to obtain the
tensorial structure in the right-hand sides of these
identities. Of course in each term, there should be present
$N$ dependence coefficients explicitly.
The tensorial structure in terms of
multiple product of
$f$ symbol, $d$ symbol and the symmetric $SO(2N)$ invariant tensor
$\delta$ of rank $2$
occurs naturally during the
explicit calculation of OPE. As one applies for
$N=2,3,4$ and $5$ cases in the $SO(2N)$ generators,
one can determine the $N$ dependence coefficients explicitly.

It turns out that the higher spin $4$ current is obtained completely
except of overall normalization factor. The eigenvalue equations
of zero mode of the higher spin $4$ current acting on several primary
states can be determined explicitly. The corresponding three-point functions
can be obtained. By choosing the overall factor correctly, one
observes the standard three-point functions
in the large $N$ 't Hooft limit from the asymptotic symmetry algebra
in the $AdS_3$ bulk theory.
From the description of adjoint fermions living in the first factor
in the numerator of (\ref{generalcoset}), one obtains the well known
${\cal N}=1$ superconformal algebra generated by the spin $2$ stress energy
tensor and its superpartner, spin $\frac{3}{2}$ current.
It turns out that the higher spin $\frac{7}{2}$ current
consists of septic, quintic, cubic and linear terms in the adjoint fermions
with appropriate  derivative terms.
The ${\cal N}=2$ superconformal algebra
is realized by two adjoint fermions living in the
two numerator factors in (\ref{generalcoset}).
In this case, the higher spin $3$ current with $U(1)$ charge
$\frac{4}{3}$ is given by
the multiple product of two fermions
contracted with $d f f$ or $f f$ tensor
without any derivative terms.
Moreover, its three partners, higher spin $\frac{7}{2}, \frac{7}{2}$,
and $4$ currents are determined.

In section $2$,
the higher spin $4$ current is obtained, the three-point functions
are given and the OPE between the higher spin $4$ current and itself
is described under some constraints.

In section $3$,
the higher spin $\frac{7}{2}$ current is obtained,
and the three OPEs between this higher spin $\frac{7}{2}$ current and
the higher spin $4$ current are described using the Jacobi identities.

In section $4$,
the lowest higher spin $3$ current is obtained, and its three other
higher spin $\frac{7}{2}, \frac{7}{2}$ and $4$ currents
are obtained which can be denoted as ${\cal N}=2$ lowest higher spin
multiplet with definite $U(1)$ charge $\frac{4}{3}$. Furthermore,
another ${\cal N}=2$ lowest higher spin multiplet with
definite $U(1)$ charge $-\frac{4}{3}$ is obtained. The OPE between these
higher spin multiplets in ${\cal N}=2$ superspace
are given using the Jacobi identities.

In section $5$, we list some future directions related to this work.

In Appendices $A$-$L$,
the technical details appearing in sections $2,3$
and $4$ are given. 

\section{The coset model with arbitrary two levels $(k_1, k_2)$}

From  the
spin $1$ currents of the coset model, one constructs the
spin $2$ stress energy tensor. By generalizing the Sugawara
construction, the higher spin $4$ current is obtained from
the quartic terms in the spin $1$ currents with the $SO(2N)$
invariant tensors of ranks $4,2$. The corresponding three-point
functions of zero mode of the higher spin $4$ current with two scalars
are described. The OPE between the higher spin $4$ current and itself
for particular $k_1$ and $N$ is obtained.

\subsection{Spin $2$ current and Virasoro algebra}

The standard stress energy tensor satisfies
the following OPE \cite{BS}
\bea
T(z)\, T(w) = \frac{1}{(z-w)^4} \, \frac{c}{2}
+ \frac{1}{(z-w)^2} \, 2 T(w) + \frac{1}{(z-w)} \, \pa T(w) + \cdots.
\label{ttope}
\eea
For the coset model in (\ref{generalcoset}), the above
stress energy tensor can be obtained by
usual Sugawara construction \cite{BS}
\bea
T(z) & = & -\frac{1}{2(k_1 +2N-2)}
J^a J^a(z)
 -\frac{1}{2(k_2 +2N-2)}
 K^a K^a(z)
 \nonu \\
 & + & \frac{1}{2(k_1 +k_2 + 2N-2)}
(J^a +K^a)(J^a +K^a)(z). 
\label{stress}
\eea

The affine Kac-Moody algebra $\hat{SO}(2N)_{k_1} \oplus
\hat{SO}(2N)_{k_2}$ in (\ref{generalcoset})
is described by the following OPEs \cite{BS}
\bea
J^a(z) \, J^b(w) & = & -\frac{1}{(z-w)^2} \, k_1 \delta^{ab}
+ \frac{1}{(z-w)} \, f^{abc} J^c(w) + \cdots,
\nonu \\
K^a(z) \, K^b(w) & = & -\frac{1}{(z-w)^2} \, k_2 \delta^{ab}
+ \frac{1}{(z-w)} \, f^{abc} K^c(w) + \cdots.
\label{jjkk}
\eea
The adjoint indices $a, b, \cdots$ corresponding to
$SO(2N)$ group run over
$a, b=1, 2, \cdots, \frac{1}{2} 2N(2N-1)$.
The Kronecker delta $\delta^{ab}$ appearing in (\ref{jjkk})
is the second rank $SO(2N)$ symmetric invariant tensor.
The structure constant $f^{abc}$
is antisymmetric as usual.
The diagonal affine Kac-Moody algebra
 $
\hat{SO}(2N)_{k_1+k_2}$ in (\ref{generalcoset})
can be obtained by adding
the above two spin $1$ currents, $J^a(z)$ and $K^a(z)$.
Of course, we have $J^a(z) \, K^b(w) = + \cdots$.

The central charge appearing the above OPE (\ref{ttope})
is given by \cite{BS}
\bea
c(k_1,k_2,N) = \frac{1}{2} 2N (2N-1) \Bigg[ \frac{k_1}{(k_1 +2N-2)}
+\frac{k_2}{(k_2+2N-2)} -\frac{(k_1+k_2)}{(k_1+k_2+2N-2)}
\Bigg]. 
\label{centralcharge}
\eea
Note that the dual Coxeter number of $SO(2N)$
is equal to $(2N-2)$ and the dimension of
$SO(2N)$ is given by $\frac{1}{2} 2N (2N-1)$. 

Then the Virasoro algebra realized in the coset model
(\ref{generalcoset}) \cite{GKOplb,GKO}
is summarized by (\ref{ttope}) together with (\ref{stress}) and
(\ref{centralcharge}).

\subsection{Higher spin $4$ current }

The $28$
$SO(8)$ generators $T^a$ are given in Appendix (\ref{SO8generators}).
Then the structure constant introduced in the above is given by
\bea
f^{abc} =-\frac{i}{2} \mbox{Tr} \Bigg[ T^c T^a T^b - T^c T^b T^a \Bigg].
\label{fdef}
\eea
Then one obtains $[T^a, T^b] = i f^{abc} T^c$.

The totally symmetric $SO(2N)$ invariant tensor of rank $4$
is defined as \cite{dMMP,Gershun}
\bea
T^a T^b T^c + T^a T^c T^b + T^c T^a T^b + T^b T^a T^c 
+ T^b T^c T^a + T^c T^b T^a = d^{abcd} \, T^d.
\label{dabcdexp}
\eea 
That is, one can express the $d$ tensor as
\footnote{\label{dabc} One can consider the rank $3$ tensor
  as $d^{abc} = \frac{1}{2}
  \mbox{Tr} \Bigg[ T^a T^b T^c + T^b T^a T^c \Bigg]$ which is
identically zero.}
         {\small
           \bea
d^{abcd} = \frac{1}{2} \mbox{Tr} 
\left[ T^d T^a T^b T^c + T^d T^a T^c T^b + T^d T^c T^a T^b + 
T^d T^b T^a T^c 
+ T^d T^b T^c T^a + T^d T^c T^b T^a\right].
\label{dabcdExp} 
\eea
}
Note that one uses $\mbox{Tr} (T^a T^b)= 2 \delta^{ab}$.

One obtains the product of the structure constants
\bea
f^{abc} f^{abd} = 2(2N-2) \delta^{cd},
\label{ffrelation}
\eea
and the triple product leads to 
\bea
f^{adb} f^{bec} f^{cfa} = -(2N-2) f^{def}.
\label{fffproduct}
\eea
Furthermore, one obtains the following nontrivial
triple product between $d$ tensor (\ref{dabcdExp}) and $f$ tensor
(\ref{fdef})
\bea
d^{adeb} f^{bfc} f^{cga}  & = & -\frac{4}{3} (N-1) d^{defg}
+ 4 \delta^{df} \delta^{eg} + 4 \delta^{dg} \delta^{ef}
- 8 \delta^{de} \delta^{fg}
\nonu \\
& - & \frac{1}{3}(2N-5) f^{dfh} f^{heg}
-\frac{1}{3}(2N-5) f^{dgh} f^{hef}.
\label{dffrelation}
\eea
By multiplying $f^{dfh}$ into
(\ref{dffrelation}) and rearranging the indices, one obtains
with (\ref{fffproduct})
\bea
d^{abcf} f^{agd} f^{bde} f^{che} = 2(2N^2-7N+11) f^{fgh}. 
\label{dfffrelation}
\eea
For the index condition $f=d$ in (\ref{dffrelation})
together with the identity (\ref{ffrelation}), one obtains
\bea
d^{aabc} =2(4N-1) \delta^{bc}.
\label{daabc}
\eea
Note that this behavior is different from the one of unitary
case where the trivial result $d^{aabc} =0$ arises \cite{BBSSfirst}. 
One also has
\bea
d^{abcd} d^{abce} = 12 \left[ N(2N-1) + 2 \right] \delta^{de}.
\label{ddproduct}
\eea

Let us describe how one can obtain the higher spin current
with the help of $d$ tensor we introduced.
For the second rank $SO(2N)$ invariant symmetric tensor $\delta^{ab}$,
one describes the stress energy tensor in (\ref{stress}).
According to the observation of footnote \ref{dabc},
there is no nontrivial third rank $SO(2N)$ invariant symmetric
tensor $d^{abc}$. Then the next nontrivial higher spin current
can be constructed from the fourth rank $SO(2N)$ invariant symmetric
tensor $d^{abcd}$ (\ref{dabcdexp}).

Let us consider the following higher spin $4$ current,
along the line of \cite{BBSSfirst,BBSSsecond,BS},
\bea
W^{(4)}(z) & = & d^{abcd} \Bigg[ 
A_1 \, J^a J^b J^c J^d + A_2 \, J^a J^b J^c K^d + A_3 \,
J^a J^b K^c K^d \nonu \\
& + &  A_4 \, J^a  K^b K^c K^d
+  A_5 \, K^a K^b K^c K^d \Bigg](z)
+  \Bigg[ A_6 \, \pa J^a \pa J^a +A_7 \, \pa^2 J^a J^a 
\nonu \\
& + &  
A_8 \, \pa K^a \pa K^a
+ A_9 \, \pa^2 K^a K^a
+ A_{10} \, \pa J^a \pa K^a + A_{11} \, \pa^2 J^a K^a  
\nonu \\
& + & A_{12} \, J^a \pa^2 K^a + A_{13} \, f^{abc} J^a \pa J^b K^c 
+ A_{14} \, f^{abc} J^a K^b \pa K^c
+ A_{15} \, J^a J^a J^b J^b \nonu \\
& + & A_{16}  \, K^a K^a K^b K^b  +  A_{17} \, J^a J^a K^b K^b 
+ A_{18} \, J^a J^a J^b K^b + A_{19} \, J^a K^a K^b K^b 
\nonu \\
& + &  A_{20} \, J^a J^b K^a K^b\Bigg](z). 
\label{w4}
\eea
One should obtain the twenty relative $(k_1,k_2,N)$-dependent
coefficients. 
The first five
     quartic terms in (\ref{w4}) can be easily understood in the
sense that they are the only possible terms from each spin $1$
current, $J^a(z)$ and $K^a(z)$
using the $d^{abcd}$ tensor.
The next seven derivative terms in (\ref{w4})
can be seen from the second derivative of stress energy tensor
$\pa^2 T(z)$. The remaining eight terms can arise in $T T(z)$.

First of all, the higher spin $4$ current should
have the regular terms with the diagonal spin $1$ current in the
coset model as follows \cite{BBSSfirst,BBSSsecond,BS}:
\bea
J^{\prime a}(z) \, W^{(4)}(w) & = & +\cdots, \qquad
J^{\prime a}(z) \equiv (J^a +K^a)(z).
\label{jpw}
\eea

Let us calculate the OPEs between the diagonal spin $1$
current and the twenty terms in (\ref{w4})
in order to use the condition (\ref{jpw}).
One can perform the various OPEs by following the procedures
done in the unitary case \cite{Ahn1111}. Let us focus on the $A_1$ term
in (\ref{w4}) which has the regular OPE with $K^a(z)$.
Then the equations $(2.22)$, $(2.23)$ and $(2.24)$
of \cite{Ahn1111} can be used.
For example, the equation $(2.24)$ of \cite{Ahn1111}
provides the information of the OPE between the $J^{\prime a}(z)$
and the above $A_1$ term. 
Using the relations (\ref{dfffrelation}) and (\ref{ffrelation}),
one can simplify the fourth-order pole in $(2.24)$ of \cite{Ahn1111}
which was given by $f^{abf} f^{fci} d^{bcde} f^{geh} f^{idg} J^h(w)$.

It turns out that we are left with $J^a(w)$ with
$N$-dependent $SO(2N)$ group theoretical factor.
The third-order pole,
{\small
\bea
f^{abf} d^{bcde} (f^{hdg} f^{fch} J^g J^e+
f^{heg} f^{fch} J^d J^g +f^{heg} f^{fdh} J^c J^g)(w) + f^{acf} d^{bcde}
f^{geh} f^{fdg} J^b J^h(w),
\label{severalterms}
\eea}
can be simplified with the help of
(\ref{dfffrelation}).
We are left with $f^{abc} J^b J^c(w)$ in (\ref{severalterms})
with
$N$ dependent coefficient factor
which is proportional to $\pa J^a(w)$.
Finally the second-order pole,
\bea
&& -4k_1 d^{abcd} J^b J^c J^d(w) \nonu \\
&& + d^{bcde} (f^{abf} f^{fcg}  J^g J^d J^e
+   f^{abf} f^{fdg}  J^c J^g J^e 
+  f^{abf} f^{feg}  J^c J^d J^g \nonu \\
&& +
f^{acf} f^{fdg}  J^b J^g J^e
+  f^{acf} f^{feg}  J^b J^d J^g
+  f^{adf} f^{feg}  J^b J^c J^g)(w),
\label{secondexpression}
\eea
can be simplified further together with
(\ref{dffrelation}).

Then we obtain the final OPE as follows:
\bea
&& J^{\prime a}(z) \, d^{bcde} J^b J^c J^d J^e(w)  = 
\frac{1}{(z-w)^4} \, 2(2N-2)(4N^2-14N+22) J^a(w) \nonu \\
&& - 
\frac{1}{(z-w)^3} \, 2(4N^2-14N+22) f^{abc} J^b J^c(w)
\nonu \\
&& +
\frac{1}{(z-w)^2} \,
\Bigg[ -(4k_1 +8(N-1)) d^{abcd} J^b J^c J^d
   - (12+(2N-2)(2N-5)) f^{abc} \pa J^b J^c
  \nonu \\
  && + (12+(2N-2)(2N-5)) f^{abc}  J^b \pa J^c
  \Bigg](w) +\cdots.
\label{eqeq}
\eea
There is no first order pole in the above (\ref{eqeq}).
One can check the second order pole in (\ref{eqeq})
from (\ref{secondexpression}).

Let us consider the $A_2$ term in (\ref{w4}) where
there exists $K^d(z)$ dependence. 
Starting from the $(2.23)$ and $(2.21)$ of \cite{Ahn1111} with the relations
(\ref{dfffrelation}) and (\ref{dffrelation}), 
one can simplify the third order pole, $f^{acf} fd^{bcde} f^{geh}
f^{fdg} J^h K^b(w)$, as $f^{abc} J^b K^c(w)$
with $N$ dependent factor.
Similarly, the second order pole,
\bea
&& -3 k_1 d^{abcd} J^c J^d K^b(w) - k_2 d^{abcd} J^b J^c J^d(w)
\nonu \\
&& + d^{bcde} (f^{acf} f^{fdg}  J^g J^e K^b
+   f^{acf} f^{feg}  J^d J^g K^b 
+  f^{adf} f^{feg}  J^c J^g K^b)(w), 
\label{intersecond}
\eea
can be 
simplified in terms of several independent terms.
It turns out that in this case also there are no first order poles.

Therefore, one obtains the following OPE corresponding to
$A_2$ term
\bea
&& J^{\prime a}(z) \, d^{bcde} J^b J^c J^d K^e(w)  = 
\frac{1}{(z-w)^3} \, (4N^2-14N+22) f^{abc} J^b K^c(w)
\nonu \\
&& +
\frac{1}{(z-w)^2} \,
\Bigg[ -(3k_1 +4(N-1)) d^{abcd} J^b J^c K^d
   - k_2 d^{abcd} J^b J^c J^d + 12 J^b J^b K^a
   \nonu \\
   &  & -12 J^a J^b K^b 
   - 12 f^{abc} \pa J^b K^c
   + (2N-5) f^{abc} f^{cde}  J^b  J^e K^d
  \Bigg](w) +\cdots.
\label{eqeq1}
\eea
One can see the second order pole in (\ref{eqeq1}) from
(\ref{intersecond}).

Let us consider the $A_3$ term in (\ref{w4}).
From the $(2.22)$ of \cite{Ahn1111},
one has the relevant OPEs.
For example, the second order pole,
{\small
\bea
(- 2k_1 d^{abcd} J^d K^b K^c + f^{adf}f^{feg} d^{bcde} J^g K^b K^c
- 2k_2 d^{abcd} K^d J^b J^c + f^{adf}f^{feg} d^{bcde} K^g J^b J^c)(w),
\label{inter2}
\eea
}
can be reexpressed in terms of various independent terms
with the help of the identity (\ref{dffrelation}).
It turns out that the relevant OPE coming from (\ref{inter2})
can be summarized as
\bea
&& J^{\prime a}(z) \, d^{bcde} J^b J^c K^d K^e(w)  = 
\frac{1}{(z-w)^2} \,
\Bigg[ -(2k_1 +\frac{4}{3}(N-1)) d^{abcd} J^b K^c K^d
  \nonu \\
&& -  (2k_2 +\frac{4}{3}(N-1)) d^{abcd} J^b J^c K^d  
+ 8 J^b K^a K^b
 + 4 f^{abc} \pa J^b  K^c
\nonu \\
&& -  8 J^a K^b K^b -4 f^{abc} J^b \pa K^c
-\frac{1}{3} (2N-5) f^{abc} f^{cde} J^d K^e K^b
\nonu \\
&& -  \frac{1}{3} (2N-5) f^{abc} f^{cde} J^d K^b K^e
- 8 J^b J^b K^a
+ 8 J^a J^b K^b
\nonu \\
& &
-\frac{1}{3} (2N-5) f^{abc} f^{cde} J^e J^b K^d
 -  \frac{1}{3} (2N-5) f^{abc} f^{cde} J^b J^e K^d\Bigg](w)
+ \cdots.
\label{eqeq2}
\eea
It is useful to realize that this OPE
remains the same after the exchange of $J^a(w)$ and $K^a(w)$
together with $k_1 \leftrightarrow k_2$.
The left hand side is invariant under this transformation
because the $d$ tensor is totally symmetric.
The twelve terms in the second order pole can be divided into
two groups and each of them has their own counterpart.

It is straightforward to complete this calculation step by step.
We summarize the remaining $17$ OPEs in Appendix $B$.
Then we have the complete expressions in (\ref{eqeq}), (\ref{eqeq1}),
(\ref{eqeq2}), and Appendix (\ref{rem17}).

The higher spin $4$ current should transform as a primary
field under the stress energy tensor (\ref{stress}).
According to the previous regular condition (\ref{jpw}),
the diagonal spin $1$ current $J^{\prime a}(z)$
does not have any singular terms in the OPE with
the higher spin $4$ current $W^{(4)}(w)$ after we use the results of
Appendix (\ref{linearequationforjprime}).
Then there are no singular terms in the OPE
between the stress energy tensor in the denominator of the coset model
(\ref{generalcoset}) and the higher spin $4$ current
because the former is given by $J^{\prime a} J^{\prime a}(z)$.
The singular terms can arise from the OPE
between the stress energy tensor in the numerator of the coset model
and the higher spin $4$ current.
Therefore, one should have the following condition 
\cite{BBSSfirst,BBSSsecond,BS}
\bea
\hat{T}(z) \, W^{(4)}(w)\Bigg|_{\frac{1}{(z-w)^n}, \,
  n=3,4,5,6} & = & 0.
\label{primaryother}
\eea
Here the stress energy tensor
in the numerator is described by
\bea
\hat{T}(z) \equiv  -\frac{1}{2(k_1 +2N-2)}
J^a J^a(z)
 -\frac{1}{2(k_2 +2N-2)}
 K^a K^a(z). 
\label{numstress}
\eea
Of course, the higher spin $4$ current has the standard OPE
(the second and first order poles)
with stress energy tensor (\ref{stress}) as usual.

Let us calculate the OPE between
the stress energy tensor
(\ref{numstress}) and the $A_1$ term in (\ref{w4}).
First of all, because the $A_1$ term does not contain the
$K^a(w)$ spin $1$ current, one can consider the OPE
between the first term of (\ref{numstress}) and the
$A_1$ term. It is known that
the spin $1$ current $J^a(w)$ transforms as a primary
field under the
the first term of (\ref{numstress}) (i.e., stress energy tensor
in the first factor of the numerator).
Then one should obtain the OPE  $J^b(z)$
$d^{bcde} J^c J^d J^e(w)$ and this turns out that
there exists a nontrivial second order pole given by
$-3 k_1 (8N-2) J^c J^c(w)$
where
the identity (\ref{daabc}) is used.
Note that the structure constant term vanishes due to the presence of
$d^{bcde}$.
Furthermore, one should calculate
the OPE
between the above stress energy tensor and
the previous expression $d^{bcde} J^c J^d J^e(w)$
where the order of the singular terms is greater than
$2$.
Then we are left with $-3 k_1 (8N-2) J^c J^c(w)$
by combing the contribution $-2 k_1 (8N-2) J^c J^c(w)$ from
the contraction between the stress energy tensor and
$J^c(w)$
and the contribution $- k_1 (8N-2) J^c J^c(w)$ from the OPE
between the stress energy tensor and  $d^{bcde} J^d J^e(w)$.
Therefore, the final total contribution is summarized by
$-6 k_1 (8N-2) J^c J^c(w)$ and we present this OPE as follows:
\bea
\hat{T}(z) \, d^{bcde} J^b J^c J^d J^e(w) & = &
-\frac{1}{(z-w)^4} \, 12 k_1 (4N-1) J^a J^a(w)
+{\cal O}(\frac{1}{(z-w)^2}).
\label{tjjjj}
\eea
This result in (\ref{tjjjj})
is different behavior from the corresponding OPE in the
unitary case because in the latter, there is no contribution
from the fourth order pole because the above
$d^{aabc}$ tensor for the $SU(N)$ group 
vanishes \cite{BBSSfirst}.

Let us move on the $A_2$ term in (\ref{w4}).
In this case, the spin $1$ current $K^d(w)$
is present.
However, the contribution in the higher singular terms of
the stress energy tensor coming from the second term
of (\ref{numstress}) vanishes.
Then one can calculate the OPE between the
stress energy tensor in the first factor of the numerator
and the $A_2$ term.
By using the previous procedure one can
obtain the contribution $-2k_1(8N-2) J^c K^c(w)$ from the
contraction with $J^b(w)$ current
and  the contribution $-k_1(8N-2) J^c K^c(w)$ from the
contraction with other remaining factor $d^{bcde} J^c J^d K^e(w)$. 
By adding these two, one obtains the following OPE 
\bea
\hat{T}(z) \, d^{bcde} J^b J^c J^d K^e(w) & = &
-\frac{1}{(z-w)^4} \, 6 k_1 (4N-1) J^a K^a(w)
+{\cal O}(\frac{1}{(z-w)^2}).
\label{tjjjk}
\eea

Now let us describe the contribution from
$A_3$ term in (\ref{w4}) where
the quadratic $K^c K^d(w)$ appears. 
In this case,
one should also calculate the contribution from
the stress energy tensor in the second term in (\ref{numstress}).
As done before,
the contribution from the contraction with
$J^b(w)$ spin $1$ current is given by
$-k_1(8N-2) K^c K^c(w)$. Similarly
the contribution from the contraction with
the remaining factor is given by
$-k_2(8N-2) J^c J^c(w)$.
Then we are left with 
\bea
\hat{T}(z) \, d^{bcde} J^b J^c K^d K^e(w) & = &
-\frac{1}{(z-w)^4} \Bigg[ 2 k_2 (4N-1) J^a J^a + 2k_1
  (4N-1) K^a K^a \Bigg](w)
\nonu \\
& + & {\cal O}(\frac{1}{(z-w)^2}).
\label{tjjkk}
\eea
One also sees the symmetry under the
transformation $J^a(z) \leftrightarrow K^a(z)$ and $k_1 \leftrightarrow
k_2$.

It is straightforward to determine other remaining
calculation step by step.
We summarize the remaining $17$ OPEs in Appendix $C$.
Then we are left with (\ref{tjjjj}), (\ref{tjjjk}), (\ref{tjjkk}),
and Appendix (\ref{17rem}).

Now one can determine the undetermined coefficient functions
$A_1, A_2, \cdots, A_{20}$ appearing in the higher spin $4$ current in
(\ref{w4}). The twenty three linear equations are given in Appendix 
(\ref{linearequationforjprime})
explicitly. The eight linear equations are given in Appendix 
(\ref{linearequationfort}). By
solving them, one obtains the final expressions in Appendix $D$.
They depend on $k_1, k_2$ and $N$.
The corresponding coefficients for $k_1=1$ are presented in Appendix $E$.
Appendix $F$ corresponds to the case where $k_1=2N-2$.

\subsection{Three-point functions \cite{GH}
  with two scalars where $k_1=1$}

The zero modes of the current satisfy the commutation
relations of the underlying finite dimensional Lie algebra $SO(2N)$.
For the state $|(v;0)>$, $ T^a$ corresponds to $i K_0^a$
and for the state $|(0;v)>$, $ T^a$ corresponds to $ i J_0^a$
as follows:
\bea
|(v;0)>:  \qquad  T^a \leftrightarrow  i K_0^a, \qquad
|(0;v)>:  \qquad  T^a  \leftrightarrow i J_0^a.
\label{generator}
\eea
Note that from the defining equation of the OPEs (\ref{jjkk}),
one obtains
\bea
[J^a_m, J^b_n] & = & -k_1 m \delta^{ab} \delta_{m+n,0} + f^{abc} J^c_{m+n},
[K^a_m, K^b_n]  = 
          -k_2 m \delta^{ab} \delta_{m+n,0} + f^{abc} K^c_{m+n}.
\label{comm}
\eea
In (\ref{comm}), the central terms for the zero modes vanish. 
Recall that our generators for the $SO(2N)$
satisfy $[T^a, T^b] = i f^{abc} T^c$ \cite{BS}.

The large $N$ 't Hooft limit is described
as \cite{Ahn1106,GV1106}
\bea
N, k_2 \rightarrow \infty, \qquad
\la \equiv \frac{2N}{2N-2+k_2} \qquad
\mbox{fixed}.
\label{limit}
\eea
The presence of numerical value $-2$ in the denominator
of (\ref{limit}) is not important
under the large $N$ 't Hooft limit \cite{CHR1209}.

Compared to the large ${\cal N}=4$ holography in
\cite{GG1305,AK1506,AKP1510}
where one can obtain the eigenvalue equations from the
several low $N$ values inside the package of \cite{Thielemans},
one should analyze both
the coefficients and zero modes of the twenty terms in
higher spin $4$ current in order to obtain the
corresponding eigenvalue equations. 

\subsubsection{Eigenvalue equation 
  of the zero mode of the higher
  spin $4$ current acting on the state $|(0;v)>$  }

Let us consider the eigenvalue equation of the zero mode
of the $A_1$ term of the higher spin $4$ current
in (\ref{w4}) acting on the primary state
$(0;v)$
\bea
d^{abcd} (J^a J^b J^c J^d)_0 |(0;v)>.
\label{some}
\eea
Using the fact that the zero mode is nothing but
the product of each zero mode but the ordering is reversed
\cite{BBSSfirst,BBSSsecond},
the above expression (\ref{some}) becomes
\bea
d^{abcd} (J^d J^c J^b J^a)_0 |(0;v)>.
\label{exp1}
\eea
Note that the ground state transforms as a vector representation
with respect to $J_0^a$ while the zero mode $K_0^a$ has
vanishing eigenvalue equation \cite{GH}
\bea
K_0^a |(0;v)> =0.
\label{vanishingK}
\eea
Then the above expression (\ref{exp1}) becomes
\bea
\frac{1}{2N} d^{abcd} (-i)^4 \mbox{Tr} (T^d T^c T^b T^a)|(0;v)>.
\label{exp2}
\eea
In order to use the previous identity in (\ref{dabcdexp}),
one can express the above $A_1$ term as follows:
\bea
\frac{1}{6} d^{abcd} (J^a J^b J^c J^d + J^b J^c J^a J^d + J^b J^a J^c J^d
+ J^c J^a J^b J^d + J^a J^c J^b J^d + J^c J^b J^a J^d),
\label{equequ}
\eea
due to the symmetric property of $d$ tensor.
Then the equivalent expression corresponding to
(\ref{exp2}) with (\ref{equequ}) can be written in terms of
{\small
\bea
&& \frac{1}{2N} \frac{1}{6} d^{abcd} \mbox{Tr} (T^d T^c T^b T^a
+ T^d T^a T^c T^b +T^d T^c T^a T^b  + T^d T^b T^a T^c  +
T^d T^b T^c T^a + T^d T^a T^b T^c )\nonu \\
&& \times |(0;v)>.
\label{exp3}
\eea}
The reason why there exists the extra $\frac{1}{2N}$ is that
one should have the eigenvalue not the trace.
Using the identity (\ref{dabcdexp}), one can reexpress
(\ref{exp3})
as
\bea
&& \frac{1}{2N} \frac{1}{3} d^{abcd} d^{cbad} =
\frac{1}{2N} \frac{1}{3} 12 \left[N(2N-1)+2 \right] \delta^{aa}
= \frac{1}{2N} \frac{1}{3} 12 \left[N(2N-1)+2 \right]
\frac{1}{2} 2N (2N-1)
\nonu \\
&& \rightarrow 8N^3.
\label{eigen}
\eea
Here the identity (\ref{ddproduct}) is used and
we take the large $N$ limit at the last result in (\ref{eigen}).

One can analyze the other $19$ terms in (\ref{w4}).
Among them, the $16$ terms which have the $K^a(z)$ spin $1$ current
do not contribute to the eigenvalue equation
because one can take the zero mode and change the ordering of the
zero modes as in (\ref{exp1}). Then one can move the rightmost zero mode
$K_0^a$ to the right and use the previous condition (\ref{vanishingK}).
On the other hand, the remaining $A_6, A_7$ and $A_{15}$ terms
can contribute to the eigenvalue equation.

The zero mode of the $A_6$ term of the higher spin $4$ current
acting on the primary state $(0;v)$
is
\bea
(\pa J^a \pa J^a)_0 |(0;v)> = (\pa J^a)_0 (\pa J^a)_0 |(0;v)>
= (-J^a_0)(-J^a_0) |(0;v)> = J^a_0 J^a_0 |(0;v)>,
\label{der}
\eea
where the zero mode of $\pa J^a$ in (\ref{der})
can be obtained from the usual
mode expansion and is given by the zero mode of $-J^a$.
Now using the correspondence (\ref{generator}), the above
expression leads to
\bea
\frac{1}{2N} \mbox{Tr} (i T^a i T^a) |(0;v)> =
-\frac{1}{2N} 2 \delta^{aa} = -\frac{1}{2N} 2 \frac{1}{2} 2N(2N-1)
\rightarrow -2N,
\label{limita6}
\eea
where the extra factor $\frac{1}{2N}$ is considered as in
(\ref{exp3}) and the large $N$ limit is taken.

Now the final contribution from the
zero mode of 
 the $A_7$ term of the higher spin $4$ current
acting on the primary state $(0;v)$
is given by
\bea
(\pa^2 J^a J^a)_0  |(0;v)> = J^a_0 (\pa^2 J^a)_0  |(0;v)>
=  J^a_0 2 J^a_0  |(0;v)>,
\label{derder}
\eea
where the zero mode of $\pa^2 J^a$ in (\ref{derder})
can be obtained from the usual
mode expansion also and is given by the zero mode of $2 J^a$.
Therefore, one can follow the previous description.
It turns out that
\bea
2 \frac{1}{2N} \mbox{Tr} (i T^a i T^a) |(0;v)> =
 -2 \frac{1}{2N} 2 \frac{1}{2} 2N(2N-1)
\rightarrow -4N.
\label{limita7}
\eea

For the $A_{15}$ term,
one has
the eigenvalue equation 
\bea
(J^a J^a J^b J^b)_0 |(0;v)> = \delta^{ab}
\delta^{cd} (J^a J^b J^c J^d)_0  |(0;v)>.
\label{above}
\eea
By following the procedure in the $A_1$ term,
one obtains that the above (\ref{above}) can be written as 
\bea
\frac{1}{2N} \frac{1}{3} \delta^{ab} \delta^{cd} d^{cbad}
= \frac{1}{2N} \frac{1}{3} 2(4N-1) \frac{1}{2} 2N (2N-1) \rightarrow
\frac{8}{3} N^2,
\label{Expp}
\eea
where the identity (\ref{daabc}) is used in (\ref{Expp}).
Furthermore, the $A_{15}$ term itself behaves as $N^0$ in Appendix
(\ref{coeffA}).
Then there is no contribution at the leading order approximation.

By combining (\ref{eigen}), (\ref{limita6}) and (\ref{limita7})
with the corresponding coefficients in the large $N$ limit of Appendix
(\ref{coeffA}),
the zero mode eigenvalue equation leads to
\bea
W_0^{(4)} |(0;v)> & = & \Bigg[ 8N^3 A_1 +(-2N)
  \left( N^2 \frac{12  (2 \lambda -9) 
  }{5  (2 \lambda -3)} A_1 \right)  +(-4N) \left(
  -N^2 \frac{8  (2 \lambda -9)}
  {5  (2 \lambda -3)} A_1
  \right)
  \Bigg] \nonu \\
& \times & |(0;v)> 
 =  N^3 \Bigg[ \frac{96  (\lambda -2)}{5 (2 \lambda -3)} \Bigg]
 A_1 |(0;v)>.
\label{eigenexpressionov}
\eea
One can also calculate the same eigenvalue equation at finite
$N$ and $k_2$ corresponding to (\ref{eigenexpressionov})
which will appear later.

\subsubsection{Eigenvalue equation of the zero mode of the higher
  spin $4$ current acting on the state $|(v;0)>$  }

Let us describe the eigenvalue equation of the zero mode
of the $A_1$ term of the higher spin $4$ current
in (\ref{w4}) acting on the primary state
$(v;0)$
\bea
d^{abcd} (J^a J^b J^c J^d)_0 |(v;0)> =d^{abcd} J^d_0 J^c_0 J^b_0 J^a_0
|(v;0)>.
\label{exp4}
\eea
Note that the ground state transforms as a vector representation
with respect to $K_0^a$ and  the singlet condition for the primary
state $(v;0)$ can be described as \cite{GH}
\bea
(J^a_0+K_0^a) |(v;0)> =0.
\label{singlet}
\eea
Then the above expression (\ref{exp4}) is equivalent to
\bea
- d^{abcd} J^d_0 J^c_0 J^b_0 K^a_0
|(v;0)> = - d^{abcd} K^a_0 J^d_0 J^c_0 J_0^b
|(v;0)>,
\label{exp6}
\eea
where the relation (\ref{singlet}) is used
and the zero mode $K^a_0$ is moved to the left.
Now the singlet condition is applied to the rightmost $J_0^b$
and we are left with
\bea
d^{abcd} K^a_0 J^d_0 J^c_0 K_0^b
|(v;0)> = d^{abcd} K^a_0 K_0^b J^d_0 J^c_0 
|(v;0)>.
\label{exp8}
\eea
One can further take the previous steps and obtains
\bea
  d^{abcd} K^a_0 K_0^b  K^c_0 K^d_0
|(v;0)>.
\label{exp5}
\eea
Then using the correspondence (\ref{generator}),
the above expression (\ref{exp5}) becomes
\bea
\frac{1}{2N} d^{abcd} (-i)^4 \mbox{Tr} (T^a T^b T^c T^d)|(0;v)>,
\label{exp7}
\eea
which leads to the previous eigenvalue in (\ref{eigen}) \footnote{
The eigenvalue equation of the zero mode
of the $A_2$ term of the higher spin $4$ current
in (\ref{w4}) acting on the primary state
$(v;0)$
can be written as
\bea
d^{abcd} (J^a J^b J^c K^d)_0 |(v;0)> =d^{abcd} K^d_0 J^c_0 J^b_0 J^a_0
|(v;0)>,
\label{jjjk}
\eea
which is equivalent to (\ref{exp6}) with an extra minus sign due to the
symmetric property of the $d$ symbol.
Then we are left with the fact that
the relation (\ref{jjjk}) is equal to
the previous result (\ref{exp7})
with minus sign.

The eigenvalue equation of the zero mode
of the $A_3$ term of the higher spin $4$ current
in (\ref{w4}) acting on the primary state
$(v;0)$
leads to
\bea
d^{abcd} (J^a J^b K^c K^d)_0 |(v;0)> =d^{abcd} K^d_0 K^c_0 J^b_0 J^a_0
|(v;0)>,
\label{JJKK}
\eea
which is equal to (\ref{exp8}) and this (\ref{JJKK}) becomes
the expression (\ref{exp7}).

Similarly,
the eigenvalue equation of the zero mode
of the $A_4$ term of the higher spin $4$ current
in (\ref{w4}) acting on the primary state
$(v;0)$
can be described as
\bea
d^{abcd} (J^a K^b K^c K^d)_0 |(v;0)> =d^{abcd} K^d_0 K^c_0 K^b_0 J^a_0
|(v;0)> = - d^{abcd} K^d_0 K^c_0 K^b_0 K^a_0
|(v;0)>,
\label{jkkk}
\eea
where the singlet condition is used and the above
expression (\ref{jkkk}) leads to
(\ref{exp7}) with an extra minus sign.}.

What happens for $A_5$ term of
higher spin $4$ current in (\ref{w4})?
According to the large $N$ behavior of the coefficient
$A_5$, this coefficient behaves as $\frac{1}{N}$
in Appendix (\ref{coeffA}) and moreover
the analysis of eigenvalue equation leads to $N^3$ behavior.
Therefore, the total power for the large $N$ behavior is
given by $N^2$ and can be ignored in this approximation.

Let us move on the $A_6$ term.
The eigenvalue equation leads to
\bea
(\pa J^a \pa J^a)_0 |(v;0)>  = J^a_0 J^a_0 |(v;0)> =
-J_0^a K_0^a  |(v;0)> =K_0^a K_0^a  |(v;0)>,
\label{exp9}
\eea
where the singlet condition (\ref{singlet}) is used.
After using the correspondence (\ref{generator}), this becomes
the previous result in (\ref{limita6}).

Similarly, the $A_7$ term
eigenvalue equation gives
\bea
(\pa^2 J^a J^a)_0  |(v;0)> 
=  J^a_0 2 J^a_0  |(v;0)> = -2  J^a_0  K^a_0  |(v;0)>
=  2 K^a_0  K^a_0  |(v;0)>,
\label{exp10}
\eea
which leads to (\ref{limita7}).

For the
$A_8$ and $A_9$ terms
of the higher spin $4$
current, these coefficients behave as $N$ from Appendix (\ref{coeffA})
in the large $N$ limit and
the corresponding eigenvalues  behave as
$N$. Then the total power of the large $N$ behavior
is given by $2$ and these terms can be ignored at the leading order
calculation \footnote{
Let us describe the
next $A_{10}$ term of the higher spin $4$ current
in (\ref{w4}).
One obtains
\bea
(\pa J^a \pa K^a)_0 |(v;0)>  = K^a_0 J^a_0 |(v;0)> =
-K_0^a K_0^a  |(v;0)>,
\label{derjderk}
\eea
where this (\ref{derjderk})
is equivalent to the previous relation (\ref{exp9}) with
an extra minus sign.
We can also calculate the eigenvalue equation
for the $A_{11}$ term
\bea
(\pa^2 J^a K^a)_0  |(v;0)> 
=  K^a_0 2 J^a_0  |(v;0)> = -2  K^a_0  K^a_0  |(v;0)>.
\label{derderjk}
\eea
This (\ref{derderjk})
is equivalent to (\ref{exp10}) with an extra minus sign.

One can continue to
calculate the eigenvalue equation corresponding to
$A_{12}$ term as follows:
\bea
( J^a \pa^2 K^a)_0  |(v;0)> 
=  2 K^a_0  J^a_0  |(v;0)> = -2  K^a_0  K^a_0  |(v;0)>.
\label{jderderk}
\eea
Then this (\ref{jderderk})
is the same contribution from $A_{11}$ term. }.

Let us consider
$A_{13}$ term of the higher spin $4$ current in (\ref{w4}). 
One can easily see that there exists
the relation
\bea
f^{abc} J^a \pa J^b K^c (z) &= &
J^a J^b J^a K^b(z) - J^a J^a J^b K^b(z),
\label{fjjk}
\eea
by writing the derivative term as the commutator of normal ordered
product.
Then the zero mode of this expression (\ref{fjjk})
is given by
\bea
(K^b_0 J^a_0 J^b_0 J^a_0 - K^b_0 J^b_0 J^a_0 J^a_0)  |(v;0)>
= -(K^b_0 K^a_0 K^b_0 K^a_0 - K^b_0 K^a_0 K^a_0 K^b_0)  |(v;0)>.
\label{this}
\eea
Then this (\ref{this}) becomes
\bea
-\frac{1}{2N} (-i)^4 \mbox{Tr}
(T^b T^a T^b T^a - T^b T^a T^a T^b) |(v;0)>.
\label{that}
\eea
Furthermore, this (\ref{that}) will reduce to
{\small
\bea
-\frac{1}{2N} (-i)^4  i f^{bac} \mbox{Tr} (T^b T^a T^c )  |(v;0)>=
-\frac{1}{2N} (-i)^4  i f^{bac} \frac{1}{2} \mbox{Tr}
(T^b T^a T^c -T^b T^c T^a)  |(v;0)>.
\label{Exp1}
\eea
}
One can use the identity (\ref{fdef})
and obtains, together with (\ref{ffrelation}),
\bea
-\frac{1}{2N} (-i)^4  i f^{bac} \frac{1}{2}
2 i f^{bac}
= \frac{1}{2N} 2(2N-2) \frac{1}{2} 2N (2N-1)
\rightarrow 4N^2.
\label{Exp2}
\eea

Let us focus on the $A_{14}$ term.
One has the relation
\bea
f^{abc} J^a  K^b \pa K^c (z) &= &
J^a K^b K^a K^b(z) - J^a K^b K^b K^a(z).
\label{fjkk}
\eea
The zero mode of (\ref{fjkk}) can be described as
\bea
(K^b_0 K^a_0 K^b_0 J^a_0 -K^a_0 K^b_0 K^b_0 J^a_0)
|(v;0)> = -(K^b_0 K^a_0 K^b_0 K^a_0 -K^a_0 K^b_0 K^b_0 K^a_0)
|(v;0)>.
\label{there}
\eea
Then this (\ref{there}) becomes
\bea
-\frac{1}{2N} (-i)^4 \mbox{Tr}
(T^b T^a T^b T^a - T^a T^b T^b T^a) |(v;0)>.
\label{here}
\eea
Furthermore, this (\ref{here}) will reduce to
\bea
-\frac{1}{2N} (-i)^4  i f^{bac} \mbox{Tr} ( T^c T^b T^a  )  |(v;0)>,
\label{fiinal}
\eea
by combining the first two generators.
This (\ref{fiinal}) is equivalent to (\ref{Exp1}) and (\ref{Exp2}).

Are there any contributions from the $A_{15}$-$A_{20}$ terms?
These coefficients
behave as $N^0$, $\frac{1}{N^2}$, $\frac{1}{N}$, $N^0$, $\frac{1}{N}$
and $N^0$ respectively from Appendix (\ref{coeffA}). 
There are no contributions.
Then one obtains the final eigenvalue equation as follows:
\bea
W_0^{(4)} |(v;0)> & = & \Bigg[
  8N^3 A_1 -8N^3 \left(\frac{4  \lambda }{(\lambda -1)} \right) A_1 +
  8N^3 \left(\frac{12  \lambda ^2}{(\lambda -1) (2 \lambda -3)} \right)
  A_1 \nonu \\
  & - & 8N^3 \left(\frac{8  \lambda ^3}{(\lambda -3) (\lambda -1)
    (2 \lambda -3)} \right)A_1 -2N \left( N^2 \frac{12
    (2 \lambda -9) }{5  (2 \lambda -3)} \right)A_1
  \nonu \\
  & - & 4N \left( -N^2 \frac{8  (2 \lambda -9) }{5
    (2 \lambda -3)} \right) A_1 +2N \left(
  \frac{ N^2 48  (\lambda -2) \lambda  }{5 (\lambda -1)
    (2 \lambda -3)}\right) A_1
  \nonu \\
  &+& 4N \left( - N^2 \frac{16  (\lambda -12) \lambda
  }{5 (\lambda -1) (2 \lambda -3)}\right) A_1 +
  4N \left( - N^2 \frac{16  \lambda  \left(\lambda ^2+15 \lambda +6\right) }{5 (\lambda -3) (\lambda -1) (2 \lambda -3)} \right) A_1
  \nonu \\
  &+& 4N^2 \left( - N \frac{24  \lambda  }{(\lambda -1) (2 \lambda -3)}\right) A_1 + 4N^2 \left( N \frac{48  \lambda ^2 }{(\lambda -3)
    (\lambda -1) (2 \lambda -3)}\right) A_1 
  \Bigg]  \nonu \\
& \times & |(v;0)>
=  - N^3 \Bigg[ \frac{96  (\lambda +1) (\lambda +2) (\lambda +3) }
  {5 (\lambda -3) (\lambda -1) (2 \lambda -3)}
  \Bigg] A_1   |(v;0)>.
\label{eigenvo}
\eea
The eigenvalue has a simple factorized form. 

With the following normalization,
\bea
A_1 = -\frac{5}{96 N^3} (\la-3)(\la-1)(2\la-3),
\label{a1value}
\eea
the two eigenvalue equations, (\ref{eigenexpressionov})
and (\ref{eigenvo}),  lead to
\bea
W_0^{(4)} |(v;0)> & = & (1+\la)(2+\la)(3+\la)  |(v;0)>, \nonu \\
W_0^{(4)} |(0;v)> & = & (1-\la)(2-\la)(3-\la)  |(0;v)>.
\label{EigenEigen}
\eea
If one takes the overall normalization factor for the $W^{(4)}(z)$ 
as $A_4$ rather than $A_1$ as in (\ref{a1value}), then $A_4$ becomes
$A_4 = -\frac{5}{12 N^3} \la^3 $.
In principle, one can calculate the OPE between $W^{(4)}(z)$
and $W^{(4)}(w)$ from the explicit twenty terms in (\ref{w4}) although
the complete computation of the eighth order singular terms is rather
involved for general $(k_2,N)$ manually.  
Then one expects that the central term, the eighth order pole of the above
OPE, is given by $A_4^2 \, f (\la, N)$ where $f(\la,N)$ is a (fractional)
function of
$\la$ and $N$ (after the large $N$ limit is taken).
That is, our normalization is given by
the central term of the OPE between the higher spin $4$ current 
and itself which behaves as $ \frac{25}{144 N^6} \la^6 \, f (\la, N)$
where $f(\la, N)$ is not known at the moment. 

The above eigenvalues are also observed in \cite{GV1106} by following the
descriptions in \cite{GGHR} where the unitary case is analyzed.

One of the primaries is given by
$(v;0) \otimes (v;0)$ and the other primary is given by
$(0;v) \otimes (0;v)$ by pairing up identical representations
on the holomorphic and antiholomorphic sectors
in the context of diagonal modular invariant \cite{CY}.
Let us denote them 
as follows:
\bea
    {\cal O}_{+} = (v;0) \otimes (v;0), \qquad
  {\cal O}_{-} = (0;v) \otimes (0;v).    
\label{TwoOO}
\eea
The ratio of the three-point functions, from (\ref{EigenEigen}),
is given by
\bea
\frac{<{\cal O}_{+} {\cal O}_{+} W^{(4)}>}{<{\cal O}_{-}
  {\cal O}_{-} W^{(4)}>} =  \frac{(1+\la)(2+\la)(3+\la)}
     {(1-\la)(2-\la)(3-\la)},
\label{ratioexpression}
\eea
in the notation of (\ref{TwoOO}).
This is the same form for the unitary case \cite{Ahn1111,AK1308}.
In the corresponding unitary bulk calculation of \cite{CY},
for $\la =\frac{1}{2}$,
this ratio for generic spin
is given by $(-1)^s (2s-1)$ with spin $s$.
One expects that the orthogonal bulk computation
will give rise to the behavior of (\ref{ratioexpression}).

\subsubsection{Eigenvalue equation of the zero mode of the higher
  spin $4$ current acting on the state $|(v;v)>$  }

For the primary $(v;v)$ with the condition 
$
J_0^a |(v;v)> =0$, one can calculate the eigenvalue equation
\cite{GG1207}.
The nontrivial contributions
arise from the $A_5$, $A_8$, and $A_9$ terms.
It turns out that
\bea
W_0^{(4)} |(v;v)> & = & -N^2 \frac{48
  \lambda ^2 (\lambda ^2+1) } 
{5 (\lambda -3) (\lambda -1) (2 \lambda -3)} A_1 |(v;v)>.
\label{w4vv}
\eea
In (\ref{w4vv}), Appendix (\ref{coeffA}) is used.

\subsubsection{Further eigenvalue equations  }
One also presents
the eigenvalue equations \cite{AK1308}
at finite $N$ and $k_2$, by using
Appendix  (\ref{reducedcoeff})
and 
Appendix (\ref{dcoeffcoeff}), 
as follows:
\bea
W_0^{(4)} |(0;v)> & = &
\frac{6 A_1}{(3 k_2+2 N-2) d(1, k_2, N)} \nonu \\
& \times &
  (32 k_2^3 N^4-64 k_2^3 N^2+96 k_2^3 N-55 k_2^3+160 k_2^2 N^5-
  168 k_2^2 N^4-268 k_2^2 N^3 \nonu \\
& + &
690 k_2^2 N^2-617 k_2^2 N+203 k_2^2+128 k_2 N^6-64 k_2 N^5-680 k_2 N^4
\nonu \\
& + & 1548 k_2 N^3-1726 k_2 N^2+987 k_2 N-220 k_2+256 N^6-1024 N^5
\nonu \\
& + & 1584 N^4-1384 N^3+900 N^2-382 N+68) |(0;v)>, 
\nonu \\
W_0^{(4)} |(v;0)> & = &
\frac{6   (k_2+2 N-1) (k_2+4 N-3) (3 k_2+8 N-5) A_1}
     {k_2 (k_2+2 N-2) (3 k_2+2 N-2) (3 k_2+4 N-4)
d(1, k_2, N)}
\nonu \\
& \times & 
(32 k_2^3 N^4-64 k_2^3 N^2+96 k_2^3 N-55 k_2^3+224 k_2^2 N^5-120 k_2^2 N^4
-500 k_2^2 N^3\nonu \\
& + & 1038 k_2^2 N^2-907 k_2^2 N+292 k_2^2+384 k_2 N^6-64 k_2 N^5-1752 k_2 N^4 \nonu \\
& + &
3636 k_2 N^3-3930 k_2 N^2+2213 k_2 N-487 k_2+1024 N^6-3328 N^5
\nonu \\
& + & 5184 N^4-5576 N^3+3976 N^2-1566 N+250) |(v;0)>.
\label{finiteresults}
\eea
Of course, these eigenvalue equations (\ref{finiteresults})
becomes (\ref{eigenexpressionov}) and (\ref{eigenvo}) respectively
under the large $N$ 't Hooft limit.
Compared to the unitary case in \cite{AK1308}, the above eigenvalues
do not have simple factorized form. This is because of the fact that
the identities between $f$ and $d$ symbols contain
rather complicated functions of $N$.

For convenience, we also present the eigenvalue equations for the
spin $2$ stress energy tensor (\ref{stress}) with $k_1=1$
\bea
T_0 |(0;v)> & = & \frac{k_2}{2 (k_2+2 N-1)} |(0;v)>
\rightarrow \frac{(1-\la)}{2}  |(0;v)>,
\nonu \\
T_0 |(v;0)> & = & \frac{(k_2+4 N-3)}{2 (k_2+2 N-2)}
 |(v;0)> \rightarrow \frac{(1+\la)}{2}  |(v;0)>.
\label{Twoeigen}
\eea
Note that the conformal dimension
of $(0;v)$ can be obtained from the formula
\cite{LF1990,LFnpb,DNS,Ahn1106,GG1207}
\bea
h(0;v)=
\frac{1}{2}(2N-1) \left[ \frac{1}{1+(2N-2)} - \frac{1}{1+k_2+(2N-2)}
\right] =\frac{k_2}{2 (k_2+2 N-1)},
\label{hov}
\eea
where the overall factor $\frac{1}{2} (2N-1)$ is the
quadratic Casimir of $SO(2N)$ vector representation.
Similarly,
 the conformal dimension
of $(v;0)$ can be obtained
\bea
h(v;0) =
\frac{1}{2} (2N-1) \left[ \frac{1}{1+(2N-2)} +\frac{1}{k_2+(2N-2)} \right]
=\frac{(k_2+4 N-3)}{2 (k_2+2 N-2)}. 
\label{hvo}
\eea
Then the two results (\ref{hov}) and (\ref{hvo})
are coincident with the ones in (\ref{Twoeigen})
\footnote{
Furthermore, 
one can write down the eigenvalue equation for the state
$|(v;v)>$
\bea
T_0 |(v;v)> & = & \frac{(2 N-1)}{2 (k_2+2 N-2) (k_2+2 N-1)}
|(v;v)>
\rightarrow \frac{\la^2}{4N}  |(v;v)>.
\label{vv}
\eea
Note that under the large $N$ 't Hooft limit
the eigenvalue (\ref{vv}) reduces to zero. 

The conformal dimension
of $(v;v)$ can be obtained
as follows:
\bea
h(v;v) & = &
\frac{1}{2} (2N-1) \left[ \frac{1}{k_2 +(2N-2)} -\frac{1}{1+k_2+(2N-2)} \right]
\nonu \\
& = & \frac{(2 N-1)}{2 (k_2+2 N-2) (k_2+2 N-1)}. 
\label{hvv}
\eea
This looks similar to the unitary case \cite{GGHR}:
the overall factor is again
the quadratic Casimir of $SO(2N)$ in vector representation.
In the denominator one has
$(k_2 +2N-2)$ and this quantity plus one.
There exists a relation together with (\ref{hov}), (\ref{hvo})
and (\ref{hvv}),
\bea
h(v;v) = h(0;v) + h(v;0) -1,
\label{hvvhovhvo}
\eea
which was also observed in \cite{Ahn1106}.
The identity in (\ref{hvvhovhvo})
is checked from (\ref{Twoeigen}) and (\ref{vv}).}.

\subsection{The OPE between the higher spin $4$ current and itself
where $k_1=1, N=4$ and $k_2$ is arbitrary}

Let us describe the OPE
between the higher spin $4$ current and itself.
Because it is rather involved to calculate this OPE
manually,
one fixes the value of $N$
and then one can compute this OPE inside the package of \cite{Thielemans}.
For fixed $N=4$ which is the lowest value one can consider nontrivially,
one obtains the following fourth order pole
of this OPE, by realizing that the right structure constants
should behave according to the known results \cite{AP1301},  as follows:
{\small
\bea
W^{(4)}(z) \, W^{(4)}(w) \Bigg|_{\frac{1}{(z-w)^4}} & = & 
\frac{3}{10} \pa^2 T(w) +\frac{42}{(5c+22)} \left( T^2 -
\frac{3}{10} \pa^2 T \right)(w) +
\sqrt{\frac{18(c+24)}{(5c+22)}} \, W^{(4)}(w)
\nonu \\
& + &  W^{(4')}(w).
\label{wwpole4}
\eea}
Here the central charge reduces to
\bea
c(k_1=1, k_2,N=4) & = & \frac{4 k_2
  (k_2+13)}{(k_2+6) (k_2+7)},
\label{centralk2}
\eea
which can be obtained from (\ref{centralcharge}) by substituting
the two values of $k_1=1$ and $N=4$. 
The overall factor can be fixed as
\bea
A_1(k_1=1,k_2,N=4) & = & \frac{k_2}{2520(k_2+7)}
\sqrt{\frac{(k_2+2)
    (k_2+4)}{3  (k_2+9) (k_2+11)}},
\label{a1fixed}
\eea
by comparing the coefficient of the first term in the right hand side
of (\ref{wwpole4}).

Let us emphasize that there exists a new primary field in (\ref{wwpole4})
which is given by
\bea
&& W^{(4')}(z)\Bigg|_{k_1=1,k_2,N=4}  = 
\frac{1}{ (k_2+2) (k_2+11)} \Bigg[-\frac{1}{9}
\,
d^{abcd} J^a J^b K^c K^d
 + 
\frac{2}{35} (k_2-1) k_2
\, \pa J^a \pa J^a 
\nonu \\
&&
-\frac{4}{105} (k_2-1) k_2
\, \pa^2 J^a J^a 
 +   
\frac{28}{15}
\, \pa J^a \pa K^a
 +
\frac{64}{105} (k_2-1)
\, \pa^2 J^a K^a  
\nonu \\
& & - 
\frac{4}{35} (k_2-1)
\, f^{abc} J^a \pa J^b K^c 
-\frac{2}{735} (k_2-1) k_2
\, J^a J^a J^b J^b - 
\frac{68}{315 }
\, J^a J^a K^b K^b 
\nonu \\
&& +
\frac{8}{105} (k_2-1)
\, J^a J^a J^b K^b
 - 
\frac{28}{45}
\, J^a J^b K^a K^b
+
\frac{1}{90} \,
d^{abef} d^{efcd} J^a J^b K^c K^d \Bigg](z).
\label{primespin4}
\eea
In other words, there exists a nonzero expression
by combining the fourth order pole with the first line of
(\ref{wwpole4}) with minus sign. Furthermore, one can express
the various nonzero terms as the one in (\ref{primespin4}). 
One can easily see that the ten operators except the last operator
appear in the previous higher spin $4$ current in (\ref{w4}).
It is straightforward to analyze the description appearing in Appendix
$B$ and Appendix $C$ for the last operator in (\ref{primespin4}).

Let us further restrict to the simplest case where
one can see the full structure of the corresponding OPE
without losing any terms in the right hand side.
In other words, in this particular limit where $k_2 \rightarrow \infty$
corresponding to $c=4$,
the structure constants
do not vanish. That is, there is no $(c-4)$ factor in the right hand
side of the OPE.

Then the higher spin $4$ current can be written in terms of
\bea
W^{(4)}(z)\Bigg|_{k_1=1, k_2\rightarrow \infty,N=4} & = &
\frac{1}{2520 \sqrt{3}} \Bigg(
d^{abcd} J^a J^bJ^c J^d +  18 \pa J^a \pa J^a 
-12
\, \pa^2 J^a J^a 
\nonu \\
& - & 
3
\, J^a J^a J^b J^b \Bigg)(z),
\label{simplestspin4}
\eea
by substituting $N=4$ and $k_2 \rightarrow \infty$ limit in Appendix
$E$.
The normalization factor is consistent with the general
form in (\ref{a1fixed}).
The field contents in (\ref{simplestspin4}) are given in terms of
the numerator spin $1$ current (having the level $k_1=1$)
of the coset model.
Of course, the stress energy tensor contains only the first term
with $k_1=1$ in
(\ref{stress}) in this limit.

Then one can obtain the corresponding higher spin $4'$ current
from (\ref{primespin4}) by taking $k_2 \rightarrow \infty$ limit
and it turns out that
\bea
 W^{(4')}(z)\Bigg|_{k_1=1, k_2\rightarrow \infty,N=4} & = & 
\frac{2}{35} \left(
 \pa J^a \pa J^a 
-\frac{2}{3}
\, \pa^2 J^a J^a 
 - 
\frac{1}{21}
\, J^a J^a J^b J^b \right)(z).
\label{otherother}
\eea
In (\ref{otherother}), there is no $d$ symbol.

Now we can calculate the OPE between the higher spin $4$
current
(\ref{simplestspin4})
and itself as follows:
\bea
&& W^{(4)}(z) \; W^{(4)}(w)\Bigg|_{k_1=1, k_2\rightarrow \infty,N=4}
=\frac{1}{(z-w)^8} \, \frac{c}{4} +\frac{1}{(z-w)^6}
\, 2 \,T(w) +\frac{1}{(z-w)^5} \, \frac{1}{2} \, 2 \, \pa T(w) 
\nonu \\
&& +\frac{1}{(z-w)^4} \, \left[ \frac{3}{20} \, 2 \, \pa^2 T + 
\frac{42}{(5c+22)} \, \left( T^2 -\frac{3}{10} \pa^2 T \right) 
+ C_{44}^{4} \, W^{(4)}
\right] (w) \nonu \\
&& +\frac{1}{(z-w)^3} \, \left[ \frac{1}{30} \, 2 \, \pa^3 T + 
\frac{1}{2} \, \frac{42}{(5c+22)} \, \pa 
\left( T^2 -\frac{3}{10} \pa^2 T \right) 
+ \frac{1}{2}  \, C_{44}^{4} \, \pa W^{(4)}  \right] (w) \nonu \\
&&  +\frac{1}{(z-w)^2} \, \left[ \frac{1}{168} \, 2 \, \pa^4 T + 
\frac{5}{36} \, 
\frac{42}{(5c+22)} \, \pa^2 \left( T^2 -\frac{3}{10} \pa^2 T \right) 
+\frac{5}{36}  \, C_{44}^{4}\, \pa^2 W^{(4)}  \right. \nonu \\
&&  + \frac{24(72c+13)}{(5c+22)(2c-1)(7c+68)} \, \left( T (T^2 -\frac{3}{10} 
\pa^2 T) -\frac{3}{5} \pa^2 T T +\frac{1}{70} \pa^4 T \right) \nonu \\
&& -\frac{(95c^2+1254c-10904)}{6(5c+22)(2c-1)(7c+68)} \, \left( \frac{1}{2}
\pa^2 (T^2 -\frac{3}{10} \pa^2 T) -\frac{9}{5} \pa^2 T T +\frac{3}{70} 
\pa^4 T  \right) \nonu \\
&& \left. + \frac{28}{3(c+24)}  \, C_{44}^{4} \, \left( T W^{(4)} - \frac{1}{6} \pa^2 W^{(4)} \right)
 + C_{44}^{6} \, W^{(6)} \right] (w) \nonu \\
&&  +\frac{1}{(z-w)} \, \left[ \frac{1}{1120} \, 2 \, \pa^5 T + 
\frac{1}{36} \, 
\frac{42}{(5c+22)} \, \pa^3 \left( T^2 -\frac{3}{10} \pa^2 T \right) 
+\frac{1}{36}  \, C_{44}^{4} \, \pa^3 W^{(4)}  \right. \nonu \\
&&  + \frac{1}{2} \, \frac{24(72c+13)}{(5c+22)(2c-1)(7c+68)}  \, \pa
\left( T (T^2 -\frac{3}{10} 
\pa^2 T) -\frac{3}{5} \pa^2 T T +\frac{1}{70} \pa^4 T \right) \nonu \\
&& -\frac{1}{2} \, 
\frac{(95c^2+1254c-10904)}{6(5c+22)(2c-1)(7c+68)}  \, \pa 
\left( \frac{1}{2}
\pa^2 (T^2 -\frac{3}{10} \pa^2 T) -\frac{9}{5} \pa^2 T T +\frac{3}{70} 
\pa^4 T  \right) \nonu \\
&& \left. + \frac{1}{2} \, \frac{28}{3(c+24)}  \, C_{44}^{4}\,  \pa 
\left( T W^{(4)} - \frac{1}{6} \pa^2 W^{(4)} \right) + \frac{1}{2} \, C_{44}^{6} \, \pa W^{(6)}
 \right] (w)
\nonu \\
&& + {\it  \frac{1}{(z-w)^4} \, W^{(4')}(w) 
 +  \frac{1}{(z-w)^3} \, \frac{1}{2} \pa W^{(4')}(w)
 +    \frac{1}{(z-w)^2} \, \frac{1}{3} \left( T W^{(4')} - \frac{1}{6}
 \pa^2 W^{(4')} \right)(w) }
 \nonu \\
 & & + {\it 
 \frac{1}{(z-w)} \, \frac{1}{2} \frac{1}{3}
 \pa \left( T W^{(4')} - \frac{1}{6}
 \pa^2 W^{(4')} \right)(w)} +
 \cdots.
\label{wwopeexpression}
\eea
Here the central charge coming from (\ref{centralk2}) is given by
\bea
c(k_1=1, k_2 \rightarrow \infty, N=4) & = & 4,
\label{centralc4}
\eea
from (\ref{centralcharge}) by substituting the right numbers.
Moreover the two structure constants
are given by
\bea
C_{44}^{4} = \sqrt{\frac{18(c+24)}{(5c+22)}}, \qquad
C_{44}^{6} = \sqrt{\frac{12(c-1)(11c+656)}{(2c-1)(7c+68)}},
\label{twoC}
\eea
together with (\ref{centralc4}).
Note that there are extra two last lines in (\ref{wwopeexpression})
associated with
the new primary higher spin $4'$ current, compared to the previous
result in \cite{AP1301}.
The expression in (\ref{twoC}) already appeared in
\cite{Hornfeck,CGKV,AP1301}.

\subsection{Next higher spin currents }

In the  second order pole of (\ref{wwopeexpression}), there exists
a primary higher spin $6$ current.
One can imagine the six product of spin $1$ current
with correct contractions of $SO(2N)$ indices.
Let us consider the higher spin $4$ current
$W^{(4)}(z)$
which contains $d^{abcd} J^a J^b J^c J^d(z)$
and the same higher spin $4$ current
which contains
$d^{d'efg} J^{d'} J^e J^f J^g(z)$.
Then one has the second order pole of this OPE,
$d^{abcd} d^{d'efg} \delta^{dd'} J^a J^b J^c J^e J^f J^g(w)$,
by considering the singular term between $J^d(z)$ and $J^{d'}(w)$.
This gives rise to the term of 
$d^{abcd} d^{defg}  J^a J^b J^c J^e J^f J^g(w)$.
Then one expects that the higher spin $6$ current contains
this term and is given by
$ W^{(6)}(z)  =  d^{abcd} d^{defg} J^a J^b J^c J^e J^f J^g(z) + \cdots$.
According to the description of \cite{dMMP,Gershun},
the tensorial structure
of $SO(2N)$ symmetric invariant tensor of rank $6$
can be determined by
the product of two rank $4$ $d$ symbols.
Therefore the above expression can be rewritten in terms of
$d$ tensor of rank $6$ and one should have 
$ W^{(6)}(z) =d^{abcefg}  J^a J^b J^c J^e J^f J^g(z) + \cdots$.
It would be interesting to observe the full expression for the
higher spin $6$ current.








\section{
Higher spin currents with 
${\cal N}=1$ supersymmetry in 
the  stringy coset model with two levels $(2N-2, k_2)$}

In the presence of adjoint fermions coming from the equality of
one of the levels and the dual Coxeter number of $SO(2N)$, one
can construct the higher spin $\frac{7}{2}$ current
which is the superpartner of the previous higher spin $4$ current.
In doing this, the role of spin $\frac{3}{2}$ current living in the
${\cal N}=1$ superconformal algebra is crucial.
The OPE between this ${\cal N}=1$ lowest higher spin multiplet
denoted by $(\frac{7}{2},4)$ is described using the Jacobi identities.

\subsection{Spin $\frac{3}{2}, 2$ currents and ${\cal N}=1$
superconformal algebra }

The spin $\frac{3}{2}$ current can be obtained from
the spin $\frac{1}{2}$ current and spin $1$ current
as follows \cite{GKO,BS}:
\bea
G(z) & = & \sqrt{\frac{4(N-1)}{(2N-2+k_2)(4N-4+k_2)}}
\, \left( \frac{k_2}{6(N-1)} \, \psi^a J^a - \psi^a K^a \right)(z).
\label{gexpression}
\eea
Here
one has
\bea
\psi^a(z) \, \psi^b(w) = -\frac{1}{(z-w)} \, \frac{1}{2} \delta^{ab}
+ \cdots.
\label{psipsi}
\eea
Furthermore,
we can express the spin $1$ current
from the above spin $\frac{1}{2}$ current
satisfying (\ref{psipsi})
as
\bea
J^a(z) \equiv f^{abc} \psi^b \psi^c(z).
\label{Jfermions}
\eea
It is easy to check this spin $1$ current satisfies
the first equation of (\ref{jjkk}) with $k_1 = (2N-2)$.

Then it is easy to see that there are only two terms in
(\ref{gexpression}) and the relative coefficients can be fixed
by using the above spin $\frac{3}{2}$ current should transform
as a primary field under the stress energy tensor (\ref{stress})
with $k_1=(2N-2)$ as follows: 
\bea
T(z) \, G(w) & = &
\frac{1}{(z-w)^2} \, \frac{3}{2} G(w) + \frac{1}{(z-w)} \, \pa G(w) +
\cdots.
\label{TGope}
\eea
In other words, the condition (\ref{TGope})
determines the relative coefficients of (\ref{gexpression}).

The overall factor in (\ref{gexpression})
can be determined by the following OPE between the
spin $\frac{3}{2}$ current and itself
\bea
G(z) \, G(w) & = & \frac{1}{(z-w)^3} \, \frac{2c}{3}
+
\frac{1}{(z-w)} \, 2 T(w) + \cdots.
\label{GGope}
\eea
Here the central charge in (\ref{GGope})
is given by (\ref{centralcharge})
with the condition $k_1 =(2N-2)$.

It is useful to write down the following OPEs which will be used
in later calculations
\bea
\hat{G}(z) \, \psi^a(w)  & = & \frac{1}{(z-w)}
\frac{1}{2} \Bigg[- \frac{k_2}{2(N-1)} J^a + K^a \Bigg](w) + \cdots,
\nonu \\
\hat{G}(z) \, J^a(w)  & = &
-\frac{1}{(z-w)^2} \, k_2 \psi^a(w) + 
\frac{1}{(z-w)}
\Bigg[- k_2 \pa \psi^a - f^{abc} \psi^b K^c \Bigg](w) + \cdots,
\nonu \\
\hat{G}(z) \, K^a(w)  & = &
\frac{1}{(z-w)^2} \, k_2 \psi^a(w) + 
\frac{1}{(z-w)}
\Bigg[ k_2 \pa \psi^a + f^{abc} \psi^b K^c \Bigg](w) + \cdots,
\label{gthreeope}
\eea
where we introduce the following quantity
\bea
\hat{G}(z) & \equiv &
\left( \frac{k_2}{6(N-1)} \, \psi^a J^a - \psi^a K^a \right)(z).
\label{ghat}
\eea
Compared to the unitary case \cite{Ahn1211,Ahn1305,ASS1991},
the behavior of relative coefficient, which is equal to
one over three times the level divided by dual Coxeter number,
occurs in (\ref{ghat}).  
See also \cite{BS,GKO}.

\subsection{Eigenvalue equation of the zero mode of the higher
  spin $4$ current  }

We also present the eigenvalue equations for the
spin $2$ stress energy tensor (\ref{stress}) with $k_1=(2N-2)$
\bea
T_0 |(0;v)> & = & \frac{k_2 (2 N-1)}{8 (N-1) (k_2+4 N-4)}
|(0;v)>
\rightarrow  \frac{(1-\la)}{4(1+\la)} |(0;v)>,
\nonu \\
T_0 |(v;0)> & = & \frac{(2 N-1) (k_2+6 N-6)}{8 (N-1)
  (k_2+2 N-2)}
 |(v;0)> \rightarrow \frac{(1+2\la)}{4}  |(v;0)>.
\label{eigeneigenrelation}
\eea
In (\ref{eigeneigenrelation}), the large $N$ 't Hooft limit is taken
at the final stage.
Note that the conformal dimension
of $(0;v)$ can be obtained from the formula
\bea
h(0;v) & = &
\frac{1}{2}(2N-1) \left[ \frac{1}{(2N-2)+(2N-2)} - \frac{1}{(2N-2)+k_2+(2N-2)}
  \right] \nonu \\
& = &  \frac{k_2 (2 N-1)}{8 (N-1) (k_2+4 N-4)},
\label{hov1}
\eea
where the overall factor $\frac{1}{2} (2N-1)$ is the
quadratic Casimir of $SO(2N)$ vector representation.
Similarly,
 the conformal dimension
of $(v;0)$ can be obtained
\footnote{
Moreover, 
the eigenvalue equation for the state
$|(v;v)>$ can be obtained as follows:
\bea
T_0 |(v;v)> & = & \frac{(N-1) (2 N-1)}{(k_2+2 N-2)
  (k_2+4 N-4)}
|(v;v)>
\rightarrow  \frac{\lambda ^2}{2 (\lambda +1)} |(v;v)>.
\label{Tvv}
\eea
The conformal dimension
of $(v;v)$ in (\ref{Tvv}) can be also obtained
as follows:
\bea
h(v;v) & = &
\frac{1}{2} (2N-1) \left[ \frac{1}{k_2 +(2N-2)} -
  \frac{1}{(2N-2)+k_2+(2N-2)} \right]
\nonu \\
& = & \frac{(N-1) (2 N-1)}{(k_2+2 N-2) (k_2+4 N-4)}. 
\label{hvv1}
\eea
There exists a relation together with (\ref{hov1}), (\ref{hvo1})
and (\ref{hvv1}),
\bea
h(v;v) = h(0;v) + h(v;0) -\frac{(2 N-1)}{4 (N-1)}.
\label{hvvrelation}
\eea
Here the last term in (\ref{hvvrelation}) 
is the ratio of 
quadratic Casimir for the vector representation
and dual Coxeter number of
$SO(2N)$.}
\bea
h(v;0) & = &
\frac{1}{2} (2N-1) \left[ \frac{1}{(2N-2)+(2N-2)} +
  \frac{1}{k_2+(2N-2)} \right]
\nonu \\
& = & \frac{(2 N-1) (k_2+6 N-6)}{8 (N-1) (k_2+2 N-2)}. 
\label{hvo1}
\eea

As done in section $2$, one obtains the following
eigenvalue equations 
\bea
W_0^{(4)} |(0;v)> & = &  -N^3 \Bigg[\frac{48  }{(2 \lambda -3)} \Bigg] A_1
|(0;v)>, \nonu \\
W_0^{(4)} |(v;0)> & = &  - N^3 \Bigg[ \frac{48  (\lambda +1)^2
  (2 \lambda +1) (4 \lambda +3) }{(\lambda -3) (\lambda -1)
  (2 \lambda -3)} \Bigg] A_1 |(v;0)>,
\nonu \\
W_0^{(4)} |(v;v)> & = & -N^3 \Bigg[
  \frac{96  \lambda ^2 \left(4 \lambda ^2+\lambda +1\right) }{(\lambda -3) (\lambda -1) (2 \lambda -3)} \Bigg]
A_1 |(v;v)>.
\label{three}
\eea
Using these relations (\ref{three}), one
can obtain the several three-point functions.
The relevant coefficients are given in
Appendix (\ref{Acoeff}).

\subsection{Higher spin $\frac{7}{2}, 4$ currents }

One way to determine the higher spin $\frac{7}{2}$ current
is to use the OPE between the spin $\frac{3}{2}$ current
and the higher spin $4$ current in previous section.
Note that the corresponding coefficients
at the critical level $k_1=(2N-2)$ are given in
Appendix $F$.
In other words, from the ${\cal N}=1$ super primary
condition \cite{Ahn1211,Ahn1305}, one should have
\bea
\hat{G}(z) \, W^{(4)}(w) \Bigg|_{\frac{1}{(z-w)^2}} & = &
\frac{1}{(z-w)^2} \, 7 \,
 \sqrt{\frac{(2N-2+k_2)(4N-4+k_2)}{4(N-1)}}
 \, W^{(\frac{7}{2})} (w) \nonu \\
 & + &  {\cal O}(\frac{1}{(z-w)}).
\label{gwexp}
\eea

In order to calculate the second order pole of
(\ref{gwexp}), one can use the three OPEs in (\ref{gthreeope}).
The explicit results are given in Appendix $G$.
Of course, this will give us the final higher spin $\frac{7}{2}$
current but it is rather nontrivial to simplify in simple form. 
Therefore,
after we identify the correct field contents for fixed $N=4$,
introduce the undetermined coefficients and
fix them using the previous methods we used in previous section.
That is, the higher spin $\frac{7}{2}$ current should
not have any singular terms with the diagonal spin $1$ current 
and transform as a primary higher spin current under the stress energy
tensor.

Then one can express the higher spin $\frac{7}{2}$ current
as follows \cite{BBSSfirst,BBSSsecond,BS}:
\bea
W^{(\frac{7}{2})}(z) & = & B_1
\, d^{abcd} \psi^a J^b J^c J^d(z)+ B_2
\, d^{abcd} f^{aef} f^{beg} J^c J^d \psi^f K^g(z)
+B_3 \, d^{abcd} J^a K^b \psi^c K^d(z) \nonu \\
&
+ & B_4 \, d^{abcd} f^{aef} f^{beg} K^c K^d \psi^f K^g(z)
+ B_5 \, J^a \psi^a J^b J^b(z)
+B_6 \, K^a K^a \psi^b K^b(z)
\nonu \\
&+& B_7 \, J^a J^a \psi^b K^b(z)
+B_8 \, J^a J^a \psi^b J^b(z)
+B_9 \, \psi^a K^a K^b K^b(z)
\nonu \\
&+& B_{10} \, f^{abc} f^{cde} K^a K^e \psi^b K^d(z)
+ B_{11} \, J^a \psi^a K^b K^b(z)
+B_{12} \, \psi^a J^b K^a K^b(z) 
\nonu \\
&+ & B_{13} \, J^a J^b \psi^a K^b(z)
+B_{14} \, f^{abc} f^{cde} J^a J^e \psi^b K^d(z)
+B_{15} \, J^a J^b K^a \psi^b(z)
\nonu \\
&+& X(k_2,N) (G T -\frac{1}{8} \pa^2 G)(z).
\label{w7half}
\eea
The $B_7$ term can be written as
$(\psi^a J^b J^b K^a -2 f^{abc} \psi^a \pa J^b K^c -(2N-2) \pa^2
\psi^a K^a)(z)$ by moving the field $\psi^b$ to the left.
Similarly, the $B_8$ term
can be described as
$(\psi^a J^a J^b J^b +(2N-2) \pa^2 \psi^a J^a -(2N-2) \psi^a \pa^2 J^a
+ 2(2N-2) \pa \psi^a \pa J^a)(z)$.
For the $B_{13}$ term one obtains
$( \psi^a J^b J^a K^b +(2N-2) \pa^2 \psi^a K^a)(z)$.
For the $B_{14}$ term
one can write down
$(3(2N-2) f^{abc} \psi^a \pa J^b K^c +(2N-2)^2 \pa^2 \psi^a K^a)(z)$.
For the $B_{15}$ term, one has
$(\psi^a J^b J^a K^b + f^{abc} \psi^a \pa J^b K^c)(z)$ by moving
$\psi^b$ to the left.
Furthermore, the $B_2$ term and the $B_4$ term
can be simplified using the identity (\ref{dffrelation}).
For the remaining other terms, the fermion $\psi^a$ can be
moved to the leftmost without any extra terms because of the properties
of $f$ and $d$ symbols.
The $B_5, B_6, B_7, B_{11}, B_{12}$, and $ B_{13}$
terms can be seen from $G T(z)$. The $B_8, B_9$, and $B_{15}$ terms
are written in terms of $B_5$, $B_6$, and $B_{13}$ terms plus derivative
terms respectively.

Note that the last term in (\ref{w7half}) is a quasiprimary field
in the sense that the OPE between the stress energy tensor
and this field does not contain the third order pole.
We realize that this term does not appear for the particular $N=4$
case.

We would like to determine the undetermined coefficients
$B_1$-$B_{15}$ and $X(k_2,N)$ in
(\ref{w7half}).
As done in (\ref{jpw}), one should have the regular condition
as follows:
\bea
J^{\prime a}(z) \, W^{(\frac{7}{2})}(w) & = & + \cdots.
\label{regconspin7half}
\eea
In Appendix $H$, we present the OPEs between the
diagonal spin $1$ current and the $15$ fields in
(\ref{w7half}).
Moreover, the higher spin $\frac{7}{2}$ current
transforms as a primary field under the stress energy
tensor. In other words,
one has
\bea
\hat{T}(z) \, W^{(\frac{7}{2})}(w)
\Bigg|_{\frac{1}{(z-w)^n}, n=3,4,5} =0,
\label{primaryspin7half}
\eea
as in (\ref{primaryother}).
One obtains the corresponding OPEs in Appendix $I$.

By solving the various linear equations on the coefficients
satisfying the above requirements (\ref{regconspin7half})
and (\ref{primaryspin7half}),
one obtains the final coefficients in Appendix $J$.
There are in Appendix (\ref{Bcoeff}), Appendix (\ref{b1a1})
and Appendix (\ref{Xexp}).

For consistency check, one can calculate
the OPE
between $\hat{G}(z)$ and
$W^{(\frac{7}{2})}(w)$. The first order pole should
behave as
$ \sqrt{\frac{(2N-2+k_2)(4N-4+k_2)}{4(N-1)}}
\, W^{(4)}(w)$.
In doing this, the OPEs in (\ref{gthreeope})
are crucial. In order to see the presence of higher spin $4$,
the rearrangement of the normal ordered product should be taken
because the above first order pole terms contain unwanted terms.
Of course, we do not have to worry about the extra contractions in the OPEs
because we are interested in the first order pole as described above.

\subsection{The OPEs between the higher spin $\frac{7}{2}, 4$ currents 
 }

It is natural to ask how the OPEs between
the higher spin $\frac{7}{2}$ current
and the higher spin $4$ current arise.
They have rather long expressions
for $N=4$ case.

Therefore, one tries to obtain the corresponding OPEs
from the Jacobi identities for the above higher spin currents
and other relevant higher spin currents.
We will consider only the three OPEs,
$W^{(\frac{7}{2})}(z) \, W^{(\frac{7}{2})}(w)$,
$W^{(\frac{7}{2})}(z) \, W^{(4)}(w)$ and
$W^{(4)}(z) \, W^{(4)}(w)$.
What kind of new primary higher spin currents  are present in
the right hand side of OPEs?
From the OPEs of $W^{(\frac{7}{2})}(z) \, W^{(4)}(w)$
or $W^{(4)}(z)\, W^{(\frac{7}{2})}(w)$, one can think of
the presence of new higher spin $\frac{13}{2}$ current
at the first order pole.
Furthermore, from  the OPE
$W^{(4)}(z) \, W^{(4)}(w)$, the new higher spin $6$ current
can appear in the second order pole of this OPE. Note that
there is no new
higher spin $7$ current in the first order pole.
The reason is as follows.
One can calculate the OPE $W^{(4)}(w) \, W^{(4)}(z)$
in the presence of the new higher spin $7$ current
at the first order pole, use the symmetry
$z \leftrightarrow w$ and
end up with the
OPE $W^{(4)}(z) \, W^{(4)}(w)$. By focusing on the
new higher spin $7$ current, one realizes that
there exists an extra minus sign. Therefore,
the new higher spin $7$ current should vanish.

Then one can assign the above two higher spin currents
as one single ${\cal N}=1$ higher spin current,
denoted by $(6', \frac{13}{2})$ where the numbers
stand for each spin.
From the OPE in 
$W^{(4)}(z) \, W^{(4)}(w)$, the second order pole
provides a new higher spin $6$ current. Then one can think of
${\cal N}=1$ higher spin current denoted by
$( \frac{11}{2},6)$. Furthermore,
from the bosonic higher spin $4'$ current in previous section, 
one can introduce its superpartner whose spin is given by
$\frac{9}{2}$. The corresponding ${\cal N}=1$ higher spin
current is characterized by $(4', \frac{9}{2})$ via
above notation.

For $N=4$, one can calculate the
OPE in $W^{(\frac{7}{2})}(z) \, W^{(\frac{7}{2})}(w)$.
By requiring that the seventh order pole should be equal to
$\frac{2c}{7}$, one can determine the  coefficient
$A_1$ as
\bea
A_1(k_1=6,k_2,N=4) & = &
\frac{k_2}{5040 (k_2+12)} \sqrt{\frac{(k_2+2) (k_2+4)}
    {6 (k_2+6) (k_2+12) (k_2+14) (k_2+16)}}.
\label{constantA_1}
\eea
The fifth order pole gives $2T(w)$
and the fourth order pole gives $\pa T(w)$.
Similar behaviors arise in (\ref{wwopeexpression}).
Let us describe the third order pole.
One can easily check that the following
quantity together with (\ref{constantA_1}), 
\bea
\frac{1}{(4 c+21) (10 c-7)} \Bigg[ 8 (37 c+3)
T T
+3 (2 c-117)  \pa G G
-\frac{3}{10} (302 c-327) \pa^2 T\Bigg](w),
\label{quasifield}
\eea
is a quasiprimary field.
The third order pole subtracted by 
both (\ref{quasifield}) and $\frac{3}{10} \pa^2 T(w)$
(which is a descendant field)
is a primary field. However this is not written in terms of
the previous higher spin $4$ current.  
This implies that there exists a new primary higher spin $4'$
current.
The structure constants appearing in (\ref{quasifield})
are obtained from the Jacobi identities.  
Because we are dealing with the extensions of ${\cal N}=1$
superconformal algebra, the $\pa G G(w)$ term appears in addition
to $TT(w)$ and $\pa^2 T(w)$.

By assuming that
the ${\cal N}=1$ OPE
between the ${\cal N}=1$ higher spin $\frac{7}{2}$
multiplet contains the ${\cal N}=1$ higher spin $4',\frac{11}{2}, 6'$
multiplets, one obtains the complete structure of these OPEs in component
approach (and ${\cal N}=1$ superspace).  They are given in Appendix $K$
in terms of the central charge and some undetermined structure constants.
It would be interesting to see whether there exist other additional
higher spin currents or not. See also the work in \cite{AK1607}
where the Jacobi identities are used.



\subsection{The OPE in the ${\cal N}=1$ superspace }

From the three OPEs in component approaches
described in Appendix $K$, one summarizes
its ${\cal N}=1$ superspace in simple notation as follows:
\bea
\left[ {\bf W}^{(\frac{7}{2})} \cdot  {\bf W}^{(\frac{7}{2})}
  \right] =\left[ {\bf I} \right] +
\left[ {\bf W}^{(\frac{7}{2})}  \right]
+\left[ {\bf W}^{(4')}  \right]
+\left[ {\bf W}^{(\frac{11}{2})}  \right]
+\left[ {\bf W}^{(6')}  \right],
\label{fusion}
\eea
where $\left[ {\bf I} \right]$ appearing in (\ref{fusion})
is the ${\cal N}=1$
superconformal family of the identity operator.
According to the field contents in \cite{KS1606} where $k_2$ is fixed
as $k_2=1$, the above OPE should not contain
the ${\cal N}=1$ higher spin integer multiplets.
See also \cite{BCGG}.
The right-hand side
should contain the first, the second and the fourth terms.
It would be interesting to observe this behavior explicitly.
First of all, the single higher spin $4$ current should exist by combining
the previous two kinds of higher spin $4$ currents under the constraint
$k_2=1$.

\section{
Higher spin currents with 
${\cal N}=2$ supersymmetry in 
the stringy coset model with two levels $(2N-2, 2N-2)$}

The additional adjoint fermions allow us to
construct the spin $1, \frac{3}{2}$ currents in the
${\cal N}=2$ superconformal algebra. 
Furthermore, the additional higher spin $3, \frac{7}{2}$ currents
can be found explicitly along the line of \cite{GHKSS}. 
The lowest higher spin $3$ current of $U(1)$ charge $\frac{4}{3}$
is obtained
and it can be written in terms of two adjoint fermions.
There exists another ${\cal N}=2$ higher spin multiplet
which consists of the above same
spin contents, $(3, \frac{7}{2}, \frac{7}{2}, 4)$ with different
$U(1)$ charges.
Finally, the OPE between these two ${\cal N}=2$ higher spin multiplets
is described.

\subsection{Spin $1, \frac{3}{2}, \frac{3}{2}, 2$ currents
and ${\cal N}=2$ superconformal algebra }

Let us introduce the second adjoint fermions
which satisfy the following OPE 
\bea
\chi^a(z) \, \chi^b(w) = -\frac{1}{(z-w)} \,
\frac{1}{2} \delta^{ab}
+ \cdots.
\label{chichi}
\eea
It is easy to see that  one can express the spin $1$ current
from the above spin $\frac{1}{2}$ current with (\ref{chichi})
as
\bea
K^a(z) \equiv f^{abc} \chi^b \chi^c(z).
\label{Kfermions}
\eea
This spin $1$ current satisfies
the second equation of (\ref{jjkk}) with $k_2 = (2N-2)$.

Then it is straightforward to construct
the four generating currents, denoted by
$(1, \frac{3}{2},\frac{3}{2}, 2)$, corresponding to
the ${\cal N}=2$ superconformal
algebra as follows \cite{BFK}:
{\small
\bea
J(z) & = & 
\frac{2}{3} \, i\, \psi^a  \chi^a(z),
\nonu \\
G^{+}(z) & = & -
\frac{1}{6\sqrt{3(2N-2)}}
\Biggr[ \psi^a  J^a -3 \, \psi^a  K^a - i \, \chi^a  K^a
+ 3 \, i \, \chi^a J^a \Biggr](z),
\nonu \\
G^{-}(z) & = &
-
\frac{1}{6\sqrt{3(2N-2)}}
\Biggr[ \psi^a  J^a -3 \, \psi^a  K^a + i \, \chi^a  K^a
- 3 \, i \, \chi^a J^a \Biggr](z),
\label{jggt} \\
T(z) & = &
 -\frac{1}{4(2N-2)} \, J^a  J^a(z) -\frac{1}{4(2N-2)} \, K^a  K^a(z) 
+ \frac{1}{6(2N-2)}  (J^a +K^a) (J^a+K^a)(z).
\nonu
\eea}
By realizing that the difference between $G^{+}(z)$ and $G^{-}(z)$
occurs in the third and fourth terms, under the
$\chi^a(z) \rightarrow -\chi^a(z)$, one sees the relation $G^{+}(z)
\leftrightarrow G^{-}(z)$.

Let us introduce the following spin $1$ current
by taking the product of two adjoint fermions 
\bea
L^a &\equiv& f^{abc} \psi^b \chi^c.
\label{Lfermions}
\eea
The central charge can be reduced to
\bea
c = \frac{1}{3} N (2N-1),
\label{n2central}
\eea
which can be obtained from (\ref{centralcharge})
by substituting the corresponding two levels.
In order to construct the higher spin currents, 
let us introduce the following intermediate spin $2$ current
\bea
M_1^a &\equiv& d^{abcd} \psi^b \chi^c J^d,
\nonu \\
M_2^a &\equiv& d^{abcd} \psi^b \chi^c K^d,
\nonu \\
M_3^a &\equiv& d^{abcd} \psi^b \chi^c L^d,
\label{threeM}
\eea
together with (\ref{Jfermions}), (\ref{Kfermions})
and (\ref{Lfermions}).
Compared to the unitary case in \cite{Ahn1604},
the contracted indices appear in the two different
adjoint fermions (because of the symmetric $d$ tensor) as well as
the spin $1$ currents. 

\subsection{Higher spin $3, \frac{7}{2}, \frac{7}{2}, 4$ currents }

From the experience of section $2$ and section $3$, there exist
the higher spin $4$ current and the ${\cal N}=1$
higher spin $\frac{7}{2}$ current denoted by
$(\frac{7}{2}, 4)$, then
there are two choices where the above ${\cal N}=1$
higher spin $\frac{7}{2}$ multiplet 
can arise from the 
lower two component currents or
higher two component currents.
Let us try to find the higher spin currents by taking the second choice.

By writing the possible candidate terms
for the higher spin $3$ current, one can think of
the product of spin $1$ currents (\ref{Jfermions}), (\ref{Kfermions})
or (\ref{Lfermions}) and the intermediate
spin $2$ currents in (\ref{threeM}).
Furthermore, one can think of
the product of each component field in the
spin $\frac{3}{2}$ currents
living in the ${\cal N}=2$ superconformal algebra.
Of course, one should consider the possible derivative terms.
Therefore, one can consider the following higher spin $3$ current
\cite{BS}
{\small
\bea
W^{(3)}_{\frac{4}{3}}(z) & = & a_1 \, J^a M^a_1(z) + a_2 \, K^a M^a_1(z) +
a_3 \, L^a M^a_1(z)+
a_4 \, J^a M^a_2(z) + a_5 \, K^a M^a_2(z) +
a_6 \, L^a M^a_2(z) \nonu \\
& + &
a_7 \, J^a M^a_3(z) + a_8 \, K^a M^a_3(z) +
a_9 \, L^a M^a_3(z) +
a_{10} \, J^a \pa J^a(z) + a_{11}\, J^a \pa K^a(z)
\nonu \\
& + & 
a_{12} \, J^a \pa L^a(z) 
+  a_{13} \, \pa J^a K^a(z) + a_{14} \, K^a \pa K^a(z) +
a_{15} \, K^a \pa L^a(z) + a_{16} \, \pa J^a L^a(z)
\nonu \\
& + & 
a_{17} \, \pa K^a L^a(z) +
a_{18} \, L^a \pa L^a(z)
+ a_{19} \, (\psi^a J^a)(\psi^b J^b)(z) +
a_{20} \, (\psi^a J^a)(\psi^b K^b)(z)
\nonu \\
& + & 
a_{21} \, (\psi^a J^a)(\chi^b J^b)(z)+
a_{22} \, (\psi^a J^a)(\chi^b K^b)(z)
+  
a_{23} \, (\psi^a K^a)(\psi^b K^b)(z)
\nonu \\
& + &   
a_{24} \, (\psi^a K^a)(\chi^b J^b)(z)
+  
a_{25} \, (\psi^a K^a)(\chi^b K^b)(z)+
a_{26} \, (\chi^a J^a)(\chi^b J^b)(z)
\nonu \\
& + & 
a_{27} \, (\chi^a J^a)(\chi^b K^b)(z)+
a_{28} \, (\chi^a K^a)(\chi^b K^b)(z).
\label{w3first}
\eea}
The $U(1)$ charge $\frac{4}{3}$ will be determined later.

As done in previous sections,
one can use two requirements in order to fix the above
coefficients. One of them is the regularity with the diagonal
spin $1$ current as follows:
\bea
J^{\prime a}(z) \,  W^{(3)}_{\frac{4}{3}}(w) & = & +\cdots.
\label{jprimew3}
\eea
Here the diagonal spin $1$ current in (\ref{jprimew3}) is the sum of
(\ref{Jfermions}) and (\ref{Kfermions}).
The other is given by the primary condition,
which can be described as follows together with (\ref{w3first}):
\bea
\hat{T}(z) \, W_{\frac{4}{3}}^{(3)}(w)
\Bigg|_{\frac{1}{(z-w)^n}, n=3,4,5} =0.
\label{tw3}
\eea
Here the stress energy tensor is given by (\ref{numstress})
substituted by (\ref{Jfermions}) and (\ref{Kfermions}).

It turns out, from (\ref{jprimew3}) and (\ref{tw3}),  that 
the above higher spin $3$ current with the explicit coefficients
is given by
\bea
W^{(3)}_{\frac{4}{3}}(z) & = & \Bigg[ -\frac{i}{4} (a_7-a_8)
\, J^a M^a_1 + \frac{i}{2} (a_7-a_8) \, K^a M^a_1 -a_8
\, L^a M^a_1 -\frac{i}{4} (a_7-a_8) \, K^a M^a_2
\nonu \\
& - & 
a_7 \, L^a M^a_2 + 
a_7 \, J^a M^a_3 + a_8 \, K^a M^a_3 +
i (a_7-a_8) \, L^a M^a_3 \nonu \\
& + & 
 \frac{3i}{4} (a_7-a_8) \, J^a \pa L^a  +   
 \frac{3i}{4} (a_7-a_8) \, K^a \pa L^a
 +  \frac{3i}{4} (a_7-a_8) \, \pa J^a L^a
 \nonu \\
 & + & 
 \frac{3i}{4} (a_7-a_8) \, \pa K^a L^a 
+ 
(-a_7 +a_8) \, (\psi^a J^a)(\psi^b K^b)+
\frac{i}{2} (a_7-a_8)\, (\psi^a J^a)(\chi^b J^b)
\nonu \\
& - &
\frac{i}{2} (a_7-a_8) \, (\psi^a J^a)(\chi^b K^b)
+  
 \frac{3i}{2} (a_7-a_8) \, (\psi^a K^a)(\chi^b J^b)
 \nonu \\
 & + &   
 \frac{i}{2} (a_7-a_8) \, (\psi^a K^a)(\chi^b K^b)
 +    
(a_7-a_8) \, (\chi^a J^a)(\chi^b K^b) \Bigg](z).
\label{interw3}
\eea
Note that there exist also $a_4, a_{11}, a_{14}, a_{20}, a_{24}$, and
$a_{28}$ dependent terms (other coefficients
depend on these six coefficients and $a_7$ and $a_8$
after the above two conditions are
used) but they are identically zero respectively.
From the definitions of (\ref{threeM}), the first eight terms
in (\ref{interw3}) contain the rank $4$ $d$ symbol.
One can see  the common nonderivative expression in
the third term and sixth term and then
one can combine them with coefficient $(a_7-a_8)$.
Similarly, the fifth term and the seventh term
share the common nonderivative term with the coefficient
$-(a_7-a_8)$. Furthermore,
the composite fields appearing in (\ref{interw3}) contain
the various derivative terms (it is obvious that
the ninth-twelfth terms do have the derivative terms and also
they can appear from the ordering for the composite fields)
but the precise coefficients
will lead to the vanishing of these derivative terms.

For the extended ${\cal N}=2$ superconformal algebra, there is
one additonal condition for the higher spin current which is the
$U(1)$ charge (i.e., the coefficient of the first
order pole of the OPE with
the spin $1$ current). That is \cite{Ahn1604}, 
\bea
J(z) \, W_q^{(3)}(w) & = & \frac{1}{(z-w)} \, q \,
W_q^{(3)}(w) + \cdots. 
\label{jw}
\eea
It turns out that 
the $U(1)$ charge is fixed and for $q=\frac{4}{3}$, there is a relation
$a_{12} =\frac{3 i}{4} (a_7-a_8)$.
This relation is used in (\ref{interw3}).
For 
$q=-\frac{4}{3}$, there is a relation
$a_{12} =-\frac{3 i}{4} (a_7-a_8)$.
It is useful to express the above higher spin $3$ current
in manifestly $U(1)$ charge symmetric way.
Let us focus on the first term in (\ref{interw3}).
If one substitutes the definition of $M_1^a$ in (\ref{threeM}),
one has $f^{abc} \psi^b \psi^c d^{adef} \psi^d \chi^e J^f(w)$
where $J^a$ is replaced by the fermions.
One substitutes for the $J^f$ using the relation (\ref{Jfermions})
and obtains  $f^{abc} d^{adef} f^{fgh} \psi^b \psi^c
\psi^d \chi^e \psi^g \psi^h(z)$.
Now move the composite field $\psi^d \chi^e$ to the right.
One obtains
$f^{abc} d^{adef} f^{fgh} \psi^b \psi^c \psi^g \psi^h  \psi^d \chi^e(z)$
which can be written in terms of
$ \frac{i}{2} f^{abc} d^{adef} f^{fgh} \psi^b \psi^c \psi^g \psi^h
(\psi^d + i \chi^d)  ( \psi^e - i \chi^e)(z)$ from the
symmetric property of $d^{adef}$.
Then the overall factor is given by $\frac{1}{8} (a_7-a_8)$
by considering the numerical factor $-\frac{i}{4}(a_7-a_8)$ \footnote{
Similarly, the second term can be analyzed also.
The relevant term can be written in terms of
$f^{abc} d^{adef} f^{fgh} \chi^b \chi^c
\psi^d \chi^e \psi^g \psi^h(z)$ which can be also expressed as
$f^{abc} d^{adef} f^{fgh} \chi^b \chi^c
\psi^g \psi^h \psi^d \chi^e(z)$.
Once again this can be described as
$- \frac{i}{2} f^{abc} d^{adef} f^{fgh} i \chi^b i \chi^c
\psi^g \psi^h (\psi^d +i \chi^d ) (\psi^e -i \chi^e)(z)$
as done before.
The overall factor of the second term is
given by $\frac{i}{2}(a_7-a_8)$.
Then the total overall factor
gives $\frac{1}{4} (a_7-a_8)$. Intentionally, we rewrite the above as
$ \frac{1}{8} (a_7-a_8)
f^{abc} d^{adef} f^{fgh} (i \chi^b i \chi^c
\psi^g \psi^h + i \psi^b i \psi^c
\chi^g \chi^h )(\psi^d +i \chi^d ) (\psi^e -i \chi^e)(z)$
using the property of $d$ symbol.
One can analyze the other terms up to the seventh term.}.
Let us describe the eighth term which is the last term which contains
the $d$ symbol.
So one has
$f^{abc} d^{adef} f^{fgh} \psi^b \chi^c
\psi^g \chi^h \psi^d \chi^e(z)$ which can be identified with
$ -\frac{i}{2} f^{abc} d^{adef} f^{fgh} \psi^b i \chi^c
\psi^g i \chi^h (\psi^d +i \chi^d) (\psi^e -i \chi^e)(z)$.
By multiplying the overall factor
$i (a_7-a_8)$, one obtains
$ \frac{1}{2} (a_7-a_8) f^{abc} d^{adef} f^{fgh} \psi^b i \chi^c
\psi^g i \chi^h (\psi^d +i \chi^d) (\psi^e -i \chi^e)(z)$. This can be
further rewritten in terms of
$ \frac{1}{8} (a_7-a_8) f^{abc} d^{adef} f^{fgh} (\psi^b i \chi^c
\psi^g i \chi^h +\psi^b i \chi^c
i \chi^g  \psi^h
+ i \chi^b  \psi^c
\psi^g i \chi^h
+
\psi^b i \chi^c
i \chi^g  \psi^h
) (\psi^d +i \chi^d) (\psi^e -i \chi^e)(z)$.
Finally, one can summarize the first eight terms
in (\ref{interw3}) are given by
$\frac{1}{8} (a_7-a_8) d^{abcd} f^{aef} f^{bgh}
(\psi^e + i \chi^e)(\psi^f + i \chi^f)(\psi^g + i \chi^g)
(\psi^h + i \chi^h)(\psi^c + i \chi^c)(\psi^d - i \chi^d)(z)$.

Now we are considering the last six terms in (\ref{interw3}).  
The first term is given by
$ -(a_7-a_8) f^{acd}  \psi^a \psi^c \psi^d f^{bef}  \psi^b
 \chi^e  \chi^f(z)$.
This can be rewritten as
$-\frac{1}{4}(a_7-a_8)
(f^{acd} \psi^a \psi^c \psi^d f^{bef} \psi^b  \chi^e  \chi^f+
3 f^{acd}  \psi^a i \chi^c  i \chi^d f^{bef} \psi^b \psi^e \psi^f)(z)$
where we use the fact that there exists a minus sign when the first three
factors $ \psi^a i \chi^c  i \chi^d$ move to the right.
Therefore, there should an overall factor $\frac{1}{4}$.
One can analyze the other four terms\footnote{
  Let us describe the last term,
which is given by
$ -(a_7-a_8) f^{acd} i \chi^a \psi^c \psi^d f^{bef} i \chi^b 
\chi^e  \chi^f(z)$.
As above, this can be written as
$ -\frac{1}{4}(a_7-a_8) ( 3 f^{acd} i \chi^a \psi^c \psi^d f^{bef} i \chi^b 
\chi^e  \chi^f -f^{acd} i \chi^a \chi^c \chi^d f^{bef} i \chi^b 
\psi^e  \psi^f )(z)$.

There are also identities as follows:
\bea
f^{abc} \psi^a \psi^b \psi^c f^{def} \psi^d \psi^e \psi^f & = & 0,
\qquad
f^{abc} \psi^a \chi^b \chi^c f^{def} \psi^d \chi^e \chi^f  =  0,
\nonu \\
f^{abc} \chi^a \psi^b \psi^c f^{def} \chi^d \psi^e \psi^f & = & 0,
\qquad
f^{abc} \chi^a \chi^b \chi^c f^{def} \chi^d \chi^e \chi^f  =  0.
\label{iden}
\eea
As explained before, this (\ref{iden}) can be checked by
moving the first three fermions to the right and
there exists a minus sign.}.
Finally, one can summarize the last six terms
in (\ref{interw3}) are given by
$-\frac{1}{4} (a_7-a_8)  f^{abc} f^{def}
(\psi^a + i \chi^a)(\psi^b + i \chi^b)(\psi^c + i \chi^c)
(\psi^d + i \chi^d)(\psi^e + i \chi^e)(\psi^f - i \chi^f)(z)$.

By putting $(a_7-a_8)=1$, one obtains the following
higher spin $3$ current with $U(1)$ charge $\frac{4}{3}$
as follows:
{\small
\bea
W^{(3)}_{\frac{4}{3}}(z) & = & 
\frac{1}{8} d^{abcd} f^{aef} f^{bgh}
(\psi^e + i \chi^e)(\psi^f + i \chi^f)(\psi^g + i \chi^g)
(\psi^h + i \chi^h)(\psi^c + i \chi^c)(\psi^d - i \chi^d)(z)
\nonu \\
& - & \frac{1}{4} f^{abc} f^{def} (\psi^a + i \chi^a)(\psi^b + i \chi^b)
(\psi^c + i \chi^c)
(\psi^d + i \chi^d)(\psi^e + i \chi^e)(\psi^f - i \chi^f)(z).
\label{w3fermions}
\eea
}
One can calculate the $U(1)$ charges for the adjoint fermions with
(\ref{jggt}) as
follows:
\bea
J(z) \, (\psi^a + i \chi^a)(w) & = & \frac{1}{(z-w)} \, \frac{1}{3} \,
(\psi^a + i \chi^a)(w) + \cdots,
\nonu \\
J(z) \, (\psi^a - i \chi^a)(w) & = & \frac{1}{(z-w)} \, (-1) \frac{1}{3} \,
(\psi^a - i \chi^a)(w) + \cdots.
\label{basicu1}
\eea
Then it is obvious that the above higher spin $3$ current  
(\ref{w3fermions}) has $U(1)$ charge $\frac{4}{3}$: there exist
five factors with $U(1)$ charge $\frac{1}{3}$ and one factor
with $U(1)$ charge $-\frac{1}{3}$ according to (\ref{basicu1}).

For the unitary case \cite{Ahn1604}, one sees
the factor $ f^{aef} f^{bgh}
(\psi^e + i \chi^e)(\psi^f + i \chi^f)(\psi^g + i \chi^g)
(\psi^h + i \chi^h)$
and the other factor is given by
$d^{abc} f^{chi} (\psi^h - i \chi^h)(\psi^i - i \chi^i)$
of $U(1)$ charge $-\frac{2}{3}$
in the nonderivative terms.
However, the orthogonal case contains the different factor
$ d^{abcd}(\psi^c + i \chi^c)(\psi^d - i \chi^d)$ of
$U(1)$ charge $0$ in (\ref{w3fermions}).

In order to obtain the other higher spin currents,
it is useful to calculate the following OPEs,
{\small
\bea
G^{+}(z) \, (\psi^a + i \chi^a)(w) & = & + \cdots,
\nonu \\
G^{+}(z) \, (\psi^a - i \chi^a)(w) & = & \frac{1}{(z-w)}
\, \frac{1}{2\sqrt{3(2N-2)}} f^{abc} (\psi^b + i \chi^b)
(\psi^c + i \chi^c)(w)+ \cdots,
\nonu \\
G^{-}(z) \, (\psi^a - i \chi^a)(w) & = & + \cdots,
\nonu \\
G^{-}(z) \, (\psi^a + i \chi^a)(w) & = & \frac{1}{(z-w)}
\, \frac{1}{2\sqrt{3(2N-2)}} f^{abc} (\psi^b - i \chi^b)
(\psi^c - i \chi^c)(w)+ \cdots.
\label{GOPEfermions}
\eea
}
We will use this property to calculate the OPEs for the particular
singular terms.
One sees the $U(1)$ charge conservation in (\ref{GOPEfermions}).

How does one determine other higher spin currents related to
the lowest one?
Let us recall that
the following OPE \cite{Ahn1604,Ahn1208,Ahn1206}
\bea
G^{+}(z) \, W_{\frac{4}{3}}^{(3)}(w) & = &
-\frac{1}{(z-w)} \, W_{\frac{7}{3}}^{(\frac{7}{2})}(w) +\cdots.
\label{g+opew}
\eea
Here the higher spin current
appears in the first order pole.
Once we have obtained the
first order pole in the above OPE, then
we obtain the higher spin current.
See also the relevant work in \cite{AK1411}.
Because the lowest higher spin $3$ current is written in terms of
adjoint fermions, it is better to
calculate the OPE between $G^{+}(z)$ and
fermions appearing in (\ref{w3fermions}).
According to the observations of (\ref{GOPEfermions}),
the spin $\frac{3}{2}$ current $G^{+}(z)$
has nontrivial OPE with spin $\frac{1}{2}$ current of
$U(1)$ charge $-\frac{1}{3}$ while
the spin $\frac{3}{2}$ current $G^{-}(z)$
has nontrivial OPE with spin $\frac{1}{2}$ current of
$U(1)$ charge $\frac{1}{3}$.
Then it is obvious that
when one calculates the left hand side of
(\ref{g+opew}),
the only nontrivial singular terms appear 
at the location of the last factors, $(\psi^d- i \chi^d)(w)$
and $(\psi^f - i \chi^f)(w)$ in (\ref{w3fermions}).
This leads to
the following higher spin $\frac{7}{2}$ current of $U(1)$ charge
$\frac{7}{3}$
\bea
W^{(\frac{7}{2})}_{\frac{7}{3}}(z) & = & 
\frac{1}{2\sqrt{3(2N-2)}} \Bigg[
  \frac{1}{8} d^{abcd} f^{aef} f^{bgh} f^{dij}
  \nonu \\
  & \times & (\psi^e + i \chi^e)(\psi^f + i \chi^f)(\psi^g + i \chi^g)
  (\psi^h + i \chi^h)(\psi^c + i \chi^c)(\psi^i + i \chi^i)
 (\psi^j + i \chi^j) 
\nonu \\
& - & \frac{1}{4} f^{abc} f^{def} f^{fgh}
\label{7halfwone} \\
& \times & (\psi^a + i \chi^a)(\psi^b + i \chi^b)
(\psi^c + i \chi^c)
(\psi^d + i \chi^d)(\psi^e + i \chi^e)(\psi^g + i \chi^g)
(\psi^h + i \chi^h) \Bigg]
(z).
\nonu
\eea
In (\ref{7halfwone}), the $N$ dependence appears in the overall factor
rather than the relative coefficients.
One easily sees that the above two expressions
preserve $U(1)$ charge by counting the $U(1)$ charge at each factor.
In other words, each factor has $U(1)$ charge of $\frac{1}{3}$.

From the OPE \cite{Ahn1604}
\bea
G^{-}(z) \, W_{\frac{4}{3}}^{(3)}(w) & = &
\frac{1}{(z-w)} \, W_{\frac{1}{3}}^{(\frac{7}{2})}(w) +\cdots,
\label{gwgwgw}
\eea
one can obtain the other higher spin $\frac{7}{2}$ current
of $U(1)$ charge $\frac{1}{3}$.
It turns out, from the first order pole of (\ref{gwgwgw}), that
{\small
\bea
W_{\frac{1}{3}}^{(\frac{7}{2})}(z) & = &
\frac{1}{2\sqrt{3(2N-2)}} 
\frac{1}{8} d^{abcd} f^{aef} f^{bgh} \Bigg[
\nonu \\
  &   + & f^{eij}
\left((\psi^i - i \chi^i)(\psi^j - i \chi^j) \right)
(\psi^f + i \chi^f)(\psi^g + i \chi^g)
  (\psi^h + i \chi^h)(\psi^c + i \chi^c) (\psi^d - i \chi^d)
\nonu \\
&- &  f^{fij}
(\psi^e + i \chi^e)
\left((\psi^i - i \chi^i)(\psi^j - i \chi^j) \right)
(\psi^g + i \chi^g)
  (\psi^h + i \chi^h)(\psi^c + i \chi^c) (\psi^d - i \chi^d)
\nonu\\
&+ &  f^{gij}
(\psi^e + i \chi^e)(\psi^f + i \chi^f)
\left((\psi^i - i \chi^i)(\psi^j - i \chi^j) \right)
  (\psi^h + i \chi^h)(\psi^c + i \chi^c) (\psi^d - i \chi^d)
\nonu \\
&- &  f^{hij}
(\psi^e + i \chi^e)(\psi^f + i \chi^f) (\psi^g + i \chi^g)
\left((\psi^i - i \chi^i)(\psi^j - i \chi^j) \right)
 (\psi^c + i \chi^c) (\psi^d - i \chi^d)
\nonu \\
&+ &  f^{cij}
(\psi^e + i \chi^e)(\psi^f + i \chi^f) (\psi^g + i \chi^g)
 (\psi^h + i \chi^h)
\left((\psi^i - i \chi^i)(\psi^j - i \chi^j) \right)
 (\psi^d - i \chi^d)
\Bigg](z)
\nonu \\
& - & \frac{1}{2\sqrt{3(2N-2)}} \frac{1}{4} 
f^{abc} f^{def} \Bigg[
  \label{7halfwtwo} \\
  &+& f^{aij}
  \left((\psi^i - i \chi^i)
 (\psi^j - i \chi^j) \right)
  (\psi^b + i \chi^b)
(\psi^c + i \chi^c)
(\psi^d + i \chi^d)(\psi^e + i \chi^e)(\psi^f - i \chi^f)
\nonu \\
&-& f^{bij}
 (\psi^a + i \chi^a)
  \left((\psi^i - i \chi^i)
 (\psi^j - i \chi^j) \right)
(\psi^c + i \chi^c)
(\psi^d + i \chi^d)(\psi^e + i \chi^e)(\psi^f - i \chi^f)
  \nonu \\
  &+& f^{cij}
 (\psi^a + i \chi^a)(\psi^b + i \chi^b)
  \left((\psi^i - i \chi^i)
 (\psi^j - i \chi^j) \right)
(\psi^d + i \chi^d)(\psi^e + i \chi^e)(\psi^f - i \chi^f)
\nonu \\
  &-& f^{dij}
 (\psi^a + i \chi^a)(\psi^b + i \chi^b)(\psi^c + i \chi^c)
  \left((\psi^i - i \chi^i)
 (\psi^j - i \chi^j) \right)
(\psi^e + i \chi^e)(\psi^f - i \chi^f)
  \nonu \\
   &+& f^{eij}
  (\psi^a + i \chi^a)(\psi^b + i \chi^b)(\psi^c + i \chi^c)
  (\psi^d + i \chi^d)
  \left((\psi^i - i \chi^i)
 (\psi^j - i \chi^j) \right)
(\psi^f - i \chi^f)
  \Bigg](z).
\nonu
\eea
}
From (\ref{GOPEfermions}),
the OPE between $G^{-}(z)$ and $(\psi^a - i \chi^a)(w)$
does not have any singular terms and the contribution from
this OPE in (\ref{7halfwtwo}) vanishes.
Note that the  big bracket stands for the normal ordered
product \cite{BBSSfirst,BBSSsecond}. Of course, one can move
those factors to the right in order to simplify further.
Each term has the $U(1)$ charge $\frac{1}{3}$
because there are four factors for the $U(1)$ charge $\frac{1}{3}$
and three factors for the $U(1)$ charge $-\frac{1}{3}$. Totally
one has $\frac{1}{3}$ $U(1)$ charge.

From the relation \cite{Ahn1604},
{\small
\bea
G^{-}(z) \, W_{\frac{7}{3}}^{(\frac{7}{2})}(w)
& = & \frac{1}{(z-w)^2} \, (-1) \frac{7}{3} W_{\frac{4}{3}}^{(3)}(w)
+ \frac{1}{(z-w)} \, \Bigg[
W_{\frac{4}{3}}^{(4)} -\frac{1}{2} \pa W_{\frac{4}{3}}^{(3)}
  \Bigg](w) + \cdots,
\label{gwopeope}
\eea
}
one obtains, by calculating the left hand side of (\ref{gwopeope})
with (\ref{jggt}) and (\ref{7halfwone})
and reading off the first order pole, 
{\small
\bea
&& (W_{\frac{4}{3}}^{(4)} -\frac{1}{2} \pa W_{\frac{4}{3}}^{(3)})(z)  =  
\frac{1}{2\sqrt{3(2N-2)}} 
  \frac{1}{8} d^{abcd} f^{aef} f^{bgh} f^{dij}
[
  \nonu \\
  &  & + f^{ekl} \left((\psi^k - i \chi^k)(\psi^l - i \chi^l)
  \right)(\psi^f + i \chi^f)(\psi^g + i \chi^g)
  (\psi^h + i \chi^h)(\psi^c + i \chi^c)(\psi^i + i \chi^i)
  (\psi^j + i \chi^j)
  \nonu \\
  && -f^{fkl} (\psi^e + i \chi^e)
  \left((\psi^k - i \chi^k)(\psi^l - i \chi^l)
  \right)(\psi^g + i \chi^g)
  (\psi^h + i \chi^h)(\psi^c + i \chi^c)(\psi^i + i \chi^i)
  (\psi^j + i \chi^j)
  \nonu \\
 && +f^{gkl} (\psi^e + i \chi^e)(\psi^f + i \chi^f)
  \left((\psi^k - i \chi^k)(\psi^l - i \chi^l)
  \right)
  (\psi^h + i \chi^h)(\psi^c + i \chi^c)(\psi^i + i \chi^i)
  (\psi^j + i \chi^j)
  \nonu \\
&& -f^{hkl} (\psi^e + i \chi^e)(\psi^f + i \chi^f) (\psi^g + i \chi^g)
  \left((\psi^k - i \chi^k)(\psi^l - i \chi^l)
  \right)
 (\psi^c + i \chi^c)(\psi^i + i \chi^i)
  (\psi^j + i \chi^j)
  \nonu \\
&& +f^{ckl} (\psi^e + i \chi^e)(\psi^f + i \chi^f) (\psi^g + i \chi^g)
 (\psi^h + i \chi^h)
  \left((\psi^k - i \chi^k)(\psi^l - i \chi^l)
  \right)
(\psi^i + i \chi^i)
  (\psi^j + i \chi^j)
  \nonu \\
&& -f^{ikl} (\psi^e + i \chi^e)(\psi^f + i \chi^f) (\psi^g + i \chi^g)
 (\psi^h + i \chi^h)(\psi^c + i \chi^c)
  \left((\psi^k - i \chi^k)(\psi^l - i \chi^l)
  \right)
  (\psi^j + i \chi^j)
  \nonu \\
&& +f^{jkl} (\psi^e + i \chi^e)(\psi^f + i \chi^f) (\psi^g + i \chi^g)
 (\psi^h + i \chi^h)(\psi^c + i \chi^c) (\psi^i + i \chi^i)
  \left((\psi^k - i \chi^k)(\psi^l - i \chi^l)
  \right)](z)
\nonu \\
&& - 
\frac{1}{2\sqrt{3(2N-2)}} 
\frac{1}{4} f^{abc} f^{def} f^{fgh}
[
\label{w4final} \\
&  & + f^{aij} \left((\psi^i - i \chi^i)(\psi^j - i \chi^j)
\right)(\psi^b + i \chi^b)
(\psi^c + i \chi^c)
(\psi^d + i \chi^d)(\psi^e + i \chi^e)(\psi^g + i \chi^g)
(\psi^h + i \chi^h)
\nonu \\
&  & -f^{bij} (\psi^a + i \chi^a)
\left((\psi^i - i \chi^i)(\psi^j - i \chi^j)
\right)
(\psi^c + i \chi^c)
(\psi^d + i \chi^d)(\psi^e + i \chi^e)(\psi^g + i \chi^g)
(\psi^h + i \chi^h)
\nonu \\
&  & +f^{cij} (\psi^a + i \chi^a)(\psi^b + i \chi^b)
\left((\psi^i - i \chi^i)(\psi^j - i \chi^j)
\right)
(\psi^d + i \chi^d)(\psi^e + i \chi^e)(\psi^g + i \chi^g)
(\psi^h + i \chi^h)
\nonu \\
&  & -f^{dij} (\psi^a + i \chi^a)(\psi^b + i \chi^b)
(\psi^c + i \chi^c)
\left((\psi^i - i \chi^i)(\psi^j - i \chi^j)
\right)
(\psi^e + i \chi^e)(\psi^g + i \chi^g)
(\psi^h + i \chi^h)
\nonu \\
&  & + f^{eij} (\psi^a + i \chi^a)(\psi^b + i \chi^b)
(\psi^c + i \chi^c)(\psi^d + i \chi^d)
\left((\psi^i - i \chi^i)(\psi^j - i \chi^j)
\right)
(\psi^g + i \chi^g)
(\psi^h + i \chi^h)
\nonu \\
&  & -f^{gij} (\psi^a + i \chi^a)(\psi^b + i \chi^b)
(\psi^c + i \chi^c)(\psi^d + i \chi^d)(\psi^e + i \chi^e)
\left((\psi^i - i \chi^i)(\psi^j - i \chi^j)
\right)
(\psi^h + i \chi^h)
\nonu \\
&  & + f^{hij} (\psi^a + i \chi^a)(\psi^b + i \chi^b)
(\psi^c + i \chi^c)(\psi^d + i \chi^d)(\psi^e + i \chi^e)
(\psi^g + i \chi^g)
\left((\psi^i - i \chi^i)(\psi^j - i \chi^j)
\right)](z).
\nonu
\eea
}
The properties in (\ref{GOPEfermions})
are used.
One can check that the $U(1)$ charge of each term is equal to
$\frac{4}{3}$ where there are six positive ones and two negative ones.  
In order to obtain the primary current,
one should consider $
(W_{\frac{4}{3}}^{(4)} -\frac{1}{9} \pa W_{\frac{4}{3}}^{(3)})(z)$
\cite{Ahn1604} which can be obtained from (\ref{w4final})
and (\ref{w3fermions}).

Then,
the higher spin $3, \frac{7}{2}, \frac{7}{2}$, and $4$
currents are summarized by
(\ref{w3fermions}), (\ref{7halfwone}), (\ref{7halfwtwo}) and
(\ref{w4final}) with addition of the derivative of (\ref{w3fermions}).

\subsection{ Other higher spin $3, \frac{7}{2}, \frac{7}{2}, 4$ currents }

In the description of
(\ref{jw}),
for the opposite
$U(1)$ charge,
there exists also
other solution for the higher spin $3$ current.
One obtains the higher spin $3$ current of $U(1)$ charge
$-\frac{4}{3}$ as follows:
\bea
W_{-\frac{4}{3}}^{(3)}(w) =W_{\frac{4}{3}}^{(3)}(w) \Bigg|_{\chi^a \rightarrow
- \chi^a}.
\label{otherw3}
\eea
More explicitly, one can read off the explicit
expression which can be obtained from (\ref{w3fermions})
by replacing the second adjoint fermions 
with those together with minus sign.
It is obvious to see that the $U(1)$ charge
$-\frac{4}{3}$ of this higher spin current
can be seen during this process: five
factors of $U(1)$ charge $-\frac{5}{3}$
and one factor with $U(1)$ charge $\frac{1}{3}$.

Let us calculate other higher spin currents.
From the known OPE \cite{Ahn1604}
\bea
G^{+}(z) \, W_{-\frac{4}{3}}^{(3)}(w) & = &
-\frac{1}{(z-w)} \, W_{-\frac{1}{3}}^{(\frac{7}{2})}(w) +\cdots,
\label{7halfgenerating}
\eea
one obtains
the higher spin $\frac{7}{2}$ current together with (\ref{otherw3}) and
(\ref{7halfgenerating}) as follows:
\bea
W_{-\frac{1}{3}}^{(\frac{7}{2})}(w) = - W_{\frac{1}{3}}^{(\frac{7}{2})}(w)
\Bigg|_{\chi^a \rightarrow -\chi^a}.
\label{other7halfone}
\eea
Note that under the change of
$\chi^a \rightarrow -\chi^a$, the original $U(1)$ charge
is changed into the negative one.
More explicitly, one can take this operation in (\ref{gwgwgw}).
Then the left hand side of (\ref{gwgwgw}) leads to the
left hand side of (\ref{7halfgenerating}) with the help of
(\ref{otherw3}) and the right hand side
of  (\ref{gwgwgw}) becomes
$W_{\frac{1}{3}}^{(\frac{7}{2})}(w)
\Bigg|_{\chi^a \rightarrow -\chi^a}$. By realizing that the first order pole
from (\ref{7halfgenerating}),
then we are left with
(\ref{other7halfone}).

Similarly,
the OPE \cite{Ahn1604} with (\ref{otherw3})
\bea
G^{-}(z) \, W_{-\frac{4}{3}}^{(3)}(w) & = &
\frac{1}{(z-w)} \, W_{-\frac{7}{3}}^{(\frac{7}{2})}(w) +\cdots,
\label{gwotherope}
\eea
provides
the following result for the higher spin current,
by considering the relation (\ref{g+opew}) where
the operation $\chi^a \rightarrow -\chi^a$
is taken and the relation
(\ref{otherw3}),
\bea
W_{-\frac{7}{3}}^{(\frac{7}{2})}(w)
&=&
-W_{\frac{7}{3}}^{(\frac{7}{2})}(w) \Bigg|_{\chi^a \rightarrow - \chi^a}.
\label{other7halftwo}
\eea
In other words, the left-hand side of (\ref{gwotherope})
is equal to the left hand side of
(\ref{g+opew}) with the additional operation
$\chi^a \rightarrow -\chi^a$. We also use the previous relation
(\ref{otherw3}).
Then the right-hand side
of (\ref{gwotherope}) can be read off from 
this relation and we arrive at (\ref{other7halftwo}).

From the relation \cite{Ahn1604},
\bea
G^{+}(z) \, W_{-\frac{7}{3}}^{(\frac{7}{2})}(w)
& = & \frac{1}{(z-w)^2} \,  \frac{7}{3} W_{-\frac{4}{3}}^{(3)}(w)
+ \frac{1}{(z-w)} \, \Bigg[
W_{-\frac{4}{3}}^{(4)} +\frac{1}{2} \pa W_{-\frac{4}{3}}^{(3)}
  \Bigg](w) + \cdots,
\label{expexpexp}
\eea
one obtains
that the first order pole of (\ref{expexpexp})
leads to
\bea
(W_{-\frac{4}{3}}^{(4)} +\frac{1}{2} \pa W_{-\frac{4}{3}}^{(3)})(w)
= -(W_{\frac{4}{3}}^{(4)} -\frac{1}{2} \pa W_{\frac{4}{3}}^{(3)})(w)\Bigg|_{
  \chi^a \rightarrow -\chi^a},
\label{otherw4rel}
\eea
where the previous relation (\ref{gwopeope}) together with
the operation $\chi^a \rightarrow -\chi^a$
is used.
Moreover, the previous relation (\ref{other7halftwo})
is used also.
As described before,
the field (\ref{otherw4rel}) is not a primary under the stress energy
tensor. The primary current is given by
$ (W_{-\frac{4}{3}}^{(4)} + \frac{1}{9}  \pa  W_{-\frac{4}{3}}^{(3)} )(w)$
which can be obtained from
$ (-W_{\frac{4}{3}}^{(4)} + \frac{1}{9}  \pa  W_{\frac{4}{3}}^{(3)} )(w)$
by changing of $\chi^a(w) \rightarrow - \chi^a(w)$.

Therefore, 
the higher spin $3, \frac{7}{2}, \frac{7}{2}, 4$ currents
are summarized by (\ref{otherw3}), (\ref{other7halfone}),
(\ref{other7halftwo}), and (\ref{otherw4rel}).
They are obtained from the higher spin currents appearing in previous
subsection by simple change of the adjoint fermions $\chi^a(z)$
up to signs.



\subsection{The OPE between the two lowest higher spin currents
  in ${\cal N}=2$ superspace }

Because the coset with the critical levels has the
${\cal N}=2$ supersymmetry, one can
describe the OPE between the two lowest higher spin multiplets
in the ${\cal N}=2$ superspace.
Let us
consider the OPE between the two ${\cal N}=2$ lowest higher spin $3$
multiplets where  they have two opposite $U(1)$ charges.
That is, 
\bea
{\bf W}_{\frac{4}{3}}^{(3)}(Z_1) \, {\bf W}_{-\frac{4}{3}}^{(3)}(Z_2),
\label{n2ope}
\eea
        where each four component current, which obtained in previous
subsection,
        is given by 
\bea
{\bf W}_{\frac{4}{3}}^{(3)} 
&\equiv& \left(W_{\frac{4}{3}}^{(3)}, \, W_{\frac{7}{3}}^{(\frac{7}{2})}, \,
W_{\frac{1}{3}}^{(\frac{7}{2})}, \,
W_{\frac{4}{3}}^{(4)} \right),
\nonu \\
{\bf W}_{-\frac{4}{3}}^{(3)} &\equiv& 
\left(W_{-\frac{4}{3}}^{(3)}, \, W_{-\frac{1}{3}}^{(\frac{7}{2})}, \,
W_{-\frac{7}{3}}^{(\frac{7}{2})}, \,
W_{-\frac{4}{3}}^{(4)} \right).
\label{twon2superfields}
\eea
In principle, in order to obtain the explicit
OPE in (\ref{n2ope}), one should calculate only the four OPEs
between the four component currents living in
${\bf W}_{\frac{4}{3}}^{(3)}(Z_1)$ in (\ref{twon2superfields})
and the lowest component
current in ${\bf W}_{-\frac{4}{3}}^{(3)}(Z_2)$ in (\ref{twon2superfields}),
due to the
${\cal N}=2 $ supersymmetry.
See also the relevant works in \cite{Ahn1311,Ahn1408,Ahn1504}
where the various ${\cal N}=2$ multiplets in different coset model
are studied.
From the four OPEs, one can realize that
the right hand sides of these OPEs
should have $U(1)$ charges, $0, 1$, or $-1$ by adding the
$U(1)$ charges.
Recall that the four currents characterized by
the ${\cal N}=2$ stress energy tensor ${\bf T} \equiv
(J, G^{+}, G^{-}, T)$
of ${\cal N}=2$ superconformal algebra
have $0, 1, -1$, and $0$ respectively. 
It is natural to consider the 
right hand side of (\ref{n2ope}) in terms of
${\cal N}=2$ stress energy tensor ${\bf T}(Z_2)$ with its various
descendant fields in minimal way.

Inside of the package of \cite{KT}, one can
introduce the OPEs, ${\bf T}(Z_1) \, {\bf T}(Z_2)$ which is the
standard OPE corresponding to the ${\cal N}=2$ superconformal algebra,
${\bf T}(Z_1) \, {\bf W}_{\frac{4}{3}}^{(3)}(Z_2)$,
which is the ${\cal N}=2$
primary condition with $U(1)$ charge $\frac{4}{3}$ 
and
${\bf T}(Z_1) \, {\bf W}_{-\frac{4}{3}}^{(3)}(Z_2)$,
which is the ${\cal N}=2$ primary condition with
$U(1)$ charge $-\frac{4}{3}$. All the coefficients appearing in these
OPEs are constants except the central charge $c$ which is a function of
$N$ in (\ref{n2central}). Then one can write down
the right hand side of OPE (\ref{n2ope}) with arbitrary coefficients
which depend on $c$ or $N$.
After using the Jacobi identities, one summarizes
the structure constants in Appendix $L$ explicitly.
See also the relevant work in \cite{AK1509}.

One expects that there should be present other higher spin multiplets
in the various OPEs. For exmaple, 
${\bf W}_{\frac{4}{3}}^{(3)}(Z_1) \, {\bf W}_{\frac{4}{3}}^{(3)}(Z_2)$
or
${\bf W}_{-\frac{4}{3}}^{(3)}(Z_1) \, {\bf W}_{-\frac{4}{3}}^{(3)}(Z_2)$
as in the unitary case \cite{Ahn1604}.
It would be interesting to obtain these higher spin multiplets explicitly
further.


\section{Conclusions and outlook }

In the coset model (\ref{generalcoset}),
we have constructed the higher spin $4$ current for general levels.
For $k_1=1$ with arbitrary $N$ and $k_2$,
the eigenvalue equations of the zero mode of the higher
spin $4$ current acting on the states are obtained.
The corresponding three-point functions are also determined.
The ${\cal N}=1$ higher spin multiplet characterized by $(\frac{7}{2},4)$
for $k_1=2N -2$ in terms of adjoint fermions and spin $1$ current
is obtained.
The  two ${\cal N}=2$ higher spin multiplets denoted by
$(3, \frac{7}{2}, \frac{7}{2}, 4)$ for $k_1=k_2=2N-2$
in terms of two adjoint fermions are determined.
Some of the OPEs in ${\cal N}=1$ or ${\cal N}=2$
coset models are given explicitly.

We consider the possible related open problems as follows:

$\bullet$
One can also try to obtain the higher spin currents 
in the following coset model
\bea
\frac{\hat{SO}(2N+1)_{k_1}
\oplus \hat{SO}(2N+1)_{k_2}}{\hat{SO}(2N+1)_{k_1+k_2}}.
\label{oddcase}
\eea
It seems that the minimum value of $N$ for the nontrivial existence
of $d$ symbol (and corresponding higher spin $4$ current) 
is given by $N=2$. In the present paper, the minimum value of
$N$ is given by $N=4$ and the number of
independent fields in the higher spin currents 
is rather big implying that it is rather nontrivial to extract the
corresponding OPEs.   
In the coset model (\ref{oddcase}), for $N=2$ or $N=3$ case, 
one expects that one can analyze the OPEs further and observe
more structures in the right-hand sides of the OPEs.

$\bullet$ Further algebraic structures

In order to observe the algebraic structures living in
the bosonic, ${\cal N}=1$ or ${\cal N}=2$ higher spin multiplets
for generic $N$ (and generic $k_2$),
one should calculate the various OPEs between them manually.
In practice, this is rather involved because
for example, the higher spin $4$ current in the bosonic
coset model consists of twenty terms and the number of OPEs is greater
than two hundreds.
In \cite{Thielemans}, one can try to
obtain the various OPEs for the fixed low $N$ values (for example,
$N=4,5,6,7, \cdots$)
and expect the $N$ dependence of structure constants appearing in the
right-hand side of the OPEs.

$\bullet$ ${\cal N}=2$ enhancement of  \cite{CHR1209}

One considers the critical level condition in \cite{CHR1209,AP1310}.
It would be interesting to observe any ${\cal N}=2$ enhancement
or not. One can easily see the breaking of adjoint representation
in $SO(2N+1)$ into the adjoint representation of $SO(2N)$ plus
the vector representation of $SO(2N)$.
The first step is to construct the ${\cal N}=2$ superconformal
algebra realization.

$\bullet$ The additional numerator factors

For example, one considers the following coset model
where the extra numerator factor exists in 
the coset
\bea
\frac{\hat{SO}(2N)_{2N-2}
  \oplus \hat{SO}(2N)_{2N-2} \oplus
\hat{SO}(2N)_{2N-2}
}{\hat{SO}(2N)_{6N-6}}.
\label{othercoset}
\eea
It is an open problem to see whether one constructs
the ${\cal N}=3$ superconformal
algebra \cite{CK}
from the three kinds of adjoint fermions or not.
It is nontrivial to obtain the three spin $\frac{3}{2}$ currents
satisfying the standard OPEs between them.
Then one can try to obtain the higher spin currents living in the
above coset model (\ref{othercoset}). 
Furthermore, one can describe another coset model
where the additional numerical factor occurs. It is an open
problem to construct the linear (or nonlinear)
${\cal N}=4$ superconformal algebra 
from the four kinds of adjoint fermions. 

$\bullet$
Further identities between $f$ and $d$ tensors of $SO(2N)$

One can analyze the various identities involving $f$ and $d$ tensors
by following the description of \cite{dMMP,Gershun}.
They will be useful in order to calculate the OPEs between the
higher spin currents in the context of section $3$ and $4$.

$\bullet$ Zero mode eigenvalue equations in other representations

There exists an adjoint representation of $SO(2N)$.
It is an open problem to describe the eigenvalue equations
for the zero mode of the higher spin $4$ current acting on
the states associated with the adjoint representation.
For the $SO(8)$ generators in the adjoint representation,
one has $28 \times 28$ matrices whose elements are given by the structure
constant. 

$\bullet$ Marginal operator

One of the motivations in section $4$ is based on
the presence of perturbing marginal operator \cite{GJL}
which breaks the higher
spin symmetry but preserving the ${\cal N}=2$ supersymmetry.
It would be interesting to obtain this operator
and calculate the mass terms with the explicit eigenvalues
along the lines of \cite{CHR1406,HR1503,CH1506,GPZ1506}.
Under the large $c$ limit, the right hand side of the OPE
has the simple linear terms.

$\bullet$ ${\cal N}=2$ superspace description for the adjoint fermions 

We obtained the two ${\cal N}=2$ higher spin multiplets. It is an
open problem to see whether one can write down the two adjoint
fermions in ${\cal N}=2$
superspace. This will allow us to write down the ${\cal N}=2$
higher spin multiplets in ${\cal N}=2$ superspace.

$\bullet$ Asymptotic quantum symmetry algebra

We have obtained the eigenvalue equations and three-point functions
at finite $N$ and $k_2$ in section $2$. 
Along the line of \cite{GG1205}, it is an open problem to
study the asymptotic quantum symmetry algebra of the higher spin theory
on the $AdS_3$ space. See also \cite{GV1106} where the brief sketch
for the large $N$ 't Hooft limit is given.

\vspace{.7cm}

\centerline{\bf Acknowledgments}

CA acknowledges warm hospitality from 
the School of  Liberal Arts (and Institute of Convergence Fundamental
Studies), Seoul National University of Science and Technology.
This research was supported by Basic Science Research Program through
the National Research Foundation of Korea  
funded by the Ministry of Education  
(No. 2016R1D1A1B03931786).

\newpage

\appendix

\renewcommand{\theequation}{\Alph{section}\mbox{.}\arabic{equation}}

\section{ The $f$ and $d$ tensors of $SO(8)$}

The $28$ generators of $SO(8)$ \cite{AP1410} are given by
{\small
\bea
T^1 & = &
\left(
\begin{array}{cccccccccc}
0 & i &0 &0 &0 &0 &0 &0  \\
-i& 0 &0 &0 &0 &0 &0 &0 \\
0 & 0 &0 &0 &0 &0 &0 &0\\
0 & 0 &0 & 0 & 0  &0 &0 &0\\
0 & 0 &0 & 0 & 0 &0 &0 &0\\
0 & 0 &0 & 0 & 0 &0 &0 &0\\
0 & 0 &0 & 0 & 0 &0 &0 &0\\
0 & 0 &0 & 0 & 0 &0 &0 &0\\
\end{array} \right),
T^2 =
\left(
\begin{array}{cccccccccc}
0 & 0 &i &0 &0 &0 &0 &0  \\
0& 0 &0 &0 &0 &0 &0 &0 \\
-i & 0 &0 &0 &0 &0 &0 &0\\
0 & 0 &0 & 0 & 0  &0 &0 &0\\
0 & 0 &0 & 0 & 0 &0 &0 &0\\
0 & 0 &0 & 0 & 0 &0 &0 &0\\
0 & 0 &0 & 0 & 0 &0 &0 &0\\
0 & 0 &0 & 0 & 0 &0 &0 &0\\
\end{array} \right),
\nonu \\
T^3  & = &
\left(
\begin{array}{cccccccccc}
0 & 0 &0 &i &0 &0 &0 &0  \\
0& 0 &0 &0 &0 &0 &0 &0 \\
0 & 0 &0 &0 &0 &0 &0 &0\\
-i & 0 &0 & 0 & 0  &0 &0 &0\\
0 & 0 &0 & 0 & 0 &0 &0 &0\\
0 & 0 &0 & 0 & 0 &0 &0 &0\\
0 & 0 &0 & 0 & 0 &0 &0 &0\\
0 & 0 &0 & 0 & 0 &0 &0 &0\\
\end{array} \right),
T^4 =
\left(
\begin{array}{cccccccccc}
0 & 0 &0 &0 &i &0 &0 &0  \\
0& 0 &0 &0 &0 &0 &0 &0 \\
0 & 0 &0 &0 &0 &0 &0 &0\\
0 & 0 &0 & 0 & 0  &0 &0 &0\\
-i & 0 &0 & 0 & 0 &0 &0 &0\\
0 & 0 &0 & 0 & 0 &0 &0 &0\\
0 & 0 &0 & 0 & 0 &0 &0 &0\\
0 & 0 &0 & 0 & 0 &0 &0 &0\\
\end{array} \right),
\nonu \\
T^5  & = &
\left(
\begin{array}{cccccccccc}
0 & 0 &0 &0 &0 &i &0 &0  \\
0& 0 &0 &0 &0 &0 &0 &0 \\
0 & 0 &0 &0 &0 &0 &0 &0\\
0 & 0 &0 & 0 & 0  &0 &0 &0\\
0 & 0 &0 & 0 & 0 &0 &0 &0\\
-i & 0 &0 & 0 & 0 &0 &0 &0\\
0 & 0 &0 & 0 & 0 &0 &0 &0\\
0 & 0 &0 & 0 & 0 &0 &0 &0\\
\end{array} \right),
T^6  =
\left(
\begin{array}{cccccccccc}
0 & 0 &0 &0 &0 &0 &i &0  \\
0& 0 &0 &0 &0 &0 &0 &0 \\
0 & 0 &0 &0 &0 &0 &0 &0\\
0 & 0 &0 & 0 & 0  &0 &0 &0\\
0 & 0 &0 & 0 & 0 &0 &0 &0\\
0 & 0 &0 & 0 & 0 &0 &0 &0\\
-i & 0 &0 & 0 & 0 &0 &0 &0\\
0 & 0 &0 & 0 & 0 &0 &0 &0\\
\end{array} \right),
\nonu \\
T^7 & = &
\left(
\begin{array}{cccccccccc}
0 & 0 &0 &0 &0 &0 &0 &i  \\
0& 0 &0 &0 &0 &0 &0 &0 \\
0 & 0 &0 &0 &0 &0 &0 &0\\
0 & 0 &0 & 0 & 0  &0 &0 &0\\
0 & 0 &0 & 0 & 0 &0 &0 &0\\
0 & 0 &0 & 0 & 0 &0 &0 &0\\
0 & 0 &0 & 0 & 0 &0 &0 &0\\
-i & 0 &0 & 0 & 0 &0 &0 &0\\
\end{array} \right),
T^{8} =
\left(
\begin{array}{cccccccccc}
0 & 0 &0 &0 &0 &0 &0 &0  \\
0& 0 &i &0 &0 &0 &0 &0 \\
0 & -i &0 &0 &0 &0 &0 &0\\
0 & 0 &0 & 0 & 0  &0 &0 &0\\
0 & 0 &0 & 0 & 0 &0 &0 &0\\
0 & 0 &0 & 0 & 0 &0 &0 &0\\
0 & 0 &0 & 0 & 0 &0 &0 &0\\
0 & 0 &0 & 0 & 0 &0 &0 &0\\
\end{array} \right),
\nonu \\
T^9 & = &
\left(
\begin{array}{cccccccccc}
0 & 0 &0 &0 &0 &0 &0 &0  \\
0& 0 &0 &i &0 &0 &0 &0 \\
0 & 0 &0 &0 &0 &0 &0 &0\\
0 & -i &0 & 0 & 0  &0 &0 &0\\
0 & 0 &0 & 0 & 0 &0 &0 &0\\
0 & 0 &0 & 0 & 0 &0 &0 &0\\
0 & 0 &0 & 0 & 0 &0 &0 &0\\
0 & 0 &0 & 0 & 0 &0 &0 &0\\
\end{array} \right),
T^{10} =
\left(
\begin{array}{cccccccccc}
0 & 0 &0 &0 &0 &0 &0 &0  \\
0& 0 &0 &0 &i &0 &0 &0 \\
0 & 0 &0 &0 &0 &0 &0 &0\\
0 & 0 &0 & 0 & 0  &0 &0 &0\\
0 & -i &0 & 0 & 0 &0 &0 &0\\
0 & 0 &0 & 0 & 0 &0 &0 &0\\
0 & 0 &0 & 0 & 0 &0 &0 &0\\
0 & 0 &0 & 0 & 0 &0 &0 &0\\
\end{array} \right),
\nonu \\
T^{11} & = &
\left(
\begin{array}{cccccccccc}
0 & 0 &0 &0 &0 &0 &0 &0  \\
0& 0 &0 &0 &0 &i &0 &0 \\
0 & 0 &0 &0 &0 &0 &0 &0\\
0 & 0 &0 & 0 & 0  &0 &0 &0\\
0 & 0 &0 & 0 & 0 &0 &0 &0\\
0 & -i &0 & 0 & 0 &0 &0 &0\\
0 & 0 &0 & 0 & 0 &0 &0 &0\\
0 & 0 &0 & 0 & 0 &0 &0 &0\\
\end{array} \right),
T^{12} =
\left(
\begin{array}{cccccccccc}
0 & 0 &0 &0 &0 &0 &0 &0  \\
0& 0 &0 &0 &0 &0 &i &0 \\
0 & 0 &0 &0 &0 &0 &0 &0\\
0 & 0 &0 & 0 & 0  &0 &0 &0\\
0 & 0 &0 & 0 & 0 &0 &0 &0\\
0 & 0 &0 & 0 & 0 &0 &0 &0\\
0 &-i &0 & 0 & 0 &0 &0 &0\\
0 & 0 &0 & 0 & 0 &0 &0 &0\\
\end{array} \right),
\nonu \\
T^{13} & = &
\left(
\begin{array}{cccccccccc}
0 & 0 &0 &0 &0 &0 &0 &0  \\
0& 0 &0 &0 &0 &0 &0 &i \\
0 & 0 &0 &0 &0 &0 &0 &0\\
0 & 0 &0 & 0 & 0  &0 &0 &0\\
0 & 0 &0 & 0 & 0 &0 &0 &0\\
0 & 0 &0 & 0 & 0 &0 &0 &0\\
0 & 0 &0 & 0 & 0 &0 &0 &0\\
0 & -i &0 & 0 & 0 &0 &0 &0\\
\end{array} \right),
T^{14} =
\left(
\begin{array}{cccccccccc}
0 & 0 &0 &0 &0 &0 &0 &0  \\
0& 0 &0 &0 &0 &0 &0 &0 \\
0 & 0 &0 &i &0 &0 &0 &0\\
0 & 0 &-i & 0 & 0  &0 &0 &0\\
0 & 0 &0 & 0 & 0 &0 &0 &0\\
0 & 0 &0 & 0 & 0 &0 &0 &0\\
0 & 0 &0 & 0 & 0 &0 &0 &0\\
0 & 0 &0 & 0 & 0 &0 &0 &0\\
\end{array} \right),
\nonu \\
T^{15} & = &
\left(
\begin{array}{cccccccccc}
0 & 0 &0 &0 &0 &0 &0 &0  \\
0 & 0 &0 &0 &0 &0 &0 &0 \\
0 & 0 &0 &0 &i &0 &0 &0\\
0 & 0 &0 & 0 & 0  &0 &0 &0\\
0 & 0 &-i & 0 & 0 &0 &0 &0\\
0 & 0 &0 & 0 & 0 &0 &0 &0\\
0 & 0 &0 & 0 & 0 &0 &0 &0\\
0 & 0 &0 & 0 & 0 &0 &0 &0\\
\end{array} \right),
T^{16} =
\left(
\begin{array}{cccccccccc}
0 & 0 &0 &0 &0 &0 &0 &0  \\
0 & 0 &0 &0 &0 &0 &0 &0 \\
0 & 0 &0 &0 &0 &i &0 &0\\
0 & 0 &0 & 0 & 0  &0 &0 &0\\
0 & 0 &0 & 0 & 0 &0 &0 &0\\
0 & 0 &-i & 0 & 0 &0 &0 &0\\
0 & 0 &0 & 0 & 0 &0 &0 &0\\
0 & 0 &0 & 0 & 0 &0 &0 &0\\
\end{array} \right),
\nonu \\
T^{17}  & = &
\left(
\begin{array}{cccccccccc}
0 & 0 &0 &0 &0 &0 &0 &0  \\
0 & 0 &0 &0 &0 &0 &0 &0 \\
0 & 0 &0 &0 &0 &0 &i &0\\
0 & 0 &0 & 0 & 0  &0 &0 &0\\
0 & 0 &0 & 0 & 0 &0 &0 &0\\
0 & 0 &0 & 0 & 0 &0 &0 &0\\
0 & 0 &-i & 0 & 0 &0 &0 &0\\
0 & 0 &0 & 0 & 0 &0 &0 &0\\
\end{array} \right),
T^{18} =
\left(
\begin{array}{cccccccccc}
0 & 0 &0 &0 &0 &0 &0 &0  \\
0 & 0 &0 &0 &0 &0 &0 &0 \\
0 & 0 &0 &0 &0 &0 &0 &i\\
0 & 0 &0 & 0 & 0  &0 &0 &0\\
0 & 0 &0 & 0 & 0 &0 &0 &0\\
0 & 0 &0 & 0 & 0 &0 &0 &0\\
0 & 0 &0 & 0 & 0 &0 &0 &0\\
0 & 0 &-i & 0 & 0 &0 &0 &0\\
\end{array} \right),
\nonu \\
T^{19}  & = &
\left(
\begin{array}{cccccccccc}
0 & 0 &0 &0 &0 &0 &0 &0  \\
0 & 0 &0 &0 &0 &0 &0 &0 \\
0 & 0 &0 &0 &0 &0 &0 &0\\
0 & 0 &0 & 0 & i  &0 &0 &0\\
0 & 0 &0 & -i & 0 &0 &0 &0\\
0 & 0 &0 & 0 & 0 &0 &0 &0\\
0 & 0 &0 & 0 & 0 &0 &0 &0\\
0 & 0 &0 & 0 & 0 &0 &0 &0\\
\end{array} \right),
T^{20} =
\left(
\begin{array}{cccccccccc}
0 & 0 &0 &0 &0 &0 &0 &0  \\
0 & 0 &0 &0 &0 &0 &0 &0 \\
0 & 0 &0 &0 &0 &0 &0 &0\\
0 & 0 &0 & 0 & 0  &i &0 &0\\
0 & 0 &0 & 0 & 0 &0 &0 &0\\
0 & 0 &0 & -i & 0 &0 &0 &0\\
0 & 0 &0 & 0 & 0 &0 &0 &0\\
0 & 0 &0 & 0 & 0 &0 &0 &0\\
\end{array} \right),
\nonu \\
T^{21} & = &
\left(
\begin{array}{cccccccccc}
0 & 0 &0 &0 &0 &0 &0 &0  \\
0 & 0 &0 &0 &0 &0 &0 &0 \\
0 & 0 &0 &0 &0 &0 &0 &0\\
0 & 0 &0 & 0 & 0  &0 &i &0\\
0 & 0 &0 & 0 & 0 &0 &0 &0\\
0 & 0 &0 & 0 & 0 &0 &0 &0\\
0 & 0 &0 & -i & 0 &0 &0 &0\\
0 & 0 &0 & 0 & 0 &0 &0 &0\\
\end{array} \right),
T^{22} =
\left(
\begin{array}{cccccccccc}
0 & 0 &0 &0 &0 &0 &0 &0  \\
0 & 0 &0 &0 &0 &0 &0 &0 \\
0 & 0 &0 &0 &0 &0 &0 &0\\
0 & 0 &0 & 0 & 0  &0 &0 &i\\
0 & 0 &0 & 0 & 0 &0 &0 &0\\
0 & 0 &0 & 0 & 0 &0 &0 &0\\
0 & 0 &0 & 0 & 0 &0 &0 &0\\
0 & 0 &0 & -i & 0 &0 &0 &0\\
\end{array} \right),
\nonu \\
T^{23} & = &
\left(
\begin{array}{cccccccccc}
0 & 0 &0 &0 &0 &0 &0 &0  \\
0 & 0 &0 &0 &0 &0 &0 &0 \\
0 & 0 &0 &0 &0 &0 &0 &0\\
0 & 0 &0 & 0 & 0  &0 &0 &0\\
0 & 0 &0 & 0 & 0 &i &0 &0\\
0 & 0 &0 & 0 & -i &0 &0 &0\\
0 & 0 &0 & 0 & 0 &0 &0 &0\\
0 & 0 &0 & 0 & 0 &0 &0 &0\\
\end{array} \right),
T^{24} =
\left(
\begin{array}{cccccccccc}
0 & 0 &0 &0 &0 &0 &0 &0  \\
0 & 0 &0 &0 &0 &0 &0 &0 \\
0 & 0 &0 &0 &0 &0 &0 &0\\
0 & 0 &0 & 0 & 0  &0 &0 &0\\
0 & 0 &0 & 0 & 0 &0 &i &0\\
0 & 0 &0 & 0 & 0 &0 &0 &0\\
0 & 0 &0 & 0 & -i &0 &0 &0\\
0 & 0 &0 & 0 & 0 &0 &0 &0\\
\end{array} \right),
\nonu \\
T^{25} & = &
\left(
\begin{array}{cccccccccc}
0 & 0 &0 &0 &0 &0 &0 &0  \\
0 & 0 &0 &0 &0 &0 &0 &0 \\
0 & 0 &0 &0 &0 &0 &0 &0\\
0 & 0 &0 & 0 & 0  &0 &0 &0\\
0 & 0 &0 & 0 & 0 &0 &0 &i\\
0 & 0 &0 & 0 & 0 &0 &0 &0\\
0 & 0 &0 & 0 & 0 &0 &0 &0\\
0 & 0 &0 & 0 & -i &0 &0 &0\\
\end{array} \right),
T^{26} =
\left(
\begin{array}{cccccccccc}
0 & 0 &0 &0 &0 &0 &0 &0  \\
0 & 0 &0 &0 &0 &0 &0 &0 \\
0 & 0 &0 &0 &0 &0 &0 &0\\
0 & 0 &0 & 0 & 0  &0 &0 &0\\
0 & 0 &0 & 0 & 0 &0 &0 &0\\
0 & 0 &0 & 0 & 0 &0 &i &0\\
0 & 0 &0 & 0 & 0 &-i &0 &0\\
0 & 0 &0 & 0 & 0 &0 &0 &0\\
\end{array} \right),
\nonu \\
T^{27} & = &
\left(
\begin{array}{cccccccccc}
0 & 0 &0 &0 &0 &0 &0 &0  \\
0 & 0 &0 &0 &0 &0 &0 &0 \\
0 & 0 &0 &0 &0 &0 &0 &0\\
0 & 0 &0 & 0 & 0  &0 &0 &0\\
0 & 0 &0 & 0 & 0 &0 &0 &0\\
0 & 0 &0 & 0 & 0 &0 &0 &i\\
0 & 0 &0 & 0 & 0 &0 &0 &0\\
0 & 0 &0 & 0 & 0 &-i &0 &0\\
\end{array} \right),
T^{28} =
\left(
\begin{array}{cccccccccc}
0 & 0 &0 &0 &0 &0 &0 &0  \\
0 & 0 &0 &0 &0 &0 &0 &0 \\
0 & 0 &0 &0 &0 &0 &0 &0\\
0 & 0 &0 & 0 & 0  &0 &0 &0\\
0 & 0 &0 & 0 & 0 &0 &0 &0\\
0 & 0 &0 & 0 & 0 &0 &0 &0\\
0 & 0 &0 & 0 & 0 &0 &0 &i\\
0 & 0 &0 & 0 & 0 &0 &-i &0\\
\end{array} \right).
\label{SO8generators}
\eea
}
Using this explicit form of Appendix (\ref{SO8generators}),
the structure constants (\ref{fdef}) and $d$ tensor
(\ref{dabcdexp}) can be written explicitly.
Inside of mathematica \cite{mathematica},
one can write down the $f$ and $d$ symbols
as follows:
\bea
 {\tt f} & = & {\tt Table[Simplify[\frac{1}{2} (-i)
      Tr[T[c1].(T[a1].T[b1]-
        T[b1].T[a1])]],}
     \nonu \\
    &&   {\tt 
      \{a1,1,28\},\{b1,1,28\},\{c1,1,28\}];}
 \nonu \\
       { \tt d} & = & { \tt Table[Simplify[\frac{1}{2}
             Tr[T[d1].(T[a1].T[b1].T[c1]+
T[a1].T[c1].T[b1]+
T[c1].T[a1].T[b1] + }\nonu \\
 & & {\tt 
T[b1].T[a1].T[c1]+
T[b1].T[c1].T[a1]+
T[c1].T[b1].T[a1])]],}
         \nonu \\
         && {\tt 
           \{a1,1,28\},
           \{b1,1,28\},
           \{c1,1,28\},
           \{d1,1,28\}];}
 \label{fdrel}
\eea
Using the command ${\tt Union[Flatten[f]]}$,
the elements of $f$ symbol in Appendix (\ref{fdrel})
is given by $-1, 0$, and $1$.
  Using the command  ${\tt Length[Position[f,*]]}$,
    the multiplicities are given by $168, 21616$, and $168$ respectively.
The sum of these is $28^3=21952$.
    For the $d$ symbol, the elements  are given by
    $-1,0,1,2$, and $6$ and their multiplicities are
    $1680, 608580,3360,1008$, and $28$.
The sum of these is $28^4=614656$.
    In particular, $d^{aaaa}=6$.
    
\section{ The OPEs between the diagonal spin $1$ current and
various spin $4$ currents}

As done in the section $2$ (the equations (\ref{eqeq}), (\ref{eqeq1})
and (\ref{eqeq2})),
the remaining
$17$ OPEs are described by
{\small
\bea
&& J^{\prime a}(z) \, d^{bcde} J^b K^c K^d K^e(w)  = 
\frac{1}{(z-w)^3} \, (4N^2-14N+22) f^{abc} K^b J^c(w)
\nonu \\
& & +
\frac{1}{(z-w)^2} \,
\Bigg[ -(3k_2 +4(N-1)) d^{abcd} K^b K^c J^d
  + (2N-5) f^{abc} f^{cde}  K^b  K^e J^d  \nonu \\
  & & - k_1 d^{abcd} K^b K^c K^d + 12 K^b K^b J^a
  -12 K^a K^b J^b  - 12 f^{abc} \pa K^b J^c
  \Bigg](w) +\cdots,
\nonu \\
&& J^{\prime a}(z) \, d^{bcde} K^b K^c K^d K^e(w)  = 
\frac{1}{(z-w)^4} \, 2(2N-2)(4N^2-14N+22) K^a(w) \nonu \\
& & - 
\frac{1}{(z-w)^3} \, 2(4N^2-14N+22) f^{abc} K^b K^c(w)
\nonu \\
& & +
\frac{1}{(z-w)^2} \,
\Bigg[ -(4k_2 +8(N-1)) d^{abcd} K^b K^c K^d
  \nonu \\
  & & - (12+(2N-2)(2N-5)) f^{abc} \pa K^b K^c
   + (12+(2N-2)(2N-5)) f^{abc}  K^b \pa K^c
  \Bigg](w) +\cdots,
\nonu \\
&& J^{\prime a}(z) \, \pa J^b \pa J^b(w)  = 
-\frac{1}{(z-w)^4} \, 2(2N-2) J^a(w)  - 
\frac{1}{(z-w)^3} \, 2((2N-2) + 2k_1) \pa J^a(w)
\nonu \\
&& +  \frac{1}{(z-w)^2} \, \Bigg[ f^{abc} \pa J^b J^c
  - f^{abc} J^b \pa J^c \Bigg](w)+ \cdots,
\nonu \\
&& J^{\prime a}(z) \, \pa^2 J^b  J^b(w)  = 
-\frac{1}{(z-w)^4} \, 2(2(2N-2)+3k_1) J^a(w)  - 
\frac{1}{(z-w)^3} \, 2(2N-2) \pa J^a(w)
\nonu \\
&& -  \frac{1}{(z-w)^2} \, \Bigg[ \frac{k_1}{(2N-2)} f^{abc}  J^b \pa J^c
  +(2+\frac{k_1}{(2N-2)}) f^{abc} \pa J^b J^c \Bigg](w)
+  \cdots,
\nonu \\
&& J^{\prime a}(z) \, \pa K^b \pa K^b(w)  = 
-\frac{1}{(z-w)^4} \, 2(2N-2) K^a(w)  - 
\frac{1}{(z-w)^3} \, 2((2N-2)+2k_2) \pa K^a(w)
\nonu \\
&& +  \frac{1}{(z-w)^2} \, \Bigg[  f^{abc}  \pa K^b  K^c
  - f^{abc} K^b \pa K^c \Bigg](w)  +  \cdots,
\nonu \\
&& J^{\prime a}(z) \, \pa^2 K^b  K^b(w)  = 
-\frac{1}{(z-w)^4} \, 2(2(2N-2)+3k_2) K^a(w)  - 
\frac{1}{(z-w)^3} \, 2(2N-2) \pa K^a(w)
\nonu \\
&& -  \frac{1}{(z-w)^2} \, \Bigg[ \frac{k_2}{(2N-2)} f^{abc}  K^b \pa K^c
  +(2+\frac{k_2}{(2N-2)}) f^{abc} \pa K^b K^c \Bigg](w)
+  \cdots,
\nonu \\
&& J^{\prime a}(z) \, \pa J^b \pa K^b(w)  = 
 -
\frac{1}{(z-w)^3} \, 2( k_1 \pa K^a + k_2 \pa J^a)(w)
\nonu \\
&& +  \frac{1}{(z-w)^2} \, \Bigg[  f^{abc}  \pa J^b  K^c
  - f^{abc} J^b \pa K^c \Bigg](w)+ \cdots,
\nonu \\
&& J^{\prime a}(z) \, \pa^2 J^b  K^b(w)  = 
-\frac{1}{(z-w)^4} \, 6k_1 K^a(w) -
\frac{1}{(z-w)^3} \, 2 f^{abc}  J^b K^c(w)
\nonu \\
&& -  \frac{1}{(z-w)^2} \, \Bigg[ \frac{k_2}{(2N-2)} f^{abc}  \pa J^b  J^c
  +\frac{k_2}{(2N-2)} f^{abc}  J^b \pa J^c
  +  2 f^{abc}  \pa J^b  K^c
  \Bigg](w)+ \cdots,
\nonu \\
&& J^{\prime a}(z) \,  J^b \pa^2 K^b(w)  = 
-\frac{1}{(z-w)^4} \, 6k_2 J^a(w) -
\frac{1}{(z-w)^3} \, 2 f^{abc}  K^b J^c(w)
\nonu \\
&& -  \frac{1}{(z-w)^2} \, \Bigg[ \frac{k_1}{(2N-2)} f^{abc}  \pa K^b  K^c
  +\frac{k_1}{(2N-2)} f^{abc}  K^b \pa K^c
  +  2 f^{abc}  \pa K^b  J^c
  \Bigg](w)+ \cdots,
\nonu \\
&& J^{\prime a}(z) \, f^{bcd} J^b \pa J^c K^d (w)  = 
-\frac{1}{(z-w)^4} \, 8k_1 (N-1) K^a(w) 
\nonu \\
&& -  \frac{1}{(z-w)^3} 2(N-1-k_1) f^{abc} J^b K^c(w)
\nonu \\
&& +  \frac{1}{(z-w)^2} \Bigg[-(2(N-1)+k_1) f^{abc} \pa J^b K^c
+ f^{abc} f^{cde} J^d J^b K^e
 -  k_2 f^{abc}  J^b \pa J^c
\Bigg](w) + \cdots,
\nonu \\
&& J^{\prime a}(z) \, f^{bcd} J^b  K^c \pa K^d (w)  = 
-\frac{1}{(z-w)^4} \, 8k_2 (N-1) J^a(w) 
\nonu \\
&& -  \frac{1}{(z-w)^3} 2(N-1-k_2) f^{abc} K^b J^c(w)
\nonu \\
&& +  \frac{1}{(z-w)^2} \Bigg[-(2(N-1)+k_2) f^{abc} \pa K^b J^c
+ f^{abc} f^{cde} K^d K^b J^e
 -  k_1 f^{abc}  K^b \pa K^c
\Bigg](w) + \cdots,
\nonu \\
&& J^{\prime a}(z) \, J^b J^b J^c J^c(w)  = 
\frac{1}{(z-w)^4} \,  4(2N-2)(2N-2+k_1) J^a(w) 
\nonu \\
&& -  \frac{1}{(z-w)^3} 4(2N-2)(2N-2+k_1) \pa J^a(w)
\nonu \\
&& +  \frac{1}{(z-w)^2} \Bigg[-4(2(N-1)+k_1)  J^a J^b J^b
   -  2(2N-2+k_1) f^{abc}  \pa J^b J^c
   \nonu \\
   && +  2(2N-2+k_1) f^{abc}  J^b \pa J^c
\Bigg](w) + \cdots,
\nonu \\
&& J^{\prime a}(z) \, K^b K^b K^c K^c(w)  =
\frac{1}{(z-w)^4} \,  4(2N-2)(2N-2+k_2) K^a(w) 
\nonu \\
&& -  \frac{1}{(z-w)^3} 4(2N-2)(2N-2+k_2) \pa K^a(w)
 +  \frac{1}{(z-w)^2} \Bigg[-4(2(N-1)+k_2)  K^a K^b K^b
  \nonu \\
  && -  2(2N-2+k_2) f^{abc}  \pa K^b K^c
 +  2(2N-2+k_2) f^{abc}  K^b \pa K^c
\Bigg](w) + \cdots,
\nonu \\
&& J^{\prime a}(z) \, J^b J^b K^c K^c(w)  = 
 -\frac{1}{(z-w)^2} \Bigg[2(2(N-1)+k_1)  J^a K^b K^b
   \nonu \\
   & & +   2(2N-2+k_2)   J^b  J^b K^a
\Bigg](w) + \cdots,
\nonu \\
&& J^{\prime a}(z) \, J^b J^b J^c K^c(w)  = 
\frac{1}{(z-w)^4} \,  2(2N-2) k_1 K^a(w) 
 +  \frac{1}{(z-w)^3} 2(2N-2-k_1) f^{abc} J^b K^c(w)
\nonu \\
&& +  \frac{1}{(z-w)^2} \Bigg[-k_2  J^a J^b J^b
-k_2 f^{abc}  \pa J^b J^c
 +  k_2 f^{abc}  J^b \pa J^c
-k_1 J^b J^b K^a
\nonu \\
&& - 2(k_1+2N-2) J^a J^b K^b
 + 2 f^{abc} f^{cde} J^b J^e K^d
\Bigg](w) + \cdots,
\nonu \\
&& J^{\prime a}(z) \, J^b K^b K^c K^c(w)  =
 -\frac{1}{(z-w)^3} 2(2N-2+k_2) f^{abc} J^b K^c(w)
\nonu \\
&& +  \frac{1}{(z-w)^2} \Bigg[-k_1  K^a K^b K^b
-k_2 J^a K^b K^b
\nonu \\
&& -  2(2N-2 +k_2)   J^b K^a K^b
+ 2(k_2+2N-2) f^{abc} J^b \pa K^c
\Bigg](w)  +  \cdots,
\nonu \\
&& J^{\prime a}(z) \, J^b J^c K^b K^c(w)  =  
 -\frac{1}{(z-w)^3} (2N-2)(k_1 \pa K^a + k_2 \pa J^a)(w)
\nonu \\
&& +  \frac{1}{(z-w)^2} \Bigg[f^{abc} f^{cde} J^e K^b K^d
+ f^{abc} f^{cde} J^b J^d K^e
\nonu \\
&& -  2 k_1   J^b K^a K^b
+ k_1 f^{abc} J^b \pa K^c
-k_2 f^{abc} \pa J^b K^c
-2 k_2 J^a J^b K^b
\Bigg](w)  +  \cdots.
\label{rem17}
\eea}
One observes that there exists an invariance under the change of
$J^a(z) \leftrightarrow K^a(z)$ and $k_1 \leftrightarrow k_2$.
For example, the first equation of Appendix (\ref{rem17})
can be obtained from (\ref{eqeq1}) by this change.

We collect each independent  field with its coefficients in (\ref{w4})
appearing at various poles in Appendix (\ref{rem17})
where the rearrangement \cite{BBSSfirst}
of the normal ordered product
is used
and the coefficients should vanish
in order to satisfy the regular condition (\ref{jpw})
\bea
\mbox{pole-4}: && \Bigg[ 2 (2 N-2) (4 N^2-14 N+22) A_1-
  2 (2 N-2) A_6-2 (2 (2 N-2)+3 k_1) A_7\nonu \\
  && -6 k_2 A_{12}-
  k_2 4 (2 N-2) A_{14}+4 (2 N-2) (k_1+2 N-2) A_{15} \Bigg] J^a(w)=0,
\nonu \\
\mbox{pole-4}: && \Bigg[ 2 (2 N-2) (4 N^2-14 N+22) A_5-
  2 (2 N-2) A_8-2 (2 (2 N-2)+3 k_2) A_9 \nonu \\
  && -6 k_1 A_{11}-
  k_1 4 (2 N-2) A_{13}+4 (2 N-2) (k_2+2 N-2) A_{16} \nonu \\
  && +2 (2 N-2) k_1 A_{18}
  \Bigg] K^a(w) =0,
\nonu \\
\mbox{pole-3}: &&
\Bigg[ -2 (4 N^2-14 N+22) (2 N-2) A_1-2 ((2 N-2)+2 k_1) A_6-
  2 (2 N-2) A_7\nonu \\
  && -2 k_2 A_{10}-4 (2 N-2) (k_1+2 N-2) A_{15}-
  (2 N-2) k_2 A_{20} \Bigg] \pa J^a(w)=0,
\nonu \\
\mbox{pole-3}: && \Bigg[-2 (2 N-2) (4 N^2-14 N+22) A_5-
  2 ((2 N-2)+2 k_2) A_8-2 (2 N-2) A_9\nonu \\
  && -2 k_1 A_{10}-
  4 (2 N-2) (k_2+2 N-2) A_{16}-(2 N-2) k_1 A_{20} \Bigg] \pa K^a(w)=0,
\nonu \\
\mbox{pole-3}: && \Bigg[ (4 N^2-14 N+22) A_2-(4 N^2-14 N+22) A_4-
  2 A_{11}+2 A_{12}\nonu \\
  && -((2 N-2)-2 k_1) A_{13}+((2 N-2)-2 k_2) A_{14}+
  2 (-k_1+2 N-2) A_{18} \nonu \\
  && -2 (k_2+2 N-2) A_{19} \Bigg] f^{abc} J^b K^c(w)=0,
\nonu \\
\mbox{pole-2}: &&   \Bigg[ ((2 N-5) (2 N-2)+12) A_1-A_6-
  \frac{k_1 }{(2 N-2)} A_7-\frac{k_2 }{(2 N-2)} A_{11}-
  k_2 A_{13}\nonu \\
  && +2 (k_1+2 N-2) A_{15}+k_2 A_{18} \Bigg] f^{abc} J^b \pa J^c(w)=0,
\nonu \\
\mbox{pole-2}: && \Bigg[ -((2 N-5) (2 N-2)+12) A_1+A_6-
  (\frac{k_1}{(2 N-2)}+2) A_7-\frac{k_2 }{(2 N-2)} A_{11}
  \nonu \\
  && -
  2 (k_1+2 N-2) A_{15}-k_2 A_{18} \Bigg] f^{abc} \pa J^b J^c(w)=0,
\nonu \\
\mbox{pole-2}: && \Bigg[ ((2 N-5) (2 N-2)+12) A_5-A_8-
  \frac{k_2 }{(2 N-2)} A_9-\frac{k_1 }{(2 N-2)} A_{12}-
  k_1 A_{14}\nonu
\\
  && +2 (k_2+2 N-2) A_{16} \Bigg] f^{abc} K^b \pa K^c(w)=0,
\nonu \\
\mbox{pole-2}: && \Bigg[-((2 N-5) (2 N-2)+12) A_5-
  (\frac{k_2}{(2 N-2)}+2) A_9 +A_8-\frac{k_1 }{(2 N-2)} A_{12}
  \nonu \\
  && -
  2 (k_2+2 N-2) A_{16} \Bigg] f^{abc} \pa K^b K^c(w)=0,
\nonu \\
\mbox{pole-2}: && \Bigg[ -4 (k_1+2 N-2) A_{15}-k_2 A_{18} \Bigg]
J^a J^b J^b(w)=0,
\nonu \\
\mbox{pole-2}: && \Bigg[ 12 A_2-8 A_3-2 (k_2+2 N-2) A_{17}-
  k_1 A_{18} \Bigg] J^b J^b K^a(w)=0,
\nonu \\
\mbox{pole-2}: && \Bigg[ -12 A_2+8 A_3-2 (k_1+2 N-2) A_{18}-
  2 k_2 A_{20} \Bigg] J^a J^b K^b(w)=0,
\nonu \\
\mbox{pole-2}: && \Bigg[ 8 A_3-12 A_4-2 (k_2+2 N-2) A_{19}-
  2 k_1 A_{20} \Bigg] J^b K^a K^b(w) =0,
\nonu \\
\mbox{pole-2}: && \Bigg[ -8 A_3+12 A_4-2 (k_1+2 N-2) A_{17}-
  k_2 A_{19} \Bigg] J^a K^b K^b(w)=0,
\nonu \\
\mbox{pole-2}: && \Bigg[ -4 (k_2+2 N-2) A_{16}-k_1 A_{19} \Bigg]
 K^a K^b K^b(w) =0,
\nonu \\
\mbox{pole-2}: && \Bigg[ -(8 (N-1)+4 k_1) A_1-k_2 A_2 \Bigg]
d^{abcd} J^b J^c J^d(w)=0,
\nonu \\
\mbox{pole-2}: && \Bigg[ -(4 (N-1)+3 k_1) A_2-
  (\frac{4 (N-1)}{3}+2 k_2) A_3 \Bigg] d^{abcd} J^b J^c K^d(w) =0,
\nonu \\
\mbox{pole-2}: && \Bigg[ -(\frac{4 (N-1)}{3}+2 k_1) A_3-
  (4 (N-1)+3 k_2) A_4 \Bigg] d^{abcd} J^b K^c K^d(w)=0,
\nonu \\
\mbox{pole-2}: && \Bigg[-k_1 A_4-(8 (N-1)+4 k_2) A_5 \Bigg] d^{abcd}
K^b K^c K^d(w)=0, 
\nonu \\
\mbox{pole-2}: && \Bigg[
   (k_2+(2 N-2)) A_{14}+2  (k_2+2 N-2) A_{19}-
  (\frac{1}{3}  (2 N-2) (2 N-5) +4) A_3 \nonu \\
  && + (12+(2 N-5) (2 N-2)) A_4-
   (k_1+ 2 N-2) A_{20}-A_{10}+2 A_{12}
  \Bigg] f^{abc} J^b \pa K^c(w) \nonu \\
&& =0,
\nonu \\
\mbox{pole-2}: && \Bigg[
  -(k_1+(2 N-2)) A_{13}- k_2 A_{20}-\frac{1}{3}  (2 N-2) (2 N-5) A_3-
   (2 N-2) A_{13} \nonu \\
  && -12 A_2+4 A_3+A_{10}-2 A_{11}
  \Bigg] f^{abc} \pa J^b K^c(w) =0,
\nonu \\
\mbox{pole-2}: && \Bigg[
(2 N-5) A_2-\frac{2}{3}   (2 N-5) A_3-A_{13}+2 A_{18}-A_{20}
  \Bigg] f^{abc} f^{cde} J^b J^e K^d(w) =0,
\nonu \\
\mbox{pole-2}: && \Bigg[
  -\frac{2}{3}   (2 N-5) A_3 +(2 N-5) A_4-A_{14}-A_{20}
  \Bigg] f^{abc} f^{cde} J^d K^e K^b(w)=0.
\label{linearequationforjprime}
\eea
The regular condition provides the vanishing of these coefficients
appearing in Appendix
(\ref{linearequationforjprime}) for the independent
composite fields.

Let us emphasize that one can understand the identity (\ref{dffrelation})
in different point of view. In Appendix (\ref{linearequationforjprime}),
the second
order poles having $A_2$ term with a single $K^a$ are given by
$J^b J^b K^a(w)$, $J^a J^b K^b(w)$, $d^{abcd} J^b J^c K^d(w)$,
$f^{abc} \pa J^b K^c(w)$,
and $f^{abc} f^{cde} J^b J^e K^d(w)$.
Of course, these are obtained from (\ref{intersecond}) with the help of
(\ref{dffrelation}). Although one does not know the tensorial structure
in the right hand side of (\ref{dffrelation}), one can figure out it
from the above field contents.
In other words, one can determine the tensorial structure from the
above requirement (regular condition). Then one can fix the coefficients
appearing in (\ref{dffrelation}) by applying for several lower
$N$ values.

\section{ The OPEs between the numerator spin $2$ current and
various spin $4$ currents}

As done in the section $2$ (the equations (\ref{tjjjj}), (\ref{tjjjk})
and (\ref{tjjkk})),
one obtains the remaining
$17$ OPEs described by
\bea
&& \hat{T}(z) \, d^{bcde} J^b K^c K^d K^e(w)  = 
-\frac{1}{(z-w)^4} \, 6 k_2 (4N-1) J^a K^a(w)
+{\cal O}(\frac{1}{(z-w)^2}),
\nonu \\
&& \hat{T}(z) \, d^{bcde} K^b K^c K^d K^e(w)  = 
-\frac{1}{(z-w)^4} \, 12 k_2 (4N-1) K^a K^a(w)
+{\cal O}(\frac{1}{(z-w)^2}),
\nonu \\
&& \hat{T}(z)
\, \pa J^a \pa J^a(w)  = 
-\frac{1}{(z-w)^6} 4 k_1 N(2N-1) 
+ \frac{1}{(z-w)^3}\,
2  \pa (J^a  J^a)(w)
+{\cal O}(\frac{1}{(z-w)^2}),
\nonu \\
&& \hat{T}(z)
\, \pa^2 J^a  J^a(w)  = 
-\frac{1}{(z-w)^6} 6 k_1 N(2N-1) 
+
\frac{1}{(z-w)^4}\,
 6 J^a J^a(w)
 \nonu \\
 && +  
\frac{1}{(z-w)^3}\,
6  \pa J^a  J^a(w)
+{\cal O}(\frac{1}{(z-w)^2}),
\nonu \\
&& \hat{T}(z)
\, \pa K^a \pa K^a(w)  = 
-\frac{1}{(z-w)^6} 4 k_2 N(2N-1) 
+
\frac{1}{(z-w)^3}\,
2  \pa (K^a  K^a)(w)
+{\cal O}(\frac{1}{(z-w)^2}),
\nonu \\
&& \hat{T}(z)
\, \pa^2 K^a  K^a(w)  = 
-\frac{1}{(z-w)^6} 6 k_2 N(2N-1) 
+
\frac{1}{(z-w)^4}\,
 6 K^a K^a(w)
 \nonu \\
 && +  
\frac{1}{(z-w)^3}\,
6  \pa K^a  K^a(w)
+{\cal O}(\frac{1}{(z-w)^2}),
\nonu \\
&& \hat{T}(z)
\, \pa J^a \pa K^a(w)  = 
\frac{1}{(z-w)^3}\,
2  \pa (J^a  K^a)(w)
+{\cal O}(\frac{1}{(z-w)^2}),
\nonu \\
&& \hat{T}(z)
\, \pa^2 J^a  K^a(w)  = 
\frac{1}{(z-w)^4}\,
 6 J^a K^a(w)
  +  
\frac{1}{(z-w)^3}\,
6  \pa J^a  K^a(w)
+{\cal O}(\frac{1}{(z-w)^2}),
\nonu \\
&& \hat{T}(z)
\,  J^a \pa^2 K^a(w)  = 
\frac{1}{(z-w)^4}\,
 6 J^a K^a(w)
  +  
\frac{1}{(z-w)^3}\,
6  J^a \pa K^a(w)
+{\cal O}(\frac{1}{(z-w)^2}),
\nonu \\
&& \hat{T}(z)
\, f^{abc} J^a \pa J^b K^c (w)  = 
\frac{1}{(z-w)^4}\,
 2(2N-2) J^a K^a(w)
 \nonu \\
 && +  
\frac{1}{(z-w)^3}\,
4(2N-2)  \pa J^a  K^a(w)
+{\cal O}(\frac{1}{(z-w)^2}),
\nonu \\
&& \hat{T}(z)
\, f^{abc} J^a  K^b \pa K^c (w)  = 
\frac{1}{(z-w)^4}\,
 2(2N-2) J^a K^a(w)
 \nonu \\
 && +  
\frac{1}{(z-w)^3}\,
4(2N-2)  J^a \pa K^a(w)
+{\cal O}(\frac{1}{(z-w)^2}),
\nonu \\
&& \hat{T}(z)
\, J^a J^a J^b J^b(w)  = 
-\frac{1}{(z-w)^4}\,
2\Bigg[ 2(2N-2+k_1) +k_1 N (2N-1)  \Bigg]J^a J^a(w)
\nonu \\
&& +  {\cal O}(\frac{1}{(z-w)^2}),
\nonu \\
&& \hat{T}(z)
\, K^a K^a K^b K^b(w)  = 
-\frac{1}{(z-w)^4}\,
2\Bigg[ 2(2N-2+k_2)+ k_2 N(2N-1) \Bigg] K^a K^a(w)
\nonu \\
&& +  {\cal O}(\frac{1}{(z-w)^2}),
\nonu \\
&& \hat{T}(z)
\, J^a J^a K^b K^b(w)  = 
 -\frac{1}{(z-w)^4}\,
N(2N-1) \Bigg[ k_2 J^a J^a + k_1 K^a K^a \Bigg](w) 
+  {\cal O}(\frac{1}{(z-w)^2}),
\nonu \\
&& \hat{T}(z)
\, J^a J^a J^b K^b(w)  = 
 -\frac{1}{(z-w)^4}\,
 \Bigg[ k_1 ( N(2N-1)+ 2 ) \Bigg] J^a K^a(w)
 \nonu \\
 && -  
\frac{1}{(z-w)^3}\,
2(2N-2) \pa J^a K^a(w)
+  {\cal O}(\frac{1}{(z-w)^2}),
\nonu \\
&& \hat{T}(z)
\, J^a K^a K^b K^b(w)  = 
-\frac{1}{(z-w)^4}\,
\Bigg[ 2(2N-2+k_2) + N(2N-1)k_2 \Bigg]J^a K^a (w)
\nonu \\
&& +  {\cal O}(\frac{1}{(z-w)^2}),
\nonu \\
&& \hat{T}(z)
\, J^a J^b K^a K^b(w)  =  -\frac{1}{(z-w)^4}\,
\Bigg[ k_2 J^a J^a + k_1 K^a K^a \Bigg](w) \nonu \\
&& +  
\frac{1}{(z-w)^3}\,
(2N-2) \pa (J^a K^a)(w)
+  {\cal O}(\frac{1}{(z-w)^2}).
\label{17rem} 
\eea
It is obvious  that there exists an invariance under the change of
$J^a(z) \leftrightarrow K^a(z)$ and $k_1 \leftrightarrow k_2$.
For example, the first equation of Appendix (\ref{17rem})
can be obtained from (\ref{tjjjk}) by this change.

We collect each independent  field with its coefficients in (\ref{w4})
appearing at various poles in Appendix (\ref{17rem})
and the coefficients should vanish
in order to be a primary under the stress energy tensor (\ref{stress})
{\small
  \bea
\mbox{pole-6}: && \Bigg[
  -4 (2 N-1) k_1 N A_6-6 k_1 N (2 N-1) A_7-4 k_2 N (2 N-1) A_8
  \nonu \\
  && -
6 k_2 N (2 N-1) A_9 \Bigg]  =  0, \nonu \\
\mbox{pole-4}: &&    \Bigg[-6 (8 N-2) k_1 A_1-k_2 (8 N-2) A_3+6 A_7\nonu \\
  && -
  2 (2 (k_1+2 N-2)+(2 N-1) k_1 N) A_{15}
  -k_2 N (2 N-1) A_{17}-k_2 A_{20} \Bigg] J^a J^a(w)  =  0,
\nonu \\
\mbox{pole-4}: && \Bigg[
  -3 (8 N-2) k_1 A_2-3 k_2 (8 N-2) A_4+6 A_{11}+6 A_{12}+
  2 (2 N-2) A_{13}\nonu \\
  && +2 (2 N-2) A_{14} -k_1 ((2 N-1) N+2) A_{18}
  \nonu \\
  && -
(2 (k_2+2 N-2)+(2 N-1) k_2 N) A_{19} \Bigg] J^a K^a(w)=0, \nonu \\
\mbox{pole-4}: &&
\Bigg[ (8 N-2) (-k_1) A_3-N (2 N-1) A_{17} k_1-
  6 k_2 (8 N-2) A_5\nonu \\
  && +6 A_9-2 (2 (k_2+2 N-2)+(2 N-1) k_2 N) A_{16}-
  k_1 A_{20} \Bigg] K^a K^a(w)=0,
\nonu \\
\mbox{pole-3}: && \Bigg[ 4 A_6+6 A_7 \Bigg] \pa J^a J^a(w)=0,
\nonu \\
\mbox{pole-3}: && \Bigg[ 4 A_8+6 A_9 \Bigg] \pa K^a K^a(w)=0,
\nonu \\
\mbox{pole-3}: && \Bigg[2 A_{10}+6 A_{11}+4 (2 N-2) A_{13}-2 (2 N-2) A_{18}+
(2 N-2) A_{20} \Bigg] \pa J^a K^a(w)=0,
\nonu \\
\mbox{pole-3}: && \Bigg[
  2 A_{10}+6 A_{12}+4 (2 N-2) A_{14}+(2 N-2) A_{20} \Bigg] J^a \pa K^a(w)=0.
\label{linearequationfort}
\eea}
The primary condition (\ref{primaryother})
leads to the vanishing of these coefficients
appearing in Appendix
(\ref{linearequationfort}) for the independent
composite fields.

\section{ The coefficients appearing in (\ref{w4})
which depend on $k_1, k_2$ and $N$}

By solving the linear equations in Appendix
(\ref{linearequationforjprime})
and Appendix
(\ref{linearequationfort}), we determine the various coefficients
which are the functions of $k_1$, $k_2$ and $N$
as follows:
{\small
\bea
A_2 &= & -\frac{4  (k_1+2 N-2)}{k_2} A_1,
\nonu \\
A_3 & = & 
\frac{6  (k_1+2 N-2) (3 k_1+4 N-4)}{k_2 (3 k_2+2 N-2)} A_1,
\nonu \\
A_4 & = & -\frac{4  (k_1+2 N-2) (3 k_1+2 N-2) (3 k_1+4 N-4)}
{k_2 (3 k_2+2 N-2) (3 k_2+4 N-4)} A_1,
\nonu \\
A_5 &=& \frac{ k_1 (k_1+2 N-2) (3 k_1+2 N-2) (3 k_1+4 N-4)}
{k_2 (k_2+2 N-2) (3 k_2+2 N-2) (3 k_2+4 N-4)} A_1,
\nonu \\
A_6 & = &
\frac{6 
 }{(3 k_2+2 N-2) D(k_1,k_2,N)} A_1 \nonu \\
&\times & 
 (224 k_1 k_2 N^6+64 N^6+200 k_1 k_2^2 N^5
-128 k_1 N^5+248 k_1^2 k_2 N^5-1344 k_1 k_2 N^5
\nonu \\
& + & 
352 k_2 N^5-480 N^5+36 k_1 k_2^3 N^4-224 k_1^2 N^4
+132 k_1^2 k_2^2 N^4-1000 k_1 k_2^2 N^4+304 
k_2^2 N^4\nonu \\
& + & 832 k_1 N^4+96 k_1^3 k_2 N^4-1240 k_1^2 
k_2 N^4+2296 k_1 k_2 N^4-2288 k_2 N^4+
2208 N^4-96 k_1^3 N^3\nonu \\
& + & 12 k_1^2 k_2^3 N^3-144 k_1 
k_2^3 N^3+72 k_2^3 N^3+1232 k_1^2 N^3+24 k_1^3 
k_2^2 N^3-528 k_1^2 k_2^2 N^3+1298 k_1 k_2^2 N^3
\nonu \\
& - & 1672 k_2^2 N^3-416 k_1 N^3+12 k_1^4 k_2 N^3-384 k_1^3 
k_2 N^3+830 k_1^2 k_2 N^3-1688 k_1 k_2 N^3
+6688 k_2 N^3\nonu \\
& - & 5312 N^3-12 k_1^4 N^2+432 k_1^3 N^2-36 k_1^2 
k_2^3 N^2+105 k_1 k_2^3 N^2-324 k_2^3 N^2
-1080 k_1^2 N^2\nonu \\
& - & 72 k_1^3 k_2^2 N^2+21 k_1^2 k_2^2 N^2
-956 k_1 k_2^2 N^2+3672 k_2^2 N^2-2464 k_1 N^2
-36 k_1^4 k_2 N^2-96 k_1^3 k_2 N^2\nonu \\
& + & 940 k_1^2 
k_2 N^2+2928 k_1 k_2 N^2-9856 k_2 N^2+6528 N^2
+42 k_1^4 N-144 k_1^3 N-45 k_1^2 k_2^3 N \nonu \\
& - & 93 k_1 
k_2^3 N+504 k_2^3 N-640 k_1^2 N-90 k_1^3 k_2^2 N
+213 k_1^2 k_2^2 N+1522 k_1 k_2^2 N-3544 k_2^2 N
\nonu \\
& + & 3488 k_1 N-45 k_1^4 k_2 N+348 k_1^3 k_2 N
+70 k_1^2 k_2 N-4600 k_1 k_2 N+7040 k_2 N
-3936 N+24 k_1^4
\nonu \\
& - & 192 k_1^3-12 k_1^2 k_2^3+150 k_1 k_2^3
-252 k_2^3+712 k_1^2-24 k_1^3 k_2^2+270 k_1^2 k_2^2
-1064 k_1 k_2^2+1240 k_2^2
\nonu \\
& - & 1312 k_1-12 k_1^4 k_2
+144 k_1^3 k_2-848 k_1^2 k_2+2184 k_1 k_2-
1936 k_2+928),
\nonu \\
A_7 & = &
\frac{4 
 }{(3 k_2+2 N-2) D(k_1,k_2,N)} A_1 \nonu \\
&\times &
 (12 k_1^4 k_2 N^3-36 k_1^4 k_2 N^2-45 k_1^4 k_2 N-12 k_1^4 k_2-12 k_1^4 N^2+42 k_1^4 N+24 k_1^4+24 k_1^3 k_2^2 N^3\nonu \\
& - & 72 k_1^3 k_2^2 N^2-90 k_1^3 k_2^2 N-24 k_1^3 k_2^2+96 k_1^3 k_2 N^4-384 k_1^3 k_2 N^3-96 k_1^3 k_2 N^2+348 k_1^3 k_2 N
\nonu \\
& + & 
144 k_1^3 k_2-96 k_1^3 N^3+432 k_1^3 N^2-144 k_1^3 N-192 k_1^3+12 k_1^2 k_2^3 N^3-36 k_1^2 k_2^3 N^2-45 k_1^2 k_2^3 N
\nonu \\
& - & 12 k_1^2 k_2^3+132 k_1^2 k_2^2 N^4-528 k_1^2 k_2^2 N^3+21 k_1^2 k_2^2 N^2+213 k_1^2 k_2^2 N+270 k_1^2 k_2^2+248 k_1^2 k_2 N^5
\nonu \\
& - & 1240 k_1^2 k_2 N^4+830 k_1^2 k_2 N^3+940 k_1^2 k_2 N^2+70 k_1^2 k_2 N-848 k_1^2 k_2-224 k_1^2 N^4+1232 k_1^2 N^3\nonu \\
& - & 1080 k_1^2 N^2-640 k_1^2 N+712 k_1^2+36 k_1 k_2^3 N^4-144 k_1 k_2^3 N^3+105 k_1 k_2^3 N^2-93 k_1 k_2^3 N\nonu \\
& + & 150 k_1 k_2^3+200 k_1 k_2^2 N^5-1000 k_1 k_2^2 N^4+1298 k_1 k_2^2 N^3-956 k_1 k_2^2 N^2+1522 k_1 k_2^2 N\nonu \\
& - & 1064 k_1 k_2^2+224 k_1 k_2 N^6-1344 k_1 k_2 N^5+2296 k_1 k_2 N^4-1688 k_1 k_2 N^3+2928 k_1 k_2 N^2\nonu \\
& - & 4600 k_1 k_2 N+2184 k_1 k_2-128 k_1 N^5+832 k_1 N^4-416 k_1 N^3-2464 k_1 N^2+3488 k_1 N\nonu \\
& - & 1312 k_1+72 k_2^3 N^3-324 k_2^3 N^2+504 k_2^3 N-252 k_2^3+304 k_2^2 N^4-1672 k_2^2 N^3+3672 k_2^2 N^2 \nonu \\
& - & 3544 k_2^2 N+1240 k_2^2+352 k_2 N^5-2288 k_2 N^4+6688 k_2 N^3-9856 k_2 N^2+7040 k_2 N \nonu \\
& - & 1936 k_2+64 N^6-480 N^5+2208 N^4-5312 N^3+6528 N^2-3936 N+928),
\nonu \\
A_8 & = & 
\frac{6 
 }{k_2 (k_2+2 N-2) (3 k_2+2 N-2) 
(3 k_2+4 N-4) D(k_1,k_2,N)} A_1 \nonu \\
&\times &
k_1 (k_1+2 N-2) (3 k_1+4 N-4) (12 k_1^3 k_2^2 N^3-36 k_1^3 k_2^2 N^2
 -  45 k_1^3 k_2^2 N-12 k_1^3 k_2^2+36 k_1^3 k_2 N^4
\nonu \\
& - & 144 k_1^3 k_2 N^3+105 k_1^3 k_2 N^2-93 k_1^3 k_2 N+150 k_1^3 k_2+72 k_1^3 N^3-324 k_1^3 N^2+504 k_1^3 N-252 k_1^3
\nonu \\
& + & 
24 k_1^2 k_2^3 N^3-72 k_1^2 k_2^3 N^2-90 k_1^2 k_2^3 N-24 k_1^2 k_2^3+132 k_1^2 k_2^2 N^4-528 k_1^2 k_2^2 N^3+21 k_1^2 k_2^2 N^2\nonu \\
& + & 213 k_1^2 k_2^2 N+270 k_1^2 k_2^2+200 k_1^2 k_2 N^5-1000 k_1^2 k_2 N^4+1298 k_1^2 k_2 N^3-956 k_1^2 k_2 N^2\nonu \\
& + & 1522 k_1^2 k_2 N-1064 k_1^2 k_2+304 k_1^2 N^4-1672 k_1^2 N^3+3672 k_1^2 N^2-3544 k_1^2 N+1240 k_1^2\nonu \\
& + & 12 k_1 k_2^4 N^3-36 k_1 k_2^4 N^2-45 k_1 k_2^4 N-12 k_1 k_2^4+96 k_1 k_2^3 N^4-384 k_1 k_2^3 N^3-96 k_1 k_2^3 N^2\nonu \\
& + & 348 k_1 k_2^3 N+144 k_1 k_2^3+248 k_1 k_2^2 N^5-1240 k_1 k_2^2 N^4+830 k_1 k_2^2 N^3+940 k_1 k_2^2 N^2+70 k_1 k_2^2 N
\nonu \\
& - & 848 k_1 k_2^2+224 k_1 k_2 N^6-1344 k_1 k_2 N^5+2296 k_1 k_2 N^4-1688 k_1 k_2 N^3+2928 k_1 k_2 N^2\nonu \\
& - & 4600 k_1 k_2 N+2184 k_1 k_2+352 k_1 N^5-2288 k_1 N^4+6688 k_1 N^3-9856 k_1 N^2+7040 k_1 N\nonu \\
& - & 1936 k_1-12 k_2^4 N^2+42 k_2^4 N+24 k_2^4-96 k_2^3 N^3+432 k_2^3 N^2-144 k_2^3 N-192 k_2^3-224 k_2^2 N^4\nonu \\
& + & 1232 k_2^2 N^3-1080 k_2^2 N^2-640 k_2^2 N+712 k_2^2-128 k_2 N^5+832 k_2 N^4-416 k_2 N^3-2464 k_2 N^2
\nonu \\
& + & 3488 k_2 N-1312 k_2+64 N^6-480 N^5+2208 N^4-5312 N^3+6528 N^2-3936 N+928),
\nonu \\
A_9 & = &
-\frac{ 4
 }{k_2 (k_2+2 N-2) (3 k_2+2 N-2) 
(3 k_2+4 N-4) D(k_1,k_2,N)} A_1 \nonu \\
&\times &
k_1 (k_1+2 N-2) (3 k_1+4 N-4) 
(12 k_1^3 k_2^2 N^3-36 k_1^3 k_2^2 N^2-45 k_1^3 k_2^2 N-12 k_1^3 k_2^2+36 k_1^3 k_2 N^4\nonu \\
& - & 144 k_1^3 k_2 N^3+105 k_1^3 k_2 N^2-93 k_1^3 k_2 N
+150 k_1^3 k_2+72 k_1^3 N^3-324 k_1^3 N^2+504 k_1^3 N-252 k_1^3
\nonu \\
& + & 24 k_1^2 k_2^3 N^3-72 k_1^2 k_2^3 N^2-90 k_1^2 k_2^3 N-24 k_1^2 k_2^3+132 k_1^2 k_2^2 N^4-528 k_1^2 k_2^2 N^3+21 k_1^2 k_2^2 N^2
\nonu \\
& + & 213 k_1^2 k_2^2 N+270 k_1^2 k_2^2+200 k_1^2 k_2 N^5-1000 k_1^2 k_2 N^4+1298 k_1^2 k_2 N^3-956 k_1^2 k_2 N^2
\nonu \\
& + & 1522 k_1^2 k_2 N-1064 k_1^2 k_2+304 k_1^2 N^4-1672 k_1^2 N^3+3672 k_1^2 N^2-3544 k_1^2 N+1240 k_1^2\nonu \\
& + & 12 k_1 k_2^4 N^3-36 k_1 k_2^4 N^2-45 k_1 k_2^4 N-12 k_1 k_2^4+96 k_1 k_2^3 N^4-384 k_1 k_2^3 N^3-96 k_1 k_2^3 N^2
\nonu \\
& + & 348 k_1 k_2^3 N+144 k_1 k_2^3+248 k_1 k_2^2 N^5-1240 k_1 k_2^2 N^4+830 k_1 k_2^2 N^3+940 k_1 k_2^2 N^2+70 k_1 k_2^2 N
\nonu \\
& - & 848 k_1 k_2^2+224 k_1 k_2 N^6-1344 k_1 k_2 N^5+2296 k_1 k_2 N^4-1688 k_1 k_2 N^3+2928 k_1 k_2 N^2\nonu \\
& - & 4600 k_1 k_2 N+2184 k_1 k_2+352 k_1 N^5-2288 k_1 N^4+6688 k_1 N^3-9856 k_1 N^2+7040 k_1 N\nonu \\
& - & 1936 k_1-12 k_2^4 N^2+42 k_2^4 N+24 k_2^4-96 k_2^3 N^3+432 k_2^3 N^2-144 k_2^3 N-192 k_2^3-224 k_2^2 N^4\nonu \\
& + & 1232 k_2^2 N^3-1080 k_2^2 N^2-640 k_2^2 N+712 k_2^2-128 k_2 N^5+832 k_2 N^4-416 k_2 N^3-2464 k_2 N^2\nonu \\
& + & 
3488 k_2 N-1312 k_2+64 N^6-480 N^5+2208 N^4-5312 N^3+6528 N^2-3936 N
+928),
\nonu \\
A_{10} & = & 
-\frac{ 12
 }{k_2  (3 k_2+2 N-2)  D(k_1,k_2,N)} A_1 \nonu \\
&\times &
(k_1+2 N-2) (3 k_1+4 N-4) (4 k_1^3 k_2 N^3-12 k_1^3 k_2 N^2-15 k_1^3 k_2 N-4 k_1^3 k_2-4 k_1^3 N^2\nonu \\
& + & 
14 k_1^3 N+8 k_1^3+8 k_1^2 k_2^2 N^3-24 k_1^2 k_2^2 N^2-30 k_1^2 k_2^2 N-8 k_1^2 k_2^2+20 k_1^2 k_2 N^4-80 k_1^2 k_2 N^3\nonu \\
& - & 71 k_1^2 k_2 N^2+161 k_1^2 k_2 N+6 k_1^2 k_2-56 k_1^2 N^3+252 k_1^2 N^2-216 k_1^2 N+20 k_1^2+4 k_1 k_2^3 N^3
\nonu \\
& - & 12 k_1 k_2^3 N^2-15 k_1 k_2^3 N-4 k_1 k_2^3+20 k_1 k_2^2 N^4-80 k_1 k_2^2 N^3-71 k_1 k_2^2 N^2+161 k_1 k_2^2 N\nonu \\
& + & 6 k_1 k_2^2+16 k_1 k_2 N^5-80 k_1 k_2 N^4-212 k_1 k_2 N^3+884 k_1 k_2 N^2-700 k_1 k_2 N+92 k_1 k_2\nonu \\
& - & 176 k_1 N^4+968 k_1 N^3-1584 k_1 N^2+968 k_1 N-176 k_1-4 k_2^3 N^2+14 k_2^3 N+8 k_2^3-56 k_2^2 N^3\nonu \\
& + & 
252 k_2^2 N^2-216 k_2^2 N+20 k_2^2-176 k_2 N^4+968 k_2 N^3-1584 k_2 N^2+968 k_2 N-176 k_2\nonu \\
& - & 160 N^5+1040 N^4-2368 N^3+2464 N^2-1184 N+208),
\nonu \\
A_{11} & = & 
\frac{ 4
 }{k_2  (3 k_2+2 N-2)  D(k_1,k_2,N)} A_1 \nonu \\
&\times &
(k_1+2 N-2) (12 k_1^4 k_2 N^3-36 k_1^4 k_2 N^2-45 k_1^4 k_2 N-12 k_1^4 k_2-12 k_1^4 N^2+42 k_1^4 N+24 k_1^4
\nonu \\
& + & 
24 k_1^3 k_2^2 N^3-72 k_1^3 k_2^2 N^2-90 k_1^3 k_2^2 N-24 k_1^3 k_2^2+156 k_1^3 k_2 N^4-624 k_1^3 k_2 N^3-141 k_1^3 k_2 N^2
\nonu \\
& + & 513 k_1^3 k_2 N+204 k_1^3 k_2-156 k_1^3 N^3+702 k_1^3 N^2-234 k_1^3 N-312 k_1^3+12 k_1^2 k_2^3 N^3-36 k_1^2 k_2^3 N^2
\nonu \\
& - & 45 k_1^2 k_2^3 N-12 k_1^2 k_2^3+252 k_1^2 k_2^2 N^4-1008 k_1^2 k_2^2 N^3-501 k_1^2 k_2^2 N^2+1245 k_1^2 k_2^2 N+120 k_1^2 k_2^2
\nonu \\
& + & 608 k_1^2 k_2 N^5-3040 k_1^2 k_2 N^4+1376 k_1^2 k_2 N^3+4468 k_1^2 k_2 N^2-2900 k_1^2 k_2 N-512 k_1^2 k_2\nonu \\
& - & 584 k_1^2 N^4+3212 k_1^2 N^3-4032 k_1^2 N^2+764 k_1^2 N+640 k_1^2+96 k_1 k_2^3 N^4-384 k_1 k_2^3 N^3\nonu \\
& - & 372 k_1 k_2^3 N^2+774 k_1 k_2^3 N-60 k_1 k_2^3+560 k_1 k_2^2 N^5-2800 k_1 k_2^2 N^4+224 k_1 k_2^2 N^3 \nonu \\
& + & 6622 k_1 k_2^2 N^2-5174 k_1 k_2^2 N+568 k_1 k_2^2+704 k_1 k_2 N^6-4224 k_1 k_2 N^5+3928 k_1 k_2 N^4 \nonu \\
& + & 9076 k_1 k_2 N^3-17808 k_1 k_2 N^2+9572 k_1 k_2 N-1248 k_1 k_2-704 k_1 N^5+4576 k_1 N^4\nonu \\
& - & 10208 k_1 N^3+10208 k_1 N^2-4576 k_1 N+704 k_1-96 k_2^3 N^3+432 k_2^3 N^2-936 k_2^3 N \nonu \\
& + & 600 k_2^3-560 k_2^2 N^4+3080 k_2^2 N^3-7056 k_2^2 N^2+7112 k_2^2 N-2576 k_2^2-800 k_2 N^5\nonu \\
& + & 5200 k_2 N^4-14480 k_2 N^3+20240 k_2 N^2-13840 k_2 N+3680 k_2-128 N^6+960 N^5\nonu \\
& - & 4416 N^4+10624 N^3-13056 N^2+7872 N-1856),
\nonu \\
A_{12} & = & 
\frac{ 4
 }{k_2 (3 k_2+2 N-2) (3 k_2+4 N-4)  
D(k_1,k_2,N)} A_1 \nonu \\
&\times &
(k_1+2 N-2) (3 k_1+4 N-4) (12 k_1^3 k_2^2 N^3-36 k_1^3 k_2^2 N^2-45 k_1^3 k_2^2 N-12 k_1^3 k_2^2+96 k_1^3 k_2 N^4
\nonu \\
&- & 384 k_1^3 k_2 N^3+492 k_1^3 k_2 N^2-630 k_1^3 k_2 N+480 k_1^3 k_2+336 k_1^3 N^3-1512 k_1^3 N^2+2088 k_1^3 N
\nonu \\
& - & 
912 k_1^3+24 k_1^2 k_2^3 N^3-72 k_1^2 k_2^3 N^2-90 k_1^2 k_2^3 N-24 k_1^2 k_2^3+252 k_1^2 k_2^2 N^4-1008 k_1^2 k_2^2 N^3
\nonu \\
& + & 
363 k_1^2 k_2^2 N^2-159 k_1^2 k_2^2 N+660 k_1^2 k_2^2+560 k_1^2 k_2 N^5-2800 k_1^2 k_2 N^4+4688 k_1^2 k_2 N^3
\nonu \\
& - & 5906 k_1^2 k_2 N^2+6922 k_1^2 k_2 N-3464 k_1^2 k_2+1456 k_1^2 N^4-8008 k_1^2 N^3+16128 k_1^2 N^2\nonu \\
& - & 14056 k_1^2 N+4480 k_1^2+12 k_1 k_2^4 N^3-36 k_1 k_2^4 N^2-45 k_1 k_2^4 N-12 k_1 k_2^4+156 k_1 k_2^3 N^4\nonu \\
& - & 624 k_1 k_2^3 N^3-141 k_1 k_2^3 N^2+513 k_1 k_2^3 N+204 k_1 k_2^3+608 k_1 k_2^2 N^5-3040 k_1 k_2^2 N^4
\nonu \\
& + & 2384 k_1 k_2^2 N^3+1012 k_1 k_2^2 N^2+1420 k_1 k_2^2 N-2384 k_1 k_2^2+704 k_1 k_2 N^6-4224 k_1 k_2 N^5
\nonu \\
& + & 7960 k_1 k_2 N^4-8780 k_1 k_2 N^3+14880 k_1 k_2 N^2-18364 k_1 k_2 N+7824 k_1 k_2+1504 k_1 N^5\nonu \\
& - & 9776 k_1 N^4+27856 k_1 N^3-39952 k_1 N^2+27920 k_1 N-7552 k_1-12 k_2^4 N^2+42 k_2^4 N\nonu \\
& + & 24 k_2^4-156 k_2^3 N^3+702 k_2^3 N^2-234 k_2^3 N-312 k_2^3-584 k_2^2 N^4+3212 k_2^2 N^3-2448 k_2^2 N^2\nonu \\
& - & 2404 k_2^2 N+2224 k_2^2-704 k_2 N^5+4576 k_2 N^4-3872 k_2 N^3-8800 k_2 N^2+14432 k_2 N\nonu \\
& - & 5632 k_2-128 N^6+960 N^5+1920 N^4-14720 N^3+24960 N^2-17472 N+4480),
\nonu \\
A_{13} & = &
-\frac{ 12
 }{k_2 (3 k_2+2 N-2)  
D(k_1,k_2,N)} A_1 \nonu \\
&\times &
(k_1+2 N-2) (20 k_1^3 k_2 N^3-60 k_1^3 k_2 N^2-75 k_1^3 k_2 N-20 k_1^3 k_2-20 k_1^3 N^2+70 k_1^3 N+40 k_1^3
\nonu \\
& + & 40 k_1^2 k_2^2 N^3-120 k_1^2 k_2^2 N^2-294 k_1^2 k_2^2 N+50 k_1^2 k_2^2+120 k_1^2 k_2 N^4-480 k_1^2 k_2 N^3-298 k_1^2 k_2 N^2
\nonu \\
& + & 878 k_1^2 k_2 N-112 k_1^2 k_2-120 k_1^2 N^3+540 k_1^2 N^2-444 k_1^2 N+24 k_1^2+20 k_1 k_2^3 N^3-60 k_1 k_2^3 N^2
\nonu \\
& - & 219 k_1 k_2^3 N+70 k_1 k_2^3+120 k_1 k_2^2 N^4-480 k_1 k_2^2 N^3-838 k_1 k_2^2 N^2+1688 k_1 k_2^2 N-544 k_1 k_2^2
\nonu \\
& + & 
160 k_1 k_2 N^5-800 k_1 k_2 N^4-256 k_1 k_2 N^3+3332 k_1 k_2 N^2-3580 k_1 k_2 N+1144 k_1 k_2\nonu \\
& - & 192 k_1 N^4+1056 k_1 N^3-2208 k_1 N^2+2016 k_1 N-672 k_1-56 k_2^3 N^2+196 k_2^3 N-284 k_2^3\nonu \\
& - & 
288 k_2^2 N^3+1296 k_2^2 N^2-2280 k_2^2 N+1272 k_2^2-384 k_2 N^4+2112 k_2 N^3-4944 k_2 N^2\nonu \\
& + & 5088 k_2 N-1872 k_2-64 N^5+416 N^4-1792 N^3+3520 N^2-3008 N+928),
\nonu \\
A_{14} & = & 
-\frac{ 12
 }{ (3 k_2+2 N-2) (3 k_2+4 N-4) 
D(k_1,k_2,N)} A_1 \nonu \\
&\times &
(k_1+2 N-2) (3 k_1+4 N-4) (20 k_1^3 k_2 N^3-60 k_1^3 k_2 N^2+69 k_1^3 k_2 N-110 k_1^3 k_2\nonu \\
& + & 88 k_1^3 N^2-308 k_1^3 N+220 k_1^3+40 k_1^2 k_2^2 N^3-120 k_1^2 k_2^2 N^2-6 k_1^2 k_2^2 N-130 k_1^2 k_2^2+120 k_1^2 k_2 N^4\nonu \\
& - & 480 k_1^2 k_2 N^3+650 k_1^2 k_2 N^2-1000 k_1^2 k_2 N+800 k_1^2 k_2+384 k_1^2 N^3-1728 k_1^2 N^2+2424 k_1^2 N\nonu \\
& - & 1080 k_1^2+20 k_1 k_2^3 N^3-60 k_1 k_2^3 N^2-75 k_1 k_2^3 N-20 k_1 k_2^3+120 k_1 k_2^2 N^4-480 k_1 k_2^2 N^3\nonu \\
& + & 38 k_1 k_2^2 N^2+62 k_1 k_2^2 N+512 k_1 k_2^2+160 k_1 k_2 N^5-800 k_1 k_2 N^4+1088 k_1 k_2 N^3\nonu \\
& - & 1276 k_1 k_2 N^2+2708 k_1 k_2 N-1880 k_1 k_2+384 k_1 N^4-2112 k_1 N^3+4944 k_1 N^2-5088 k_1 N\nonu \\
& + & 1872 k_1-20 k_2^3 N^2+70 k_2^3 N+40 k_2^3-120 k_2^2 N^3+540 k_2^2 N^2+84 k_2^2 N-504 k_2^2-192 k_2 N^4\nonu \\
& + & 1056 k_2 N^3-96 k_2 N^2-2208 k_2 N+1440 k_2-64 N^5+416 N^4+320 N^3-2816 N^2\nonu \\
& + & 3328 N-1184),
\nonu \\
A_{15} & = & 
-\frac{ 18
 }{ (3 k_2+2 N-2)  
D(k_1,k_2,N)} A_1 \nonu \\
&\times &
(24 k_1^2 k_2^2 N-15 k_1^2 k_2^2+28 k_1^2 k_2 N^2-68 k_1^2 k_2 N+52 k_1^2 k_2+44 k_1^2 N-44 k_1^2+24 k_1 k_2^3 N\nonu \\
& - & 15 k_1 k_2^3+124 k_1 k_2^2 N^2-224 k_1 k_2^2 N+112 k_1 k_2^2+112 k_1 k_2 N^3-384 k_1 k_2 N^2+524 k_1 k_2 N\nonu \\
& - & 252 k_1 k_2+176 k_1 N^2-352 k_1 N+176 k_1+12 k_2^3 N^2-42 k_2^3 N+42 k_2^3+56 k_2^2 N^3-252 k_2^2 N^2\nonu \\
& + & 392 k_2^2 N-196 k_2^2+64 k_2 N^4-352 k_2 N^3+824 k_2 N^2-848 k_2 N+312 k_2+176 N^3-528 N^2\nonu \\
& + & 528 N-176),
\nonu \\
A_{16} & = &
-\frac{ 18
 }{ k_2 (k_2+2 N-2) (3 k_2+2 N-2) 
(3 k_2+4 N-4)  
D(k_1,k_2,N)} A_1 \nonu \\
&\times &
 k_1 (k_1+2 N-2) (3 k_1+4 N-4) (24 k_1^3 k_2 N-15 k_1^3 k_2+12 k_1^3 N^2-42 k_1^3 N+42 k_1^3\nonu \\
& + & 24 k_1^2 k_2^2 N-15 k_1^2 k_2^2+124 k_1^2 k_2 N^2-224 k_1^2 k_2 N+112 k_1^2 k_2+56 k_1^2 N^3-252 k_1^2 N^2+392 k_1^2 N\nonu \\
& - & 196 k_1^2+28 k_1 k_2^2 N^2-68 k_1 k_2^2 N+52 k_1 k_2^2+112 k_1 k_2 N^3-384 k_1 k_2 N^2+524 k_1 k_2 N\nonu \\
& - & 252 k_1 k_2+64 k_1 N^4-352 k_1 N^3+824 k_1 N^2-848 k_1 N+312 k_1+44 k_2^2 N-44 k_2^2+176 k_2 N^2
\nonu \\
& - & 352 k_2 N +  176 k_2+176 N^3-528 N^2+528 N-176),
\nonu \\
A_{17} & = & 
-\frac{ 36
 }{ k_2  (3 k_2+2 N-2)   
D(k_1,k_2,N)} A_1 \nonu \\
&\times &
(k_1+2 N-2) (3 k_1+4 N-4) (8 k_1^2 k_2 N-5 k_1^2 k_2+22 k_1^2+8 k_1 k_2^2 N-5 k_1 k_2^2+28 k_1 k_2 N^2\nonu \\
& - & 38 k_1 k_2 N+50 k_1 k_2+88 k_1 N-88 k_1+22 k_2^2+88 k_2 N-88 k_2+88 N^2-176 N+88),
\nonu \\
A_{18} & = & 
\frac{ 72
 }{ k_2  (3 k_2+2 N-2)   
D(k_1,k_2,N)} A_1 \nonu \\
&\times &
(k_1+2 N-2) (24 k_1^2 k_2^2 N-15 k_1^2 k_2^2+28 k_1^2 k_2 N^2-68 k_1^2 k_2 N+52 k_1^2 k_2+44 k_1^2 N-44 k_1^2\nonu \\
& + & 24 k_1 k_2^3 N-15 k_1 k_2^3+124 k_1 k_2^2 N^2-224 k_1 k_2^2 N+112 k_1 k_2^2+112 k_1 k_2 N^3-384 k_1 k_2 N^2\nonu \\
& + & 524 k_1 k_2 N-252 k_1 k_2+176 k_1 N^2-352 k_1 N+176 k_1+12 k_2^3 N^2-42 k_2^3 N+42 k_2^3+56 k_2^2 N^3\nonu \\
& - & 252 k_2^2 N^2+392 k_2^2 N-196 k_2^2+64 k_2 N^4-352 k_2 N^3+824 k_2 N^2-848 k_2 N+312 k_2\nonu \\
& + & 176 N^3-528 N^2+528 N-176),
\nonu \\
A_{19} & = & 
\frac{ 72
 }{ k_2  (3 k_2+2 N-2) (3 k_2+4 N-4)
D(k_1,k_2,N)} A_1 \nonu \\
&\times &
 (k_1+2 N-2) (3 k_1+4 N-4) (24 k_1^3 k_2 N-15 k_1^3 k_2+12 k_1^3 N^2-42 k_1^3 N+42 k_1^3+24 k_1^2 k_2^2 N\nonu \\
& - & 15 k_1^2 k_2^2+124 k_1^2 k_2 N^2-224 k_1^2 k_2 N+112 k_1^2 k_2+56 k_1^2 N^3-252 k_1^2 N^2+392 k_1^2 N-196 k_1^2\nonu \\
& + & 28 k_1 k_2^2 N^2-68 k_1 k_2^2 N+52 k_1 k_2^2+112 k_1 k_2 N^3-384 k_1 k_2 N^2+524 k_1 k_2 N-252 k_1 k_2\nonu \\
& + & 64 k_1 N^4-352 k_1 N^3+824 k_1 N^2-848 k_1 N+312 k_1+44 k_2^2 N-44 k_2^2+176 k_2 N^2-352 k_2 N\nonu \\
& + & 176 k_2+176 N^3-528 N^2+528 N-176),
\nonu \\
A_{20} & = &
-\frac{ 72
 }{ k_2  (3 k_2+2 N-2)
D(k_1,k_2,N)} A_1 \nonu \\
&\times & 
(k_1+2 N-2) (3 k_1+4 N-4) (8 k_1^2 k_2 N-5 k_1^2 k_2+6 k_1^2 N^2-21 k_1^2 N+10 k_1^2+8 k_1 k_2^2 N\nonu \\
& - & 5 k_1 k_2^2+40 k_1 k_2 N^2-80 k_1 k_2 N+26 k_1 k_2+28 k_1 N^3-126 k_1 N^2+130 k_1 N-32 k_1+6 k_2^2 N^2\nonu \\
& - & 21 k_2^2 N+10 k_2^2+28 k_2 N^3-126 k_2 N^2+130 k_2 N-32 k_2+32 N^4-176 N^3+280 N^2\nonu \\
& - & 160 N+24).
\label{coeffcoeff}
\eea
}
Here one introduces the common factor which appears in the denominator
in Appendix (\ref{coeffcoeff})
as 
\bea
D(k_1,k_2,N) & \equiv & 
(40 k_1 
k_2 N^3+176 N^3+10 k_1 k_2^2 N^2+176 k_1 N^2+10 k_1^2 
k_2 N^2-60 k_1 k_2 N^2\nonu \\
& - & 528 N^2+
44 k_1^2 N-5 k_1 k_2^2 N+44 k_2^2 N-352 k_1 N-
5 k_1^2 k_2 N+152 k_1 k_2 N\nonu \\
& - & 352 k_2 N
+  528 N
-44 k_1^2+22 k_1 k_2^2-44 k_2^2+176 k_1+
22 k_1^2 k_2-132 k_1 k_2\nonu \\
& - & 176 +176 k_2 N^2 +176 k_2).
\label{Dcoeff}
\eea
In this paper, we consider the two cases where $k_1=1$ and $k_1=2N-2$
with arbitrary $k_2$,
although the most general case is an interesting subject.

\section{ 
 The coefficients appearing in (\ref{w4}),
which depend on $k_2$ and $N$ when $k_1 = 1$
}

The above coefficients in Appendix (\ref{coeffcoeff}) and
Appendix (\ref{Dcoeff}),
for the critical level $k_1=1$,  are given by 
\bea
A_2 & = & -\frac{4  (2 N-1)}{k_2} A_1,
\nonu \\
A_3 & = & \frac{6  (2 N-1) (4 N-1)}{k_2 (3 k_2+2 N-2)} A_1,
\nonu \\
A_4 & = & -\frac{4  (2 N-1) (2 N+1) (4 N-1)}
{k_2 (3 k_2+2 N-2) (3 k_2+4 N-4)} A_1,
\nonu \\
A_5 & = &
\frac{ (2 N-1) (2 N+1) (4 N-1)}
     {k_2 (k_2+2 N-2) (3 k_2+2 N-2) (3 k_2+4 N-4)}
     A_1,
     \nonu \\
     A_6 & = &
     \frac{6}{(3 k_2+2 N-2) d(1,k_2,N)} A_1
     (18 k_2^3 N^3-21 k_2^3 N^2-138 k_2^3 N+
     114 k_2^3+100 k_2^2 N^4 \nonu \\
     & - & 232 k_2^2 N^3-555 k_2^2 N^2
     +1055 k_2^2 N-422 k_2^2+112 k_2 N^5-316 k_2 N^4
     \nonu \\
     & - & 726 k_2 N^3+2366 k_2 N^2-1877 k_2 N+
     468 k_2+32 N^5\nonu \\
     & - & 288 N^4+1264 N^3-1664 N^2+870 N-160),
     \nonu \\
     A_7 & = & -\frac{4}{(3 k_2+2 N-2) d(1,k_2,N)} A_1
     (18 k_2^3 N^3-21 k_2^3 N^2-138 k_2^3 N
     +114 k_2^3+100 k_2^2 N^4
     \nonu \\
     & - & 232 k_2^2 N^3-555 k_2^2 N^2+1055 k_2^2 N-
     422 k_2^2+112 k_2 N^5\nonu \\
     & - & 316 k_2 N^4-726 k_2 N^3+2366 k_2 N^2-
     1877 k_2 N+468 k_2\nonu \\
     & + & 32 N^5-288 N^4+1264 N^3-1664 N^2+870 N-160),
     \nonu \\
     A_8 & = &
     \frac{6 (2 N-1) (2 N+1) (4 N-1)}{k_2 (k_2+2 N-2) (3 k_2+2 N-2)
       (3 k_2+4 N-4) d(1,k_2,N)}   A_1
     \nonu \\
     & \times & (3 k_2^4 N-12 k_2^4+24 k_2^3 N^2
     -114 k_2^3 N+72 k_2^3 \nonu \\
     & + & 62 k_2^2 N^3-333 k_2^2 N^2+402 k_2^2 N
     -122 k_2^2+56 k_2 N^4 \nonu \\
     & - & 318 k_2 N^3+555 k_2 N^2-317 k_2 N+
     42 k_2+16 N^4-32 N^3+60 N^2-64 N+20),
     \nonu \\
     A_9 & = &
     -\frac{4 (2 N-1) (2 N+1) (4 N-1)}{k_2 (k_2+2 N-2) (3 k_2+2 N-2)
       (3 k_2+4 N-4) d(1,k_2,N)}
     A_1 
     \nonu \\
     & \times & (3 k_2^4 N-12 k_2^4+24 k_2^3 N^2-
     114 k_2^3 N+72 k_2^3+62 k_2^2 N^3\nonu \\
     & - & 333 k_2^2 N^2+402 k_2^2 N-122 k_2^2+
     56 k_2 N^4-318 k_2 N^3\nonu \\
     & + & 555 k_2 N^2-317 k_2 N+42 k_2+16 N^4-32 N^3+
     60 N^2-64 N+20),
     \nonu \\
     A_{10} & = & -\frac{12  (2 N-1) (4 N-1)}{k_2 (3 k_2+2 N-2)d(1,k_2,N)} A_1
     \nonu \\
     & \times & (2 k_2^3 N^2-7 k_2^3 N-4 k_2^3+
     10 k_2^2 N^3-59 k_2^2 N^2+49 k_2^2 N
     \nonu \\
     & - & 18 k_2^2+8 k_2 N^4-114 k_2 N^3+283 k_2 N^2
     -250 k_2 N+82 k_2\nonu \\
     & - & 80 N^4+392 N^3-532 N^2+298 N-60), \nonu \\
     A_{11} & = &
     \frac{4}{k_2 (3 k_2+2 N-2) d(1,k_2,N)}
     A_1 (96 k_2^3 N^4-468 k_2^3 N^3+24 k_2^3 N^2
     \nonu \\
     & - & 207 k_2^3 N+528 k_2^3+560 k_2^2 N^5-
     3108 k_2^2 N^4+2320 k_2^2 N^3\nonu \\
     & - & 1007 k_2^2 N^2+3093 k_2^2 N-1912 k_2^2+
     704 k_2 N^6-4416 k_2 N^5\nonu \\
     & + & 6244 k_2 N^4-4640 k_2 N^3+6723 k_2 N^2
     -6700 k_2 N+2112 k_2\nonu \\
     & - & 128 N^6+256 N^5-424 N^4+3472 N^3-6190 N^2+3868 N-800),
     \nonu \\
     A_{12} & = &
     \frac{4  (2 N-1) (2 N+1) (4 N-1)}
          {k_2 (3 k_2+2 N-2) (3 k_2+4 N-4) d(1,k_2,N)} A_1
     \nonu \\
     & \times & (3 k_2^4 N-12 k_2^4+39 k_2^3 N^2
     -189 k_2^3 N+132 k_2^3\nonu \\
     & + & 152 k_2^2 N^3-843 k_2^2 N^2+1188 k_2^2 N
     -488 k_2^2+176 k_2 N^4\nonu \\
     & - & 1092 k_2 N^3+2502 k_2 N^2-2360 k_2 N+
     792 k_2-32 N^4\nonu \\
     & + & 616 N^3-1608 N^2+1520 N-496),
     \nonu \\
     A_{13} & = & -\frac{12}{k_2 (3 k_2+2 N-2) d(1,k_2,N)} A_1
     \nonu \\
     & \times & (20 k_2^3 N^3-116 k_2^3 N^2-
     23 k_2^3 N-214 k_2^3+120 k_2^2 N^4\nonu \\
     & - & 728 k_2^2 N^3+338 k_2^2 N^2-886 k_2^2 N+
     778 k_2^2+160 k_2 N^5\nonu \\
     & - & 1064 k_2 N^4+1396 k_2 N^3-1970 k_2 N^2+
     2311 k_2 N-860 k_2\nonu \\
     & - & 64 N^5+224 N^4-856 N^3+1832 N^2-1366 N+320),
     \nonu \\
     A_{14} & = &
     -\frac{12   (2 N-1) (2 N+1) (4 N-1)}{k_2 (3 k_2+2 N-2) (3 k_2+4 N-4)
     d(1,k_2,N)} A_1
\nonu \\
& \times &  (5 k_2^3 N-20 k_2^3+30 k_2^2 N^2-
140 k_2^2 N+122 k_2^2+40 k_2 N^3
\nonu \\
& - & 218 k_2 N^2+431 k_2 N-250 k_2-16 N^3+
200 N^2-356 N+172),
     \nonu \\
     A_{15} & = & -\frac{18}{(2 N-1) (3 k_2+2 N-2) d(1,k_2,N)} A_1
     (12 k_2^3 N^2-18 k_2^3 N+27 k_2^3
     \nonu \\
     & + & 56 k_2^2 N^3-128 k_2^2 N^2+192 k_2^2 N
     -99 k_2^2+64 k_2 N^4\nonu \\
     & - & 240 k_2 N^3+468 k_2 N^2-392 k_2 N+
     112 k_2+176 N^3\nonu \\
     & - & 352 N^2+220 N-44),
     \nonu \\
     A_{16} & = & -\frac{18   (2 N-1) (2 N+1) (4 N-1)}
     {k_2 (k_2+2 N-2) (3 k_2+2 N-2) (3 k_2+4 N-4)
       d(1,k_2,N)} A_1
    \nonu \\
     & \times & (7 k_2^2+28 k_2 N-21 k_2+16 N^2-30 N
     +18),
     \nonu \\
     A_{17} & = & -\frac{36 (4 N-1)}{k_2 (3 k_2+2 N-2) d(1,k_2,N)} A_1
      \nonu \\
     & \times & (8 k_2^2 N+17 k_2^2+28 k_2 N^2+
     58 k_2 N-43 k_2+88 N^2-88 N+22),
     \nonu \\
     A_{18} & = & \frac{72}{k_2 (3 k_2+2 N-2) d(1,k_2,N)} A_1
     (12 k_2^3 N^2-18 k_2^3 N+27 k_2^3\nonu \\
     & + & 56 k_2^2 N^3-128 k_2^2 N^2+192 k_2^2 N
     -99 k_2^2+64 k_2 N^4\nonu \\
     & - & 240 k_2 N^3+468 k_2 N^2-392 k_2 N
     +112 k_2+176 N^3\nonu \\
     & - & 352 N^2+220 N-44),
     \nonu \\
     A_{19} & = &
     \frac{72  (2 N-1) (2 N+1) (4 N-1)}
          {k_2 (3 k_2+2 N-2) (3 k_2+4 N-4) d(1,k_2,N)}
     A_1
     \nonu \\
     & \times &
     (7 k_2^2+28 k_2 N-21 k_2+16 N^2-30 N+18),
     \nonu \\
     A_{20} & = &
     -\frac{72 (2 N-1) (4 N-1)}{k_2 (3 k_2+2 N-2) d(1,k_2,N)} A_1 
     \nonu \\
     & \times & (3 k_2^2 N-5 k_2^2+14 k_2 N^2
     -36 k_2 N+11 k_2+16 N^3-66 N^2+47 N-2).
     \label{reducedcoeff}
\eea
The simplified notation is used
\bea
d(1,k_2,N) & = &
(5 k_2^2 N+22 k_2^2+20 k_2 N^2+73 k_2 N-66 k_2+88 N^2-132 N+44).
\label{dcoeffcoeff}
\eea
For the computation of the three-point functions with
finite $N$ and $k_2$ corresponding to (\ref{finiteresults}),
the above expression is needed.

Under the large 't Hooft limit (\ref{limit}),
the coefficients in Appendix (\ref{reducedcoeff})  with
(\ref{dcoeffcoeff}) become
\bea
A_2 & = & \frac{4  \lambda }{(\lambda -1)} A_1, 
\qquad
A_3  =  \frac{12  \lambda ^2}{(\lambda -1) (2 \lambda -3)} A_1,
\nonu \\
A_4 & = & \frac{8  \lambda ^3}{(\lambda -3) (\lambda -1)
  (2 \lambda -3)} A_1,
\qquad
A_5  =  -\frac{1}{N} \,
\frac{ \lambda^4}{ (\lambda -3) (2 \lambda -3)
  (\lambda  -1)} A_1,
\nonu \\
A_6 & = &  N^2 \, \frac{12 
  (2 \lambda -9)}
{5  (2 \lambda -3)} A_1,
\qquad
A_7  =  - N^2 \, \frac{8  (2 \lambda -9)
  }{5  (2 \lambda -3)} A_1,
\nonu \\
A_8 & = & - N \, \frac{12  \lambda ^2 \left(\lambda ^2+6\right)
  }
{5 (\lambda -3) (\lambda -1) (2 \lambda -3)} A_1,
\qquad
A_9  =  -N \, \frac{12  \lambda ^2 \left(\lambda ^2+6\right)
  }
{5 (\lambda -3) (\lambda -1) (2 \lambda -3)} A_1,
\nonu \\
A_{10} & = & N^2 \, \frac{48  (\lambda -2) \lambda
  }{5 (\lambda -1) (2 \lambda -3)} A_1,
\qquad
A_{11}  =  -N^2 \, \frac{16  (\lambda -12)
  \lambda  }
{5 (\lambda -1) (2 \lambda -3)} A_1,
\nonu \\
A_{12} & = & -N^2 \, \frac{16  \lambda
  \left(\lambda ^2+15 \lambda +6\right)
  }
{5 (\lambda -3) (\lambda -1) (2 \lambda -3)} A_1,
\qquad
A_{13}  =  -N \, \frac{24 
  \lambda  }
{(\lambda -1) (2 \lambda -3)} A_1,
\nonu \\
A_{14} & = & N \, \frac{48  \lambda ^2
  }
{(\lambda -3) (\lambda -1) (2 \lambda -3)} A_1,
\qquad
A_{15}  =  \frac{36  (\lambda +3)}
{5 (\lambda +1) (2 \lambda -3)} A_1,
\nonu \\
A_{16} & = & \frac{1}{N^2} \,
\frac{18  \lambda ^4 \left(3 \lambda ^2-7\right)}
{5 (\lambda -3) (\lambda +1) (2 \lambda -3) (\lambda -1)^2}
A_1,
\nonu \\
A_{17} & = & -\frac{1}{N} \, \frac{72  \lambda ^2
  (3 \lambda +4)}
{5 (\lambda +1) (2 \lambda -3) (\lambda  -1)}
A_1,
\qquad
A_{18}  =  \frac{144  \lambda  (\lambda +3)}
{5 (\lambda -1) (\lambda +1) (2 \lambda -3)} A_1,
\nonu \\
A_{19} & = & - \frac{1}{N} \,
\frac{144  \lambda ^3 \left(3 \lambda ^2-7\right)}
{5 (\lambda -3) (\lambda -1)^2 (\lambda +1) (2 \lambda -3)
  }
A_1,
\nonu \\
A_{20} & = & \frac{144  \lambda ^2
  (\lambda +3)}{5 (\lambda -1)^2 (\lambda +1) (2 \lambda -3)}
A_1.
\label{coeffA}
\eea
The nineteen coefficients are given by $A_1$ coefficient.  
By recognizing the explicit $N$ behavior of the above coefficients,
one can extract the leading behavior of the zero mode of the
higher spin $4$ current by considering the $N$ behavior of
the various zero modes consisting of the higher spin $4$ current.
For example, (\ref{eigenexpressionov}) and (\ref{eigenvo}).

\section{ 
 The coefficients appearing in (\ref{w4}),
which depend on $k_2$ and $N$ when $k_1 = 2N-2$
}

The above coefficients
in Appendix (\ref{coeffcoeff}) and Appendix (\ref{Dcoeff}),
for the critical level $k_1=2N-2$,
are given by
{\small
\bea
A_2 & = & -\frac{16  (N-1)}{k_2} A_1, \nonu \\
A_3 & = & \frac{240  (N-1)^2}{k_2 (3 k_2+2 N-2)} A_1, \nonu \\
A_4 & = &
-\frac{1280  (N-1)^3}{k_2 (3 k_2+2 N-2) (3 k_2+4 N-4)} A_1, \nonu \\
A_5 & = & \frac{640  (N-1)^4}
{k_2 (k_2+2 N-2) (3 k_2+2 N-2) (3 k_2+4 N-4)} A_1, \nonu \\
A_6 & = &
\frac{6} {(3 k_2+2 N-2) d(k_2,N)} A_1
\nonu \\
& \times &
(60 k_2^3 N^4-240 k_2^3 N^3+123 k_2^3 N^2-153 k_2^3 N+300 k_2^3
+560 k_2^2 N^5  \nonu \\
& - & 2800 k_2^2 N^4+2860 k_2^2 N^3-920 k_2^2 N^2+2620 k_2^2 N
-2320 k_2^2+1200 k_2 N^6 \nonu \\
& - & 7200 k_2 N^5+10476 k_2 N^4-1788 k_2 N^3+468 k_2 N^2
-8676 k_2 N+5520 k_2 \nonu \\
& - & 
1024 N^5+6656 N^4-9664 N^3-704 N^2+8896 N-4160),
\nonu \\
A_7 & = &
-\frac{4}{(3 k_2+2 N-2) d(k_2, N)}  A_1
(60 k_2^3 N^4-240 k_2^3 N^3+123 k_2^3 N^2-153 k_2^3 N
 \nonu \\
& + & 300 k_2^3+560 k_2^2 N^5-2800 k_2^2 N^4+2860 k_2^2 N^3
-920 k_2^2 N^2 \nonu \\
& + & 2620 k_2^2 N-2320 k_2^2+1200 k_2 N^6-7200 k_2 N^5
+10476 k_2 N^4\nonu \\
& - & 1788 k_2 N^3+468 k_2 N^2-8676 k_2 N+5520 k_2-1024 N^5
+6656 N^4\nonu \\
& - &  9664 N^3-704 N^2+8896 N-4160),
\nonu \\
A_8 & = & 
\frac{480}{k_2
  (k_2+2 N-2) (3 k_2+2 N-2) (3 k_2+4 N-4) d(k_2,N)} A_1
\nonu \\
&\times &
(N-1)^2 (12 k_2^4 N^4-48 k_2^4 N^3-15 k_2^4 N^2
 \nonu \\
& + & 54 k_2^4 N+24 k_2^4+144 k_2^3 N^5-720 k_2^3 N^4
+396 k_2^3 N^3 \nonu \\
& + & 828 k_2^3 N^2-360 k_2^3 N-288 k_2^3+560 k_2^2 N^6
-3360 k_2^2 N^5+4772 k_2^2 N^4 \nonu \\
& + & 24 k_2^2 N^3-1932 k_2^2 N^2-1856 k_2^2 N+1792 k_2^2
+768 k_2 N^7 \nonu \\
& - & 5376 k_2 N^6+13152 k_2 N^5-16176 k_2 N^4+17424 k_2 N^3
-22224 k_2 N^2 \nonu \\
& + & 18000 k_2 N-5568 k_2+1280 N^6-9600 N^5+31488 N^4-55552 N^3
\nonu \\
& + &  54528 N^2-28032 N+5888),
\nonu \\
A_9 & = &
-\frac{320  (N-1)^2}{k_2
  (k_2+2 N-2) (3 k_2+2 N-2)
  (3 k_2+4 N-4) d(k_2,N)} A_1
\nonu \\
& \times & (12 k_2^4 N^4-48 k_2^4 N^3-15 k_2^4 N^2
+54 k_2^4 N+24 k_2^4+144 k_2^3 N^5  \nonu \\
& - & 
720 k_2^3 N^4+396 k_2^3 N^3+828 k_2^3 N^2-360 k_2^3 N
-288 k_2^3\nonu \\
& + & 560 k_2^2 N^6-3360 k_2^2 N^5+4772 k_2^2 N^4+24 k_2^2 N^3
-1932 k_2^2 N^2\nonu \\
& - & 1856 k_2^2 N+1792 k_2^2+768 k_2 N^7-5376 k_2 N^6
+13152 k_2 N^5 \nonu \\
& - & 16176 k_2 N^4+17424 k_2 N^3-22224 k_2 N^2+18000 k_2 N
-5568 k_2\nonu \\
& + & 
1280 N^6-9600 N^5+31488 N^4-55552 N^3+54528 N^2-28032 N
+5888),
\nonu \\
A_{10} & = &
-\frac{480  (N-1)}{k_2 (3 k_2+2 N-2) d(k_2,N)} A_1
  (4 k_2^3 N^4-16 k_2^3 N^3-5 k_2^3 N^2  \nonu \\
  & + & 18 k_2^3 N+8 k_2^3+36 k_2^2 N^5-180 k_2^2 N^4
  +33 k_2^2 N^3 \nonu \\
  & + & 414 k_2^2 N^2-291 k_2^2 N-12 k_2^2+72 k_2 N^6
  -432 k_2 N^5\nonu \\
  & + & 130 k_2 N^4+2030 k_2 N^3-3234 k_2 N^2+1586 k_2 N
  -152 k_2\nonu \\
  & - & 
  384 N^5+2496 N^4-5472 N^3+5280 N^2-2208 N+288),
  \nonu \\
  A_{11} & = &
  \frac{32(N-1)}{k_2 (3 k_2+2 N-2) d(k_2,N)} A_1 
  (60 k_2^3 N^4-240 k_2^3 N^3-219 k_2^3 N^2
  \nonu \\
  & + & 504 k_2^3 N-168 k_2^3+580 k_2^2 N^5-2900 k_2^2 N^4
  +635 k_2^2 N^3\nonu \\
  & + & 5855 k_2^2 N^2-4930 k_2^2 N+760 k_2^2+1320 k_2 N^6-7920 k_2 N^5
  \nonu \\
  & + & 9102 k_2 N^4+9084 k_2 N^3-20970 k_2 N^2+9960 k_2 N-576 k_2
  \nonu \\
  & - & 
  1328 N^5+8632 N^4-17384 N^3+13640 N^2-3016 N-544),
  \nonu \\
  A_{12} & = &
  \frac{160  (N-1)}{k_2 (3 k_2+2 N-2)
    (3 k_2+4 N-4) d(k_2,N)} A_1
  (12 k_2^4 N^4-48 k_2^4 N^3-15 k_2^4 N^2
   \nonu \\
  & + & 54 k_2^4 N+24 k_2^4+204 k_2^3 N^5-1020 k_2^3 N^4+
  561 k_2^3 N^3\nonu \\
  & + & 1173 k_2^3 N^2-510 k_2^3 N-408 k_2^3+1160 k_2^2 N^6-
  6960 k_2^2 N^5\nonu \\
  & + & 10790 k_2^2 N^4-3540 k_2^2 N^3+1614 k_2^2 N^2-7928 k_2^2 N
  +4864 k_2^2\nonu \\
  & + & 2208 k_2 N^7-15456 k_2 N^6+41256 k_2 N^5-64032 k_2 N^4
  +85488 k_2 N^3\nonu \\
  & - & 99360 k_2 N^2+69384 k_2 N-19488 k_2+5696 N^6-42720 N^5
  \nonu \\
  & + & 
  136320 N^4-232000 N^3+219840 N^2-109536 N+22400),
  \nonu \\
  A_{13} &= &
  -\frac{48}{k_2 (3 k_2+2 N-2) d(k_2,N)} A_1
  (20 k_2^3 N^4-80 k_2^3 N^3-187 k_2^3 N^2
   \nonu \\
  & + & 387 k_2^3 N-212 k_2^3+200 k_2^2 N^5-1000 k_2^2 N^4-530 k_2^2 N^3
  \nonu \\
  & + & 4210 k_2^2 N^2-4160 k_2^2 N+1280 k_2^2+480 k_2 N^6-2880 k_2 N^5
  \nonu \\
  & + & 2576 k_2 N^4+6652 k_2 N^3-14136 k_2 N^2+9532 k_2 N-2224 k_2
  \nonu \\
  & - & 
  544 N^5+3536 N^4-8368 N^3+9328 N^2-4976 N+1024),
  \nonu \\
  A_{14} &=&
  -\frac{480  (N-1)}{k_2 (3 k_2+2 N-2) (3 k_2+4 N-4) d(k_2,N)} A_1
  (20 k_2^3 N^4-80 k_2^3 N^3
   \nonu \\
  & -& 25 k_2^3 N^2+90 k_2^3 N+40 k_2^3+200 k_2^2 N^5
  -1000 k_2^2 N^4\nonu \\
  & + & 1006 k_2^2 N^3-182 k_2^2 N^2+1000 k_2^2 N-1024 k_2^2+480 k_2 N^6
  \nonu \\
  & - & 2880 k_2 N^5+6488 k_2 N^4-9464 k_2 N^3+13224 k_2 N^2
  -12488 k_2 N\nonu \\
  & + &  4640 k_2+1472 N^5-9568 N^4+25376 N^3-33440 N^2+
  21664 N-5504),
  \nonu \\
  A_{15} & = &
  -\frac{36} {(N-1)(3 k_2+2 N-2) d(k_2,N)} A_1
  (15 k_2^3 N^2-30 k_2^3 N+18 k_2^3
   \nonu \\
  & + & 100 k_2^2 N^3-300 k_2^2 N^2+320 k_2^2 N-120 k_2^2+100 k_2 N^4
  \nonu \\
  & - &  460 k_2 N^3+876 k_2 N^2-772 k_2 N+256 k_2+176 N^3
  -  528 N^2+528 N-176),
  \nonu \\
  A_{16} & = &
  -\frac{2880  (N-1)^3 }{k_2 (k_2+2 N-2) (3 k_2+2 N-2)
    (3 k_2+4 N-4) d(k_2,N)} A_1
  \nonu \\
& \times &
  (38 k_2^2 N^2-73 k_2^2 N+52 k_2^2+228 k_2 N^3-666 k_2 N^2
   \nonu \\
  & + &  750 k_2 N-312 k_2+112 N^4-616 N^3+1376 N^2-1352 N
  +480),
  \nonu \\
  A_{17} & = &
  -\frac{1440 (N-1)}{k_2 (3 k_2+2 N-2) d(k_2,N)} A_1
  (8 k_2^2 N^2-13 k_2^2 N+16 k_2^2  \nonu \\
  & +&  44 k_2 N^3-108 k_2 N^2+168 k_2 N-104 k_2+176 N^2-352 N
  +176),
  \nonu \\
   A_{18} & = &
  \frac{144  (N+2)}{ k_2 (3 k_2+2 N-2)} A_1
  \nonu \\
   & \times &
   \frac{1}{
    (
     30 k_2^2 N^2+7 k_2^2 N+44 k_2^2+120 k_2 N^3+88 k_2 N^2+190 k_2 N+88 k_2+
     88 N^3+264 N^2-352)} 
\nonu \\
 & \times & (6 k_2^3 N^2+51 k_2^3 N-24 k_2^3+28 k_2^2 N^3+
246 k_2^2 N^2-44 k_2^2 N  \nonu \\
 & - &  32 k_2^2+32 k_2 N^4+160 k_2 N^3-236 k_2 N^2-76 k_2 N+336 k_2+
88 N^3+264 N^2-352),
  \nonu \\
  A_{19} & = &
  \frac{5760  (N-1)^2}{k_2 (3 k_2+2 N-2)
    (3 k_2+4 N-4) d(k_2,N)} A_1
  (38 k_2^2 N^2-73 k_2^2 N+52 k_2^2  \nonu \\
  & + &  228 k_2 N^3-666 k_2 N^2+750 k_2 N-312 k_2+112 N^4
  \nonu \\
  & - &  616 N^3
  +1376 N^2-1352 N+480),
  \nonu \\
  A_{20} & = &
  -\frac{1440  (N-1)}{k_2 (3 k_2+2 N-2) d(k_2,N)} A_1
  (22 k_2^2 N^2-47 k_2^2 N+20 k_2^2+140 k_2 N^3  \nonu \\
  & - &  450 k_2 N^2+414 k_2 N-104 k_2+
  112 N^4-616 N^3+1024 N^2-648 N+128),
  \label{othercoeffexp}
\eea }
where the simplified notation
\bea
&& d(k_2,N) =
\nonu \\
&&
(10 k_2^2 N^2-5 k_2^2 N+44 k_2^2+60 k_2 N^3-90 k_2 N^2+294 k_2 N-264 k_2
+352 N^2\nonu \\
&& -704 N+352),
\label{dcoeffexp}
\eea
is used.
Also one can use these coefficients for the finite $N$ behavior
in the ${\cal N}=1$ coset model corresponding to (\ref{three}).

The coefficients in Appendix (\ref{othercoeffexp}) and
Appendix (\ref{dcoeffexp})
under the large 't Hooft limit (\ref{limit}) lead to
\bea
A_2 & = &  \frac{8  \lambda }{(\lambda -1)} A_1, 
\qquad
A_3  =   \frac{60  \lambda ^2}{(\lambda -1) (2 \lambda -3)}
A_1,
\nonu \\
A_4 & = &  \frac{160  \lambda ^3}
{(\lambda -3) (\lambda -1) (2 \lambda -3)} A_1,
\qquad
A_5  =   -\frac{40 \lambda ^4}
{(\lambda -3) (\lambda -1) (2 \lambda -3)} A_1,
\nonu \\
A_6 & = &  -N^2 \, \frac{12  (2 \lambda +3)
  }{ (2 \lambda -3)} A_1,
\qquad
A_7  =   N^2 \, \frac{8  (2 \lambda +3)
  }{ (2 \lambda -3)} A_1,
\nonu \\
A_8 & = &  -N^2 \, \frac{48  \lambda ^2
  \left(2 \lambda ^2+3 \lambda +3\right)
  }{(\lambda -3) (\lambda -1) (2 \lambda -3)}
A_1,
\qquad
A_9  =   N^2 \, \frac{32  \lambda ^2
  \left(2 \lambda ^2+3 \lambda +3\right)
  }{(\lambda -3) (\lambda -1) (2 \lambda -3)}
A_1,
\nonu \\
A_{10} & = &  -N^2 \, \frac{48  \lambda  (\lambda +2)
  }{(\lambda -1) (2 \lambda -3)} A_1,
\qquad
A_{11}  =   N^2 \, \frac{16  \lambda  (5 \lambda +6)
  }{(\lambda -1) (2 \lambda -3)} A_1,
\nonu \\
A_{12} & = &  -N^2 \,
\frac{16  \lambda  \left(19 \lambda ^2+21 \lambda +6\right)
  }
{(\lambda -3) (\lambda -1) (2 \lambda -3)}
A_1,
\qquad
A_{13}  =   - N \, \frac{48  \lambda  (\lambda +1)
  }{(\lambda -1) (2 \lambda -3)} A_1,
\nonu \\
A_{14} & = &  N \, \frac{240  \lambda ^2 (\lambda +1)
  }{(\lambda -3) (\lambda -1) (2 \lambda -3)} A_1,
\qquad
A_{15}  =   - \frac{1}{N} \,
\frac{18  \left(2 \lambda ^2-4 \lambda -3\right)}
     {(2 \lambda -3) (2 \lambda +1) } A_1,
\nonu \\
A_{16} & = & \frac{1}{N} \, \frac{36  \lambda ^4
  \left(24 \lambda ^2-19 \lambda -19\right)}
{(\lambda -3) (\lambda -1)^2 (2 \lambda -3) (2 \lambda +1)
  }
A_1,
\nonu \\
A_{17} & = & -\frac{1}{N} \,
\frac{72  \lambda ^2 (7 \lambda +4)}{(2 \lambda -3) (2 \lambda +1)
  (\lambda  -1)}
A_1,
\qquad
A_{18}  =   - \frac{1}{N} \,
\frac{144  \lambda  \left(2 \lambda ^2-4 \lambda -3\right)}{(2 \lambda -3) (2 \lambda +1) (\lambda  -1)} A_1,
\nonu \\
A_{19} & = & -\frac{1}{N} \,
\frac{144  \lambda ^3 \left(24 \lambda ^2-19 \lambda -19\right)}{(\lambda -3) (\lambda -1)^2 (2 \lambda -3) (2 \lambda +1)
  }
A_1,
\nonu \\
A_{20} & = & -\frac{1}{N} \, \frac{72  \lambda ^2
  \left(10 \lambda ^2-13 \lambda -11\right)}
{(\lambda -1)^2 (2 \lambda -3) (2 \lambda +1)}
A_1.
\label{Acoeff}
\eea
The nineteen coefficients are given by $A_1$ coefficient.
One can analyze the $N$ behavior on these coefficients explicitly
as done before and they will contribute to the various
eigenvalue equations (and three-point functions) in (\ref{three}).

\section{ The OPEs between the spin $\frac{3}{2}$ current and
various spin $4$ currents}

One presents the various OPEs
between the spin $\frac{3}{2}$ current in (\ref{ghat})
and various higher spin $4$ currents in (\ref{w4})
{\small
\bea
&& \hat{G}(z) \, d^{abcd} J^a J^b J^c J^d(w) \Bigg|_{\frac{1}{(z-w)^2}}
 =  d^{abcd} \Bigg(-k_2 \psi^a J^b J^c J^d + f^{bb'd'} f^{ab'c'}
(\psi^{d'} K^{c'})(J^c J^d)  
\nonu \\
& & + f^{cb'd'} f^{ab'c'} J^b (\psi^{d'} K^{c'}) J^d
+ f^{bc'd'} f^{dc'e} J^a J^c \psi^e K^{d'}
 +   f^{db'd'} f^{ab'c'} J^b J^c \psi^{d'} K^{c'}
-k_2 J^a \psi^b J^c J^d
\nonu \\
& & + f^{bc'd'} f^{cc'e} J^a (\psi^e K^{d'}) J^d
-2 k_2 J^a J^b \psi^c J^d
 + f^{cd'e'} f^{dd'f} J^a J^b \psi^f K^{e'}\Bigg),
\nonu \\
&&
\hat{G}(z) \, d^{bcde} J^b J^c J^d K^e(w) \Bigg|_{\frac{1}{(z-w)^2}}  = 
d^{abcd} \Bigg( k_2 \psi^d J^a J^b J^c -
f^{aed'} f^{def} (\psi^{d'} K^f)(J^b J^c)
\nonu \\
& & -   f^{bed'} f^{def} J^a (\psi^{d'} K^f) J^c
-  f^{ced'} f^{def} J^a J^b \psi^{d'} K^f
 - k_2  K^d \psi^a J^b J^c +  f^{ac'd'} f^{bc'e} K^d (\psi^e
K^{d'}) J^c
\nonu \\
& & +   f^{ac'd'} f^{cc'e} K^d J^b \psi^e K^{d'}
-2k_2  K^d J^a \psi^b J^c
+  f^{bd'e'} f^{cd'f} K^d J^a \psi^f K^{e'} \Bigg),
\nonu \\
&& \hat{G}(z) \, d^{bcde} J^b J^c K^d K^e(w)  \Bigg|_{\frac{1}{(z-w)^2}}  = 
d^{abcd}  \Bigg(
-k_2 \psi^a J^b K^c K^d +  f^{aef} f^{beg} (\psi^g K^f)
(K^c K^d) 
\nonu \\
& & +  f^{cfg} f^{aef} J^b (\psi^e K^g) K^d
+  f^{aef} f^{dfg} J^b K^c \psi^e K^g 
 -  k_2 J^a \psi^b K^c K^d +d^{abcd} f^{bef} f^{cfg} J^a (\psi^e K^g)
K^d
\nonu \\
& & +  f^{bef} f^{dfg} J^a K^c \psi^e K^g
+ 2  k_2 J^a J^b \psi^c K^d
-   f^{cef} f^{dfg} J^a J^b \psi^e K^g \Bigg),
\nonu \\
&& \hat{G}(z) \, d^{bcde} J^b K^c K^d K^e(w)  \Bigg|_{\frac{1}{(z-w)^2}}  = 
d^{abcd} \Bigg(-k_2 \psi^a K^b K^c K^d +  f^{aef} f^{bfg}
(\psi^e K^g)(K^c K^d) 
\nonu \\
& & +  f^{cfg} f^{aef} K^b (\psi^e K^g) K^d
+  f^{aef} f^{dfg} K^b K^c \psi^e K^g
 +  k_2 J^a \psi^b K^c K^d - f^{bef} f^{cfg} J^a
(\psi^e K^g) K^d 
\nonu \\
& & -  f^{bef} f^{dfg} J^a K^c \psi^e K^g
+2  k_2 J^a K^b \psi^c K^d
 -  f^{cef} f^{dfg} J^a K^b \psi^e K^g \Bigg), 
\nonu \\
&& \hat{G}(z) \, d^{bcde} K^b K^c K^d K^e(w)  \Bigg|_{\frac{1}{(z-w)^2}}  = 
d^{abcd} \Bigg(
k_2 \psi^a K^b K^c K^d -  f^{aef} f^{bfg}
(\psi^e K^g)(K^c K^d) 
\nonu \\
& & -   f^{aef} f^{cfg} K^b (\psi^e K^g) K^d
- f^{aef} f^{dfg} K^b K^c \psi^e K^g
 +  k_2 K^a \psi^b K^c K^d -f^{bef} f^{cfg} K^a (\psi^e K^g) K^d
\nonu \\
& & -  f^{bef} f^{dfg} K^a K^c \psi^e K^g
+2  k_2 K^a K^b \psi^c K^d
 -  f^{cef} f^{dfg} K^a K^b \psi^e K^g \Bigg),
\nonu \\
&& \hat{G}(z)
\, \pa J^a \pa J^a(w)  \Bigg|_{\frac{1}{(z-w)^2}}  = 
-2k_2 \pa \psi^a \pa J^a -f^{abc} (\psi^b K^c) \pa J^a
 - 2 k_2 \pa J^a \pa \psi^a -f^{abc} \pa J^a \psi^b K^c,
\nonu \\
&& \hat{G}(z)
\, \pa^2 J^a  J^a(w)  \Bigg|_{\frac{1}{(z-w)^2}}  = 
-3k_2 \pa^2 \psi^a J^a -2 f^{abc} \pa (\psi^b K^c) J^a
 - k_2 \pa^2 J^a \psi^a,
\nonu \\
&& \hat{G}(z)
\, \pa K^a \pa K^a(w)  \Bigg|_{\frac{1}{(z-w)^2}}  = 
2k_2 \pa \psi^a \pa K^a + f^{abc} (\psi^b K^c) \pa K^a
 + 2k_2 \pa K^a \pa \psi^a + f^{abc} \pa K^a \psi^b K^c,
\nonu \\
&& \hat{G}(z)
\, \pa^2 K^a  K^a(w)  \Bigg|_{\frac{1}{(z-w)^2}}  = 
3k_2 \pa^2 \psi^a K^a + 2 f^{abc} \pa (\psi^b K^c) K^a
 + k_2 \pa^2 K^a \psi^a,
\nonu \\
&& \hat{G}(z)
\, \pa J^a \pa K^a(w)  \Bigg|_{\frac{1}{(z-w)^2}}  = 
-2k_2 \pa \psi^a \pa K^a -f^{abc} (\psi^b K^c) \pa K^a
 + 2k_2 \pa J^a \pa \psi^a + f^{abc} \pa J^a \psi^b K^c, 
\nonu \\
&& \hat{G}(z)
\, \pa^2 J^a  K^a(w)  \Bigg|_{\frac{1}{(z-w)^2}}  = 
- 3k_2 \pa^2 \psi^a K^a - 2 f^{abc} \pa (\psi^b K^c) K^a
+ k_2 \pa^2 J^a \psi^a,
\nonu \\
&& \hat{G}(z)
\,  J^a \pa^2 K^a(w)  \Bigg|_{\frac{1}{(z-w)^2}}  = 
-k_2 \psi^a \pa^2 K^a -2 (2N-2) \pa^2 (\psi^a K^a)
+ 3k_2 J^a \pa^2 \psi^a \nonu \\
&& +2 f^{abc} J^a \pa (\psi^b K^c),
\nonu \\
&& \hat{G}(z)
 \, f^{abc} J^a \pa J^b K^c (w)  \Bigg|_{\frac{1}{(z-w)^2}}  = 
-k_2 f^{abc} \psi^a \pa J^b K^c -(2N-2) f^{ecf} \pa (\psi^f K^e) K^c
\nonu \\
&& + (2N-2) f^{dfb} \pa J^b \psi^d K^f
\nonu \\
 && -2k_2 f^{abc} J^a \pa \psi^b K^c  
- f^{bde} f^{abc} J^a (\psi^d K^e) K^c
+ k_2 f^{abc} J^a \pa J^b \psi^c
 + 2 k_2 (2N-2) \pa J^a \pa \psi^a,
\nonu \\
&& \hat{G}(z)
 \, f^{abc} J^a  K^b \pa K^c (w)  \Bigg|_{\frac{1}{(z-w)^2}}  = 
-k_2 f^{abc} \psi^a K^b \pa K^c - (2N-2) f^{abc} (\psi^a K^b) \pa K^c  
\nonu \\
& & + (2N-2) f^{abc} K^a \pa (\psi^b K^c)
+ k_2 f^{abc} J^a \psi^b \pa K^c 
 + 2k_2 (2N-2) J^a \pa^2 \psi^a  + 2k_2 (2N-2) K^a \pa^2 \psi^a
\nonu \\
& & + (2N-2) f^{abc} J^a \pa (\psi^b K^c) + 2k_2 f^{abc} J^a K^b \pa \psi^c
 +  f^{abc} f^{cde} J^a K^b \psi^d K^e -2k_2 (2N-2) \pa \psi^a
\pa K^a,
\nonu \\
&& \hat{G}(z)
 \, J^a J^a J^b J^b(w)  \Bigg|_{\frac{1}{(z-w)^2}}  = 
-k_2 \psi^a J^a J^b J^b + 2(2N-2) (\psi^a K^a) (J^b J^b)
+ 2 (2N-2) J^a J^a \psi^b K^b
\nonu \\
& & + 2 f^{acd} f^{bce} J^a (\psi^e K^d) J^b
+ 2 f^{acd} f^{bce} J^a J^b \psi^e K^d 
 - k_2 J^a \psi^a J^b J^b - k_2 J^a J^a \psi^b J^b
  -k_2 J^a J^a J^b \psi^b, 
\nonu \\
&& \hat{G}(z)
 \, K^a K^a K^b K^b(w)  \Bigg|_{\frac{1}{(z-w)^2}}  = 
k_2 \psi^a K^a K^b K^b + 2(2N-2) (\psi^a K^a) (K^b K^b)
\nonu \\
& & -4 k_2 f^{abc} K^a \pa \psi^b K^c - f^{abc} f^{cde} K^d (\psi^e K^a) K^b
- 2 f^{abc} f^{cde} K^a K^e \psi^b K^d
+k_2 K^a \psi^a K^b K^b 
\nonu \\
& & - f^{abc} f^{cde} K^a (\psi^b K^d) K^e
+ 2(k_2 +2N-2) K^a K^a \psi^b K^b,
\nonu \\
&& \hat{G}(z)
 \, J^a J^a K^b K^b(w)  \Bigg|_{\frac{1}{(z-w)^2}}  = 
-2k_2 \psi^a J^a K^b K^b + 2(2N-2) (\psi^a K^a) (K^b K^b)
\nonu \\
& & +
4 k_2 f^{abc} J^a \pa \psi^b K^c + 2 f^{acd} f^{bde} J^a (\psi^c K^e) K^b
+2 f^{acd} f^{bde} J^a K^b \psi^c K^e
\nonu \\
&& +2(k_2 +2N-2) J^a J^a \psi^b K^b,
\nonu \\
&& \hat{G}(z)
 \, J^a J^a J^b K^b(w)  \Bigg|_{\frac{1}{(z-w)^2}}  = 
-2k_2  \psi^a J^a J^b K^b + 2(2N-2) (\psi^a K^a) (J^b K^b)
\nonu \\
& & + 2 f^{acd} f^{bce} J^a (\psi^e K^d) K^b +
2k_2 f^{acb} J^a J^b \pa \psi^c  
 + 2 f^{acd} f^{bde} J^a J^b \psi^c K^e
-k_2 J^a J^a \psi^b K^b
\nonu \\
& & - 2(2N-2) J^a J^a \psi^b K^b + k_2 J^a J^a J^b \psi^b, 
\nonu \\
&& \hat{G}(z)
 \, J^a K^a K^b K^b(w)  \Bigg|_{\frac{1}{(z-w)^2}}  = 
-k_2 \psi^a K^a K^b K^b -2 (2N-2) (\psi^a K^a) (K^b K^b)
\nonu \\
& & + k_2 f^{acb} K^a \pa \psi^c K^b + f^{acd} f^{bde} K^a (\psi^c K^e) K^b
+ k_2 f^{acb} K^a K^b \pa \psi^c + f^{acd} f^{bde} K^a K^b \psi^c K^e 
\nonu \\
& & + k_2 J^a \psi^a K^b K^b -k_2 f^{abc} J^a \pa \psi^b K^c
 - f^{acd} f^{bde} J^a (\psi^c K^e) K^b + k_2 f^{abc} J^a K^b \pa \psi^c 
\nonu \\
& & - f^{acd} f^{bde} J^a K^b \psi^c K^e + 2(k_2 +2N-2) J^a K^a \psi^b
K^b,
\nonu \\
&& \hat{G}(z)
 \, J^a J^b K^a K^b(w)  \Bigg|_{\frac{1}{(z-w)^2}}  =  
-k_2 \psi^a J^b K^a K^b + f^{acd} f^{bce} (\psi^e K^d) (K^a K^b)
\nonu \\
& & - 2(2N-2) J^a (\psi^b K^b) K^a
+2k_2 f^{acb} J^b K^a \pa \psi^c
 + f^{acd} f^{bde} J^b K^a \psi^c K^e
-k_2 J^a \psi^b K^a K^b
\nonu \\
& & + f^{bcd} f^{ade} J^a (\psi^c K^e) K^b
-2(2N-2)J^a K^a \psi^b K^b
\nonu \\
& & + k_2 J^a J^b \psi^a K^b 
-k_2 f^{acb} J^a J^b \pa \psi^c
 - f^{acd} f^{bde} J^a J^b \psi^c K^e
+ k_2 J^a J^b K^a \psi^b.
\label{gopespin4}
\eea
}
Note that
the field $d^{abcd} \psi^a J^b J^c J^d(w)$
has the coefficient $(-4k_2 A_1 +k_2 A_2)$ which becomes
$-4(-4+k_2+4N) A_1$ using Appendix (\ref{othercoeffexp}).
Further arrangement for the normal ordered product
is needed to express singular terms in simple form.
In principle, one can obtain the higher spin $\frac{7}{2}$ current
(\ref{gwexp})
by simplifying the above results in Appendix (\ref{gopespin4}).

\section{ The OPEs between the diagonal spin $1$ current and
various spin $\frac{7}{2}$ currents}

The OPE between diagonal spin $1$ current (\ref{Jfermions})
and the various higher spin $\frac{7}{2}$ currents
in (\ref{w7half}) can be written as
{\small
\bea
&& J^{\prime a}(z) \, d^{bcde} \psi^b J^c  J^d J^e(w)  =
\frac{1}{(z-w)^4} \, 2(4N^2-14N+22)(2N-2) \psi^a(w)
\nonu \\
& & -\frac{1}{(z-w)^3} \, 4(4N^2-14N+22)(2N-2) \pa \psi^a(w) 
 + 
\frac{1}{(z-w)^2} \, \Bigg[ 2(2N-2)(2N^2-7N+11) \pa^2 \psi^a
  \nonu \\
  & & -  14(N-1) d^{abcd} \psi^b J^c J^d
-12 \psi^a J^b J^b
 + 12 \psi^b J^b J^a
+ 6(2N^2-7N+13) f^{abc} \psi^b \pa J^c
\Bigg](w)+\cdots,
\nonu \\
&& J^{\prime a}(z)
\, d^{bcde} f^{bfg} f^{cfh} J^d J^e \psi^g K^h(w)
 = 
-\frac{1}{(z-w)^3} \, 4(N-1)(2N^2-7N+11) f^{abc} \psi^b K^c(w) 
\nonu \\
& & + \frac{1}{(z-w)^2} \, \Bigg[
\frac{8}{3}(2N-2)(N-1) d^{abcd} \psi^b J^c K^d
 +
\frac{4}{3} k_2 (N-1) d^{abcd} \psi^b J^c J^d 
\nonu \\
& & - \frac{4}{3}  (N-1) d^{cdeb} f^{baf} f^{fgc} \psi^g J^d K^e 
 - \frac{4}{3}  (N-1) d^{dceb} f^{baf} f^{fhd} J^c \psi^h  K^e 
\nonu \\
& & - 
\frac{4}{3}  (N-1) d^{bcde} f^{acf} f^{fdg} \psi^b J^g K^e
-  4k_2 (N-1)(2N^2-7N+11) \pa^2 \psi^a 
\nonu \\
& & +  4 f^{abc} f^{cde} \psi^e J^b K^d
 -  \frac{44}{3}(N-1)(2N^2-7N+11) f^{abc} \pa \psi^b K^c 
\nonu \\
& & - 8(2N-2) \psi^b J^b K^a 
- 24(2N-2) \psi^a J^b K^b
 -  8 k_2 \psi^b J^a J^b
 -  \frac{10}{3} k_2 (2N^2-7N+11) f^{abc} \psi^b \pa J^c
\nonu \\
& & +  32(2N-2) \psi^b J^a K^b
+ 8k_2 \psi^a J^b J^b
 + 
\frac{2}{3}(10N^2-35N+49) f^{abc} f^{cde} \psi^d J^b K^e
\Bigg](w)+ \cdots,
\nonu \\
&& J^{\prime a}(z)
\, d^{bcde} J^b K^c \psi^d K^e(w)  = 
\frac{1}{(z-w)^2} \, \Bigg[
d^{bcde} f^{fcg} f^{abf} \psi^g K^d K^e
\nonu \\
& & -  (2N-2) d^{abcd} \psi^b K^c K^d
+ d^{bcde} f^{adf} f^{feg} \psi^b J^c K^g 
 -  2k_2 d^{abcd} \psi^b J^c K^d
\Bigg](w) + \cdots,
\nonu \\
&& J^{\prime a}(z)
\, d^{bcde} f^{bfg} f^{cfh} K^d K^e \psi^g K^h(w)
 = 
-\frac{1}{(z-w)^4} \, 8 k_2 (N-1)(2N^2-7N+11)
\, \psi^a
\nonu \\
& & + \frac{1}{(z-w)^3} \,
4(N-k_2-1)(2N^2-7N+11) f^{abc} \psi^b K^c
\nonu \\
& & +  \frac{1}{(z-w)^2} \, \Bigg[
  4k_2(N-1) d^{abcd} \psi^b K^c K^d
   - \frac{4}{3}(N-1) d^{bcde} f^{acf} f^{fdg} \psi^b K^g K^e
   \nonu \\
   && -  \frac{8}{3} (N-1) d^{bcde} f^{adf} f^{feg} \psi^b K^c K^g 
   - 4(2k_2 +2(2N-2)) \psi^b K^b K^a
  \nonu \\
  & & + (4(k_2+2N-2) +(2N-5)(2N-2)(k_2+2N-2)
   -  \frac{2}{3}(2N-5)(2N-2)^2) f^{abc} \psi^b \pa K^c
  \nonu \\
  & & +  (4+ (2N-2)(2N-5)) f^{abc} f^{cde} \psi^e K^d K^b 
   +  4(2 k_2 +2(2N-2)) \psi^b K^a K^b 
  \nonu \\
  & & -  8 f^{abc} f^{cde} \psi^d K^b K^e
  \Bigg](w) +\cdots,
\nonu \\
&& J^{\prime a}(z) \, J^b \psi^b J^c J^c(w)
 = 
\frac{1}{(z-w)^4} \, 4 (2N-2)^2 \psi^a(w)
- \frac{1}{(z-w)^3} \, 8(2N-2)^2 \pa \psi^a(w) 
\nonu \\
& & + \frac{1}{(z-w)^2} \, \Bigg[
-3(2N-2) \psi^a J^b J^b + 2(2N-2) f^{abc} \psi^b \pa J^c
\nonu \\
& & +  2(2N-2)^2 \pa^2 \psi^a -4(2N-2) \psi^b J^b J^a 
\Bigg](w) + \cdots,
\nonu \\
&& J^{\prime a}(z)
 \, K^b K^b \psi^c K^c(w)
 =  
\frac{1}{(z-w)^4} \, 2 k_2 (2N-2) \psi^a(w)
\nonu \\
&& + 
\frac{1}{(z-w)^3} \, ( 2k_2 - 2(2N-2)) f^{abc} \psi^b K^c(w) 
\nonu \\
& & + \frac{1}{(z-w)^2} \, \Bigg[
-(2 k_2+ 2(2N-2)) \psi^b K^a K^b + 2 f^{abc} f^{cde} \psi^d K^b K^e
 -  k_2 \psi^a K^b K^b  
\Bigg](w) + \cdots,
\nonu \\
&& J^{\prime a}(z) \, J^b J^b \psi^c K^c(w)
  = 
\frac{1}{(z-w)^3} \, 2(2N-2) f^{abc} \psi^b K^c(w) 
\nonu \\
& & + \frac{1}{(z-w)^2} \, \Bigg[
  -4(2N-2) \psi^b J^a K^b
  -k_2 \psi^a J^b J^b
 +  k_2(2N-2)  \pa^2 \psi^a
 \nonu \\
 && +2(2N-2) f^{abc} \pa \psi^b K^c
 - 2 f^{abc} f^{cde} \psi^d J^b K^e
+ 2 k_2 f^{abc} \psi^b \pa J^c
\Bigg](w) + \cdots,
 \nonu \\
&& J^{\prime a}(z) \, J^b J^b \psi^c J^c(w)
  = 
 \frac{1}{(z-w)^4} \,
 6(2N-2)^2 \psi^a(w)
  -
 \frac{1}{(z-w)^3} \, 12(2N-2)^2 \pa \psi^a(w) 
\nonu \\
& & + \frac{1}{(z-w)^2} \, \Bigg[
  -3(2N-2) \psi^a J^b J^b
  + 3(2N-2)^2 \pa^2 \psi^a 
\nonu \\
  & & -  4(2N-2)  \psi^b J^b J^a
-2(2N-2) f^{abc} \psi^b \pa J^c
\Bigg](w) + \cdots,
 \nonu \\
&& J^{\prime a}(z) \, \psi^b K^b K^c K^c(w)
  = 
- \frac{1}{(z-w)^3} \, (2k_2+2(2N-2)) f^{abc} \psi^b K^c(w) 
\nonu \\
& & + \frac{1}{(z-w)^2} \, \Bigg[
-k_2 \psi^a K^b K^b - (2k_2+2(2N-2))  \psi^b  K^b K^a
\Bigg](w)  +  \cdots,
\nonu \\
&& J^{\prime a}(z) \, f^{bcd} f^{def} K^b K^f \psi^c K^e(w)
 = 
-\frac{1}{(z-w)^4} \, 4(2N-2)^2 k_2 \psi^a(w) 
\nonu
\\
& & + \frac{1}{(z-w)^3} \, (2N-2)(-2k_2+(2N-2)) f^{abc} \psi^b K^c(w) 
\nonu \\
& & + \frac{1}{(z-w)^2} \, \Bigg[
  (2N-2)(k_2+(2N-2)) f^{abc} \psi^b  \pa K^c
 +  (2N-2)
  f^{abc} f^{cde} \psi^e  K^d K^b
\Bigg](w) + \cdots,
\nonu \\
&& J^{\prime a}(z)
\, J^b \psi^b K^c K^c(w)
 = 
 \frac{1}{(z-w)^2} \, \Bigg[
   -3(2N-2) \psi^a K^b K^b
   -  (2k_2+2(2N-2))  \psi^b  J^b K^a
   \Bigg](w) \nonu \\
 && + \cdots,
\nonu \\
&& J^{\prime a}(z) \, \psi^b J^c K^b K^c(w)
 = 
 \frac{1}{(z-w)^2} \, \Bigg[
f^{abc} f^{cde} \psi^e K^b K^d - (2N-2)  \psi^b  K^b K^a
\nonu \\
& & - k_2 \psi^a J^b K^b + f^{abc} f^{cde} \psi^b J^d K^e
 - k_2 \psi^b J^a K^b
\Bigg](w) + \cdots,
\nonu \\
&& J^{\prime a}(z) \, J^b J^c \psi^b K^c(w)
 = 
  \frac{1}{(z-w)^2} \, \Bigg[
    -k_2 (2N-2) \pa^2 \psi^a + f^{abc} f^{cde}
    \psi^e J^b K^d \nonu \\
    & & - 
    4(2N-2) f^{abc} \pa \psi^b K^c -
    (2N-2)  \psi^b  J^b K^a
    - 3(2N-2) \psi^a J^b K^b
    -k_2 \psi^b J^a J^b
\Bigg](w) + \cdots,
\nonu \\
&& J^{\prime a}(z) \, f^{bcd} f^{def} J^b J^f \psi^c K^e(w)
 = 
 \frac{1}{(z-w)^3} \, (2N-2)^2 f^{abc} \psi^b K^c(w) 
\nonu \\
& & + \frac{1}{(z-w)^2} \, \Bigg[
  -5 (2N-2)^2 f^{abc} \pa \psi^b K^c
   +  3 (2N-2)  f^{abc}f^{cde}
  \psi^d  J^b K^e
\nonu \\
& & - 3k_2 (2N-2) f^{abc} \psi^b \pa J^c
-(2N-2)^2 k_2 \pa^2 \psi^a
  \Bigg](w) + \cdots,
\nonu \\
&& J^{\prime a}(z) \, J^b J^c K^b \psi^c(w)
=  
\frac{1}{(z-w)^3} \, 3(2N-2) f^{abc} \psi^b K^c(w) 
\nonu \\
& & + \frac{1}{(z-w)^2} \, \Bigg[
-k_2 f^{abc} \psi^b \pa J^c - 3(2N-2)  \psi^a  J^b K^b
 - (2N-2) \psi^b J^b K^a
 \nonu \\
 && -3 (2N-2) f^{abc} \pa \psi^b K^c
 -  k_2 \psi^b J^a J^b
\Bigg](w) + \cdots.
\label{opewriting}
\eea}
It is nontrivial to express these OPEs in terms of independent terms
with the rearrangement of the normal ordered product
\cite{BBSSfirst,BBSSsecond,Fuchs}.

We describe each independent terms with their coefficients
in (\ref{w7half})
appearing
in the various poles of Appendix (\ref{opewriting})
where the rearrangement \cite{BBSSfirst}
of the normal ordered product
is used
as follows:
\bea
\mbox{pole-4}: && \Bigg[
  -4 B_{10} k_2 (2 N-2)^2+2 B_6
  k_2 (2 N-2)-B_4 8 k_2 (N-1)(2N^2-7N+11)\nonu \\
  && +
  2 B_1 (4 N^2-14 N+22) (2 N-2)+
  4 B_5 (2 N-2)^2+6 B_8 (2 N-2)^2
\Bigg] \psi^a(w) =0,
\nonu \\
\mbox{pole-3}: && \Bigg[
  -4 B_1 (4 N^2-14 N+22) (2 N-2)-
  8 B_5 (2 N-2)^2 \nonu \\
  && -12 B_8 (2 N-2)^2
  \Bigg] \pa \psi^a(w) =0,
\nonu \\
\mbox{pole-3}: && \Bigg[
  B_4 4(N-k_2-1)(2N^2-7N+11)+
  B_{10} (2 N-2) (-2 k_2+2 N-2)\nonu \\
  && +
  B_6 (2 k_2-2 (2 N-2))-
  B_9 (2 k_2+2 (2 N-2))-
  B_2 4(N-1)(2N^2-7N+11)\nonu \\
  && +
  B_{14} (2 N-2)^2+2 B_7 (2 N-2)+3 B_{15} (2 N-2) \Bigg]
f^{abc} \psi^b K^c(w) =0,
\nonu \\
\mbox{pole-2}: && \Bigg[
\frac{4}{3} B_2 k_2 (N-1)-14 B_1 (N-1)
\Bigg] d^{abcd} \psi^b J^c J^d =0,
\nonu \\
\mbox{pole-2}: && 
\Bigg[ 8 B_2 k_2-B_7 k_2-
3 B_5 (2 N-2)-3 B_8 (2 N-2)-12 B_1 \Bigg]
\psi^a J^b J^b(w) =0,
\nonu \\
\mbox{pole-2}: && \Bigg[
-8 B_2 k_2-B_{13} k_2-
B_{15} k_2-4 B_5 (2 N-2)-4 B_8 (2 N-2)
\nonu \\
&& +
12 B_1 \Bigg] \psi^b J^b J^a(w)=0, 
\nonu \\
\mbox{pole-2}: && \Bigg[
-B_{14} k_2 (2 N-2)^2+B_7 k_2 (2 N-2)-
B_{13} k_2 (2 N-2) \nonu \\
&& -
B_2 4k_2(N-1)(2N^2-7N+11)+
2 B_5 (2 N-2)^2+3 B_8 (2 N-2)^2
\nonu \\
&& +
B_1 2(2N^2-7N+11) (2 N-2) \Bigg] \pa^2 \psi^a(w) =0,
\nonu \\
\mbox{pole-2}: && \Bigg[
-B_2 \frac{10}{3} k_2 (2N^2-7N+11)-
3 B_{14} k_2 (2 N-2)+2 B_7 k_2-
B_{13} k_2\nonu \\
&& -2 B_{15} k_2+
B_1 (3 (2 N-5) (2 N-2)+48)
+2 B_5 (2 N-2)\nonu \\
&& -2 B_8 (2 N-2)
\Bigg] f^{abc} \psi^b \pa J^c(w) =0,
\nonu \\
\mbox{pole-2}: && \Bigg[
  -B_6 k_2-B_9 k_2-\frac{1}{3} 32 B_4 (N-1)-
  \frac{16}{3} B_4 (N-1)-3 B_{11} (2 N-2) \nonu \\
  && -8 B_3
  \Bigg] \psi^a K^b K^b(w) =0,
\nonu \\
\mbox{pole-2}: && \Bigg[
  B_4 (4 k_2 (N-1)+\frac{16}{3} (N-1)^2)-
  \frac{1}{3} 4 B_3 (N-1)\nonu \\
  && -B_3 (2 N-2)
\Bigg] d^{abcd} \psi^b K^c K^d=0,
\nonu \\
\mbox{pole-2}: && \Bigg[
  -4 B_4 (2 k_2+2 (2 N-2))-
  4 B_4 (-2 k_2-2 (2 N-2))+
  B_6 (-2 k_2-2 (2 N-2))\nonu \\
  && -
  B_9 (2 k_2+2 (2 N-2))+
  \frac{32}{3} B_4 (N-1)-\frac{32}{3} B_4 (N-1)+
  \frac{4}{3} 16 B_4 (N-1)\nonu \\
  && -\frac{16}{3} B_4 (N-1)-
  B_{12} (2 N-2)+4 B_3+4 B_3 \Bigg]
\psi^b K^b K^a=0,
\nonu \\
\mbox{pole-2}: && \Bigg[
  -\frac{8}{9} B_4 (N-1) (2 N-5)-
  \frac{4}{9} B_4 (N-1) (2 N-5)-B_4 ((2 N-5) (2 N-2)+4)
  \nonu \\
  && -
  B_{10} (2 N-2)-
  8 B_4+2 B_6-B_{12} -\frac{2}{3}(2N-5)B_3
\Bigg] f^{abc} f^{cde} \psi^d K^b K^e =0,
\nonu \\
\mbox{pole-2}: && \Bigg[
  -2 B_3 k_2+\frac{1}{3} 16 B_2 (N-1)^2-
  \frac{4}{3} B_3 (N-1)\nonu \\
  && +\frac{8}{3} B_2 (2 N-2) (N-1)
  \Bigg] d^{abcd} \psi^b J^c K^d(w)=0,
\nonu \\
\mbox{pole-2}: && \Bigg[
  -B_{12} k_2+\frac{2}{3} 32 B_2 (N-1)-
  \frac{16}{3} B_2 (N-1)-
  24 B_2 (2 N-2)\nonu \\
  && -3 B_{13} (2 N-2)-
  3 B_{15} (2 N-2)+4 B_3
  \Bigg] \psi^a J^b K^b(w) =0, 
\nonu \\
\mbox{pole-2}: && \Bigg[
  -B_{12} k_2+32 B_2 (2 N-2)-
  4 B_7 (2 N-2)+4 B_3
  \Bigg] \psi^b J^a K^b (w) =0,
\nonu \\
\mbox{pole-2}: && \Bigg[
  -B_{11} (2 k_2+2 (2 N-2))-8 B_2 (2 N-2)-16 B_2 (N-1)-
  B_{13} (2 N-2)\nonu \\
  && -B_{15} (2 N-2)-8 B_3
\Bigg] \psi^b J^b K^a(w) =0,
\nonu \\
\mbox{pole-2}: && \Bigg[
  -\frac{1}{3} B_3 (2 N-5)-
  \frac{4}{9} B_2 (N-1) (2 N-5)-
  \frac{4}{9} B_2 (N-1) (2 N-5)\nonu \\
  && -
  \frac{4}{9} B_2 (N-1) (2 N-5)+
  \frac{1}{3} B_3 (-(2 N-5))-
  B_2 (\frac{5}{3} (2 N-5) (2 N-2)+16)
  \nonu \\
  && -
  3 B_{14} (2 N-2)+4 B_2+2 B_7-B_{12}+B_{13}
\Bigg] f^{abc} f^{cde} \psi^e J^b K^d(w) =0,
\nonu \\
\mbox{pole-2}: && \Bigg[
  -\frac{4}{3} (-1) B_2 (N-1) (4 N^2-14 N+22)-
  5 B_{14} (2 N-2)^2+2 B_7 (2 N-2)
\nonu \\
  && -4 B_{13} (2 N-2)-
3 B_{15} (2 N-2) \nonu \\
&& +B_2
  (-44 (2 N-2)-\frac{11}{3} (2 N-5) (2 N-2)^2)
\Bigg] f^{abc} \pa \psi^b K^c(w) =0,
\nonu \\
\mbox{pole-2}: && \Bigg[
  B_{10} (2 N-2) (k_2+2 N-2)+
  B_4 ((2 N-5) (2 N-2) (k_2+2 N-2)
  \nonu \\
  && +4 (k_2+2 N-2)-
  \frac{2}{3} (2 N-5) (2 N-2)^2)-
  4 B_4 (-2 k_2-2 (2 N-2))\nonu \\
  && +
  B_6 (-2 k_2-2 (2 N-2))-
  \frac{1}{9} 8 B_4 (N-1) (2 N-5) (2 N-2)
  \nonu \\
  && +
  \frac{1}{3} B_3 (-(2 N-5)) (2 N-2)-
  8 B_4 (2 N-2)+2 B_6 (2 N-2)
  \nonu \\
  && -B_{12} (2 N-2)-
  \frac{32}{3} B_4 (N-1)+
  \frac{1}{9} B_4 (-4 (N-1) (2 N-5) (2 N-2))
  \nonu \\
  && +
  \frac{32}{3} B_4 (N-1)+4 B_3
  \Bigg] f^{abc} \psi^b \pa K^c(w)=0. 
\label{pricoeffrelated}
\eea
Here one should obtain the various independent terms 
from Appendix (\ref{opewriting}).

\section{ The OPEs between the numerator spin $2$ current and
various spin $\frac{7}{2}$ currents}

The primary condition (\ref{primaryspin7half})
for the higher spin $\frac{7}{2}$ current
in (\ref{w7half})
can be described as
\bea
&& \hat{T}(z) \,  d^{abcd} \psi^a J^b  J^c J^d(w)  = 
-\frac{1}{(z-w)^4} \, 3(2N-2)(8N-2) \psi^a J^a(w)
+ {\cal O} (\frac{1}{(z-w)^2}), 
\nonu \\
&& \hat{T}(z)
\, d^{abcd} f^{aef} f^{beg} J^c J^d \psi^f K^g(w)
 = 
\frac{1}{(z-w)^4} \, 16(N-1)^2(4N-1) \psi^a K^a(w)
\nonu \\
& &+\frac{1}{(z-w)^3} \, 16(N-1)(2N^2-7N+11) \pa \psi^a K^a(w)
 +  {\cal O} (\frac{1}{(z-w)^2}),
\nonu \\
&& \hat{T}(z)
\, d^{abcd} J^a K^b \psi^c K^d(w)  = 
-\frac{1}{(z-w)^4} \, k_2 (8N-2) \psi^a J^a(w)
+ {\cal O} (\frac{1}{(z-w)^2}),
\nonu \\
&& \hat{T}(z)
\, d^{abcd} f^{aef} f^{beg} K^c K^d \psi^f K^g(w)
 = 
\frac{1}{(z-w)^4} \, 8k_2 (N-1)(4N-1) \psi^a K^a(w)
\nonu \\
& & + \frac{1}{(z-w)^3} \, 8(N-1)(2N^2-7N+11)  \psi^a \pa K^a(w)
 +  {\cal O} (\frac{1}{(z-w)^2}),
\nonu \\
&& \hat{T}(z) \, J^a \psi^a J^b J^b(w)
= 
-\frac{1}{(z-w)^4} \, (3+N(2N-1))(2N-2) \psi^a J^a(w)
 + {\cal O} (\frac{1}{(z-w)^2}),
\nonu \\
&& \hat{T}(z)
 \, K^a K^a \psi^b K^b(w)
 =  
-\frac{1}{(z-w)^4} \, k_2(N(2N-1)+2) \psi^a K^a(w)
\nonu \\
& & -  \frac{1}{(z-w)^3} \, 2(2N-2) \psi^a \pa K^a(w) 
+ {\cal O} (\frac{1}{(z-w)^2}),
 \nonu \\
&& \hat{T}(z) \, J^a J^a \psi^b K^b(w)
  = 
-\frac{1}{(z-w)^4} \, (2N-2)N(2N-1) \psi^a K^a(w)
\nonu \\
& & - \frac{1}{(z-w)^3} \, 4(2N-2) \pa \psi^a K^a(w)  
+ {\cal O} (\frac{1}{(z-w)^2}),
 \nonu \\
&& \hat{T}(z) \, J^a J^a \psi^b J^b(w)
  = 
-\frac{1}{(z-w)^4} \, (6+N(2N-1))(2N-2) \psi^a J^a(w)
 +  {\cal O} (\frac{1}{(z-w)^2}),
 \nonu \\
&& \hat{T}(z) \, \psi^a K^a K^b K^b(w)
  = 
-\frac{1}{(z-w)^4} \, (2(k_2 +2N-2)+k_2 N(2N-1)) \psi^a K^a(w)
\nonu \\
& & +  {\cal O} (\frac{1}{(z-w)^2}),
 \nonu \\
&& \hat{T}(z) \, f^{abc} f^{cde} K^a K^e \psi^b K^d(w)
 = 
\frac{1}{(z-w)^4} \, 2(2N-2)^2 \psi^a K^a(w)
\nonu \\
& & + \frac{1}{(z-w)^3} \, 4(2N-2)^2 \psi^a \pa K^a(w)
+ {\cal O} (\frac{1}{(z-w)^2}),
\nonu \\
&& \hat{T}(z)
\, J^a \psi^a K^b K^b(w)
 = 
-\frac{1}{(z-w)^4} \, k_2 N(2N-1) \psi^a J^a(w)
+ {\cal O} (\frac{1}{(z-w)^2}),
\nonu \\
&& \hat{T}(z) \, \psi^a J^b K^a K^b(w)
=  
-\frac{1}{(z-w)^4} \, k_2 \psi^a J^a(w)
+ {\cal O} (\frac{1}{(z-w)^2}),
\nonu \\
&& \hat{T}(z) \, J^a J^b \psi^a K^b(w)
 = 
\frac{1}{(z-w)^4} \, (2N-2) \psi^a K^a(w)
 +
\frac{1}{(z-w)^3} \, 4(2N-2) \pa \psi^a K^a(w)
\nonu \\
&& + {\cal O} (\frac{1}{(z-w)^2}),
\nonu \\
&& \hat{T}(z) \, f^{abc} f^{cde} J^a J^e \psi^b K^d(w)
 = 
\frac{1}{(z-w)^3} \, 4(2N-2)^2 \pa \psi^a K^a(w)
+ {\cal O} (\frac{1}{(z-w)^2}),
\nonu \\
&& \hat{T}(z) \, J^a J^b K^a \psi^b(w)
 =  
-\frac{1}{(z-w)^4} \, 3(2N-2) \psi^a K^a(w)
+ {\cal O} (\frac{1}{(z-w)^2}).
\label{priexp}
\eea
One also uses the various identities between $f$ and $d$ symols as in
section $2$.

Now one presents each independent terms with coefficients in (\ref{w7half})
in Appendix (\ref{priexp})
\bea
\mbox{pole-4}: && \Bigg[-B_3 k_2 (8 N-2)-B_{11} k_2 N (2 N-1)-
  B_{12} k_2-
  3 B_1 (2 N-2) (8 N-2)\nonu \\
  && -
  B_5 ((2 N-1) N (2 N-2)+3 (2 N-2))\nonu \\
  && -B_8 ((2 N-1) N (2 N-2)+6 (2 N-2))
\Bigg] \psi^a J^a(w) =0,
\nonu \\
\mbox{pole-4}: && \Bigg[
  B_4  8k_2(N-1) (4N-1)+
   B_6 (-k_2) ((2 N-1) N+2) \nonu \\
   && -
  B_9 (2 (k_2+2 N-2)+k_2 (2 N-1) N)+
  2 B_{10} (2 N-2)^2 \nonu \\
  && -
  B_7 N (2 N-1) (2 N-2)+
  B_{13} (2 N-2)-3
  B_{15} (2 N-2) \nonu \\
  && +
  B_2 16 (N-1)^2(4N-1) \Bigg] \psi^a K^a(w) =0,
\nonu \\
\mbox{pole-3}: && \Bigg[
  4 B_{10} (2 N-2)^2-2
  B_6 (2 N-2)\nonu \\
  && +B_4 8(N-1)(2N^2-7N+11)
  \Bigg] \psi^a \pa K^a(w) =0,
\nonu \\
\mbox{pole-3}: && \Bigg[
  4 B_{14} (2 N-2)^2-4 B_7 (2 N-2)+4 B_{13} (2 N-2)
  \nonu \\
  && +
  B_2 16 (N-1)(2N^2-7N+11)
  \Bigg] \pa \psi^a K^a =0.
\label{coeffprirelated}
\eea
One should have vanishing coefficients in order to have primary
higher spin $\frac{7}{2}$ current.
Without $X(K_2,N)$ term in (\ref{w7half}),
the coefficients $B_1$-$B_{15}$ do not satisfy the two
equations corresponding to the fourth order pole in (\ref{coeffprirelated}).
The coefficient of fourth order pole
is given by
\bea
\frac{48(N-4)(N-1)(N+1)(2N-3)(2N+1)(4N-4+k_2)(6N-6+k_2)}{
(2N-2+3k_2)(4N-4+3k_2)} \hat{G}(w).
\label{fractionexpexp}
\eea
This implies that the higher spin $\frac{7}{2}$ current
without $X(K_2,N)$ term in (\ref{w7half})
is a quasiprimary field. Therefore, we should add the other
quasiprimary field of spin $\frac{7}{2}$ with arbitrary coefficient
$X(k_2,N)$ which will appear later.

\section{ 
 The coefficients appearing in (\ref{w7half})
which depend on $k_2$ and $N$ when $k_1 = 2N-2$
}

By solving the linear equations Appendix (\ref{pricoeffrelated}) and
Appendix (\ref{coeffprirelated}), one has   
{\small
\bea
B_2 &=& \frac{21 }{2 k_2} B_1,
\nonu \\
B_3 & = &
\frac{168  (N-1)^2}{k_2 (3 k_2+2 N-2)} B_1,
\nonu \\
B_4 & = &
\frac{420  (N-1)^2}{k_2 (3 k_2+2 N-2)
  (3 k_2+4 N-4)} B_1,
\nonu \\
B_5 & = &
-\frac{4  (6 k_2^2 N-15 k_2^2+12 k_2 N^2-42 k_2 N+30 k_2-4 N^3+18 N^2+18 N-32)}
{(3 k_2+2 N-2) (3 k_2+4 N-4)}
B_1,
\nonu \\
B_6 & = &
\frac{504  (N-1)^2
  (2 N^2-7 N+11)}
     {k_2 (3 k_2+2 N-2) (3 k_2+4 N-4)} B_1,
\nonu \\
B_7 &=&
\frac{42
  (2 k_2^2 N^2-7 k_2^2 N+23 k_2^2+4 k_2 N^3-
  18 k_2 N^2+60 k_2 N-46 k_2+24 N^2-48 N+24)}
     {k_2 (3 k_2+2 N-2) (3 k_2+4 N-4)}
B_1,
\nonu \\
B_8 & = &
-\frac{1}{(N-1) (3 k_2+2 N-2) (3 k_2+4 N-4)}
(-10 k_2^2 N^2+35 k_2^2 N-7 k_2^2-20 k_2 N^3  \nonu \\
& + & 
90 k_2 N^2-84 k_2 N+14 k_2+16 N^4-88 N^3+72 N^2+56 N-56)
B_1,
\nonu \\
B_9 & = &
-\frac{168 (N-1)^2 (2 N^2-7 N+11)}{k_2 (3 k_2+2 N-2)
  (3 k_2+4 N-4)} B_1,
\nonu \\
B_{10} & = &
-\frac{84  (N-1)
  (2 N^2-7 N+11)}{k_2 (3 k_2+2 N-2) (3 k_2+4 N-4)} B_1,
\nonu \\
B_{11} & = &
-\frac{56  (N-1) (2 k_2 N^2-7 k_2 N+23 k_2+36 N-36)}{k_2 (3 k_2+2 N-2) (3 k_2+4 N-4)}
B_1
\nonu \\
B_{12} &=&
-\frac{336
  (N-1)^2 (2 k_2 N-5 k_2+4 N^2-14 N+4)}{k_2 (3 k_2+2 N-2) (3 k_2+4 N-4)}
B_1,
\nonu \\
B_{13} & = &
-\frac{21}{k_2 (3 k_2+2 N-2) (3 k_2+4 N-4)}
(2 k_2^2 N^2-7 k_2^2 N+23 k_2^2  \nonu \\
& + & 
4 k_2 N^3-18 k_2 N^2+60 k_2 N-46 k_2+8 N^4-44 N^3+72 N^2-44 N+8)
B_1,
\nonu \\
B_{14} & = &
-\frac{21
  (6 k_2^2 N-15 k_2^2+12 k_2 N^2-42 k_2 N+30 k_2+4 N^3-18 N^2+30 N-16)}{k_2 (3 k_2+2 N-2) (3 k_2+4 N-4)} B_1,
\nonu \\
B_{15} & = &  
\frac{7}{k_2 (3 k_2+2 N-2) (3 k_2+4 N-4)}
(22 k_2^2 N^2-77 k_2^2 N+37 k_2^2
\label{Bcoeff} \\
& + & 
44 k_2 N^3-198 k_2 N^2+228 k_2 N-74 k_2+24 N^4-132 N^3+216 N^2-132 N+
24) B_1.
\nonu
\eea}
From Appendix $G$, one determines the overall coefficient
$B_1$ in terms of $A_1$ (the description of Appendix $G$)
appearing in the higher spin $4$ current
with (\ref{gwexp}) as follows:
\bea
B_1 & = & -4  (k_2+4 N-4) \frac{1}{7}
\sqrt{\frac{4(N-1)}{(2N-2+k_2)(4N-4+k_2)}} A_1.
\label{b1a1}
\eea
The quantity $A_1$ in (\ref{b1a1}) appears in Appendix $F$.
Furthermore, the quantity $X$ appearing in (\ref{w7half})
can be determined by
{\small
  \bea
&& X =
\frac{384
(N-4) (N-1) (N+1) (2 N-3) (2 N+1)
(k_2+2 N-2) (k_2+4 N-4)^2  (k_2+6 N-6)
}{(3 k_2+2 N-2) (3 k_2+4 N-4)}
\nonu \\
& \times & \frac{A_1}{
  (4 k_2^2 N^2-2 k_2^2 N+21 k_2^2+24 k_2 N^3-36 k_2 N^2+138 k_2 N-126 k_2+168 N^2-336 N+168)}.
\label{Xexp}
\eea
}
One can check that the fourth order pole
of the OPE
$T(z) \, (G T -\frac{1}{8} \pa^2 G)(w)$
is given by
$\frac{1}{8}(4c+21) G(w)$.
In other words, the nonzero term in (\ref{fractionexpexp})
is canceled by the above $X$ term.
Note that there exists $(N-4)$ factor in Appendix (\ref{Xexp}).
In other words, for $N=4$, there is no last term in
 (\ref{w7half}).

It is straightforward to calculate the coefficients under the large
$N$ 't Hooft limit as done before. Then they will be used in the
eigenvalue equations involving the higher spin $\frac{7}{2}$ current.

\section{ 
  The OPEs between the higher spin $\frac{7}{2}$ current
  and the higher spin $4$ current and its ${\cal N}=1$ superspace
  description
}

The OPE between the lowest higher spin $\frac{7}{2}$
current and itself in section $3$ is summarized as
{\small
\bea
&& W^{(\frac{7}{2})}(z) \, W^{(\frac{7}{2})}(w)
 = 
\frac{1}{(z-w)^7} \, \frac{2c}{7} +
\frac{1}{(z-w)^5} \, 2 T(w)
+
\frac{1}{(z-w)^4} \, \pa T(w) \nonu \\
& & + 
\frac{1}{(z-w)^3}  \Bigg[ C_{\frac{7}{2} \frac{7}{2}}^{4}
  W^{(4)} +  C_{\frac{7}{2} \frac{7}{2}}^{4'} W^{(4')}
  + \frac{1}{(4 c+21) (10 c-7)} \Bigg( 8 (37 c+3) T T 
  \nonu \\
 & &+ 3 (2 c-117) 
  \pa G G +
  6 (2 c^2-6 c+9) 
\pa^2 T \Bigg)
  \Bigg](w)
\nonu \\
& &+ \frac{1}{(z-w)^2}  \Bigg[
 \frac{1}{2} C_{\frac{7}{2} \frac{7}{2}}^{4}
 \pa W^{(4)} +  \frac{1}{2} C_{\frac{7}{2} \frac{7}{2}}^{4'} \pa
 W^{(4')}
 + \frac{1}{(4 c+21) (10 c-7)} \Bigg( 8 (37 c+3)  \pa T T
 \nonu \\
 & & + 
\frac{3}{2} (2 c-117)   \pa^2 G G
+ \frac{4}{3} (2 c^2-6 c+9) 
\pa^3 T \Bigg)
\Bigg](w) \nonu \\
& & + \frac{1}{(z-w)} \Bigg[
 C_{\frac{7}{2} \frac{7}{2}}^{6}
  W^{(6)} +  C_{\frac{7}{2} \frac{7}{2}}^{6'} W^{(6')}
  + 
  \frac{1}{(2 c+61) (3 c+20)}
 C_{\frac{7}{2} \frac{7}{2}}^{4'}
  \Bigg( \frac{1}{2} (4 c-229)  
  G W^{(\frac{9}{2})}
\nonu \\
& &
 + (50 c+589) 
 T  W^{(4')}
+ \frac{1}{12} (2 c+9) (5 c+112)
  \pa^2  W^{(4')} \Bigg)
 \nonu \\
 && + \frac{1}{(c-1) (2 c+53)}   C_{\frac{7}{2} \frac{7}{2}}^{4}
 \Bigg(-
\frac{1}{2} (c-101) 
G \pa W^{(\frac{7}{2})}
+ \frac{1}{3} 50 (c+1)  
  T W^{(4)}
\nonu \\
& &
 +\frac{7}{6} (c-101) 
  \pa G  W^{(\frac{7}{2})}
 + \frac{1}{36} (10 c^2+157 c-567)
  \pa^2 W^{(4)} \Bigg)
  \nonu \\
  && +  \frac{1}{(c+11) (4 c+21) (10 c-7) (14 c+11)}
 \Bigg( 24 (450 c^2+1199 c+1)
  T T T 
 \nonu \\
 & & +  36 (18 c^2-1037 c+169)
       T \pa G G
 + 2 (520 c^3+5946 c^2+10067 c+867)
 \pa T \pa T 
\nonu \\
& & +
(10 c^3-303 c^2-14968 c+9111)
     \pa^2 G \pa G
     \nonu \\
    & & + 
     2 (620 c^3+2580 c^2-5459 c-891) 
          \pa^2 T T
\nonu \\
& & + \frac{3}{2} (8 c^3-482 c^2+619 c-6745)
\pa^3 G G
\nonu \\
& & +
\frac{1}{3} (20 c^4+100 c^3-2477 c^2+2159 c-1302)
\pa^4 T \Bigg)
\Bigg](w)
+ \cdots.
\label{7half7half}
\eea
}
One can check that all the nonlinear terms vanish
under the large $c \rightarrow \infty$.

The OPE between the lowest higher spin $\frac{7}{2}$
current and the higher spin $4$ current  is described as
{\small
\bea
&& W^{(\frac{7}{2})}(z) \, W^{(4)}(w)  = 
\frac{1}{(z-w)^6} \, 3 G(w)
+ \frac{1}{(z-w)^5} \, \pa G(w)
\nonu \\
&& +  \frac{1}{(z-w)^4} \Bigg[
  7 C_{\frac{7}{2} \frac{7}{2}}^{4} W^{(\frac{7}{2})}
  +  \frac{(2 c-57) }{2 (4 c+21)}
  \pa^2 G +\frac{90}{(4 c+21)} T G
\Bigg](w)
  \nonu \\
  && + \frac{1}{(z-w)^3} \Bigg[
    -\frac{1}{2}  C_{\frac{7}{2} \frac{7}{2}}^{4'}
  W^{(\frac{9}{2})}
  + 3 C_{\frac{7}{2} \frac{7}{2}}^{4}
  \pa W^{(\frac{7}{2})}
  +\frac{1 }{(4 c+21) (10 c-7)}  \Bigg( 2 (154 c-339) 
   T \pa G \nonu \\
  & & +  12 (37 c+3) 
  \pa T G
  + \frac{1}{2} (2 c-117) (2 c-3) 
\pa^3 G \Bigg)
  \Bigg](w)
  \nonu \\
  & & + \frac{1}{(z-w)^2} \Bigg[
  11 C_{\frac{7}{2} \frac{7}{2}}^{6} W^{(\frac{11}{2})} 
   +  \frac{1}{(3 c+20)}  C_{\frac{7}{2} \frac{7}{2}}^{4'}
    \Bigg( \frac{59 }{2 }
    G W^{(4')}
    -\frac{1}{3} (2 c+33)
    \pa W^{(\frac{9}{2})} \Bigg)
  \nonu \\
  && + \frac{1}{(2 c+53)}   C_{\frac{7}{2} \frac{7}{2}}^{4}
  \Bigg( \frac{51  }{2 }
 G W^{(4)}
   + 119 
    T W^{(\frac{7}{2})}
+  \frac{3}{4} (2 c+19)
    \pa^2 W^{(\frac{7}{2})} \Bigg)
    \nonu \\
    && +
    \frac{1}{(c+11) (4 c+21) (10 c-7)} \Bigg( 36 (99 c+31)
    T T G
    +
   3 (26 c^2-667 c-2019)  
   T \pa^2 G
   \nonu \\
   &&+ 
   2 (19 c+71) (4c+21)
   \pa T \pa G
   +3 (11 c-17) (4c+21)
   \pa^2 T G
   \nonu \\
   &&+ \frac{1}{24} (4c+21)(2 c^2-167 c+1233)
\pa^4 G \Bigg)
   \Bigg](w)
  \nonu \\
  && +  \frac{1}{(z-w)} \Bigg[-\frac{1}{2}
C_{\frac{7}{2} \frac{7}{2}}^{6'} W^{(\frac{13}{2})}
+  5
C_{\frac{7}{2} \frac{7}{2}}^{6} \pa W^{(\frac{11}{2})}
\nonu \\
&&
+ \frac{1}{(c-1) (2 c+53)}  C_{\frac{7}{2} \frac{7}{2}}^{4}
\Bigg( \frac{25}{2} (c+1) 
G \pa W^{(4)}
  +
 \frac{5}{3} (31 c-71) 
   T \pa W^{(\frac{7}{2})}
  + \frac{5}{6} (11 c-91) 
    \pa G W^{(4)} 
\nonu \\
&& + \frac{175}{3} (c+1) 
  \pa T W^{(\frac{7}{2})}
+\frac{5}{36} (2 c^2-31 c+69) 
\pa^3 W^{(\frac{7}{2})} \Bigg)
\nonu \\
&& + 
\frac{1}{(2 c+61) (3 c+20)} C_{\frac{7}{2} \frac{7}{2}}^{4'}
\Bigg( \frac{1}{2} (61 c+1685) 
 G \pa W^{(4')}
 -9 (3c+20)
 T W^{(\frac{9}{2})}
 \nonu \\
  & & +
  \frac{1}{2}(34 c+1505)
    \pa G W^{(4')}
 -  \frac{1}{3}(c^2+44 c+675) 
\pa^2 W^{(\frac{9}{2})} \Bigg)
\nonu \\
& & + 
\frac{1}{(c+11) (4 c+21) (10 c-7) (14 c+11)}
\Bigg( 36 (486 c^2-875 c+339)
T T \pa G
\nonu \\
& &+ 5 (44 c^3-2484 c^2-3191 c-6069)
T \pa^3 G
+
144 (342 c^2+640 c+43) 
\pa T T G \nonu \\
&& +
6 (90 c^3-2269 c^2-4405 c+2034) 
\pa T \pa^2 G
\nonu \\
& & +
(632 c^3+1266 c^2+31333 c+1119)
\pa^2 T \pa G
\nonu \\
& & +  3 (136 c^3+698 c^2-2631 c-1153)
\pa^3 T G
\nonu \\
& & +
\frac{1}{24} (16 c^4-1692 c^3+27416 c^2+86847 c+103113)
 \pa^5 G \Bigg)
\Bigg](w)
  + \cdots.
\label{7half4}
\eea
}
In this case, all the nonlinear terms become zero
when the large $c \rightarrow \infty$ is taken.
In principle, one can obtain the OPE 
$W^{(4)}(z)\, W^{(\frac{7}{2})}(w)$ from Appendix (\ref{7half4}).
In the ${\cal N}=1$ description, the latter is more
useful to the former.

The OPE between the higher spin $4$ current
and itself is given by
{\small
\bea
&& W^{(4)}(z) \, W^{(4)}(w)  = 
\frac{1}{(z-w)^8} \, 2c +
\frac{1}{(z-w)^6} \, 16 T(w)
+\frac{1}{(z-w)^5} \, 8 \pa T(w)
\nonu \\
& & +
\frac{1}{(z-w)^4} \Bigg[
10  C_{\frac{7}{2} \frac{7}{2}}^{4} W^{(4)}
+ 3  C_{\frac{7}{2} \frac{7}{2}}^{4'} W^{(4')}
+ \frac{1}{(4 c+21) (10 c-7)} \Bigg(
\nonu \\
& & +  36 (13 c-38)
\pa G G
+  6 (c-1) (16 c-69) 
\pa^2 T
+12 (224 c-99)  T T \Bigg)
\Bigg](w)
\nonu \\
& & + 
\frac{1}{(z-w)^3} \Bigg[
  5 C_{\frac{7}{2} \frac{7}{2}}^{4} \pa W^{(4)}
  +\frac{3}{2} C_{\frac{7}{2} \frac{7}{2}}^{4'} \pa W^{(4')}
  \nonu \\
 & & +  \frac{1}{(4 c+21) (10 c-7)} \Bigg ( 18 (13 c-38) 
       \pa^2 G G+
       12 (224 c-99)  \pa T T 
  \nonu \\
  & & + \frac{4}{3} (c-1) (16 c-69) 
  \pa^3 T \Bigg)
  \Bigg](w)
\nonu \\
& & + 
\frac{1}{(z-w)^2} \Bigg[
12  C_{\frac{7}{2} \frac{7}{2}}^{6}
W^{(6)} + C_{\frac{7}{2} \frac{7}{2}}^{6'} W^{(6')}
 + 
\frac{1}{ (2 c+61) (3 c+20)}
C_{\frac{7}{2} \frac{7}{2}}^{4'} 
\Bigg( 12 (14 c+349) 
   T W^{(4')}
\nonu \\
&& +  \frac{1}{4}(10 c^2+285 c+92)  
\pa^2 W^{(4')}
  -
3 (19 c+638)  
G W^{(\frac{9}{2})} \Bigg)
\nonu \\
& & + \frac{1}{(c-1) (2 c+53)} C_{\frac{7}{2} \frac{7}{2}}^{4}
\Bigg(
  \frac{20}{3} (28 c-23)  
  T W^{(4)}
+
\frac{14}{3} (13 c-38)  
 \pa G W^{(\frac{7}{2})}
\nonu \\
& & - 
2 (13 c-38) 
 G \pa W^{(\frac{7}{2})}
+
\frac{1}{18} (50 c^2+767 c-1017)
\pa^2 W^{(4)} \Bigg)
\nonu \\
  & & +  
  \frac{1}{(c+11) (4 c+21) (10 c-7) (14 c+11)}
  \Bigg(
  216 (234 c^2+81 c+85) T \pa G G
  \nonu \\
  & & + 
  3 (190 c^3+7237 c^2+12743 c+12530)
   \pa^2 G \pa G
  \nonu \\
  & & +  
  3 (326 c^3-1591 c^2-13109 c-12926)
  \pa^3 G G\nonu \\
  & & +  
  2 (4720 c^3+51330 c^2+86729 c+35121) \pa T \pa T \nonu \\
  & & +
  2 (5632 c^3+8310 c^2-101995 c-77847)
     \pa^2 T T
  \nonu \\
  & & + 
  \frac{4}{3}
  (40 c^4+70 c^3-8911 c^2-8443 c-2256)
  \pa^4 T
  \nonu \\
  & & +  
  192 (576 c^2+721 c+128)
        T T T \Bigg)
  \Bigg](w)
\nonu \\
& & + 
\frac{1}{(z-w)} \Bigg[
  6 
  C_{\frac{7}{2} \frac{7}{2}}^{6} \pa W^{(6)}+
  \frac{1}{2} 
C_{\frac{7}{2} \frac{7}{2}}^{6'} \pa W^{(6')}
\nonu \\
&& +
\frac{1}{(c-1) (2 c+53)}  C_{\frac{7}{2} \frac{7}{2}}^{4}
\Bigg( \frac{4}{3} (13 c-38)  
           \pa G \pa W^{(\frac{7}{2})}+
            \frac{7}{3} (13 c-38) 
            \pa^2 W^{(\frac{7}{2})}\nonu \\
&  & - 
            (13 c-38) 
            G \pa^2 W^{(\frac{7}{2})}
+
            \frac{10}{3} (28 c-23) T \pa W^{(4)}
            \nonu \\
  & & +  
  \frac{10}{3} (28 c-23)   \pa T W^{(4)}
 +
       \frac{1}{18}(10 c^2+c-111) \pa^3
       W^{(4)} \Bigg)
       \nonu \\
       && + \frac{1}{(2 c+61) (3 c+20)}
        C_{\frac{7}{2} \frac{7}{2}}^{4'} \Bigg( -
            \frac{3}{2} (19 c+638) 
            G \pa W^{(\frac{9}{2})} - 
            \frac{3}{2} (19 c+638) 
            \pa G W^{(\frac{9}{2}) } \nonu \\
       && +
       6 (14 c+349)  T \pa
       W^{(4')} + 
       6 (14 c+349) \pa T
       W^{(4')} 
       + 
       \frac{1}{4} (2 c^2+31 c-564) \pa^3
       W^{(4')} \Bigg)
\nonu \\
& & +  
            \frac{1}{(c+11) (4 c+21) (10 c-7) (14 c+11)}\nonu \\
            & & \times  \Bigg( \frac{1}{5}
            (40 c^4+70 c^3-8911 c^2-8443 c-2256) \pa^5 T 
  \nonu \\
  & & + 
  108 (234 c^2+81 c+85)
       T \pa^2 G G  +  
       108 (234 c^2+81 c+85)
             \pa T \pa G G
            \nonu \\
            & & + 
            288 (576 c^2+721 c+128)
             \pa T T T
             + 
            18 (12 c^3-267 c^2-701 c-694)
            \pa^4 G G\nonu
            \\ & & + 
            12 (472 c^3+354 c^2+3263 c+2361)
             \pa^2 T \pa T
+ 
            6 (c+11) (38 c^2+187 c+200)
             \pa^3 G \pa G
             \nonu \\
            & & + 
            6 (416 c^3+766 c^2-12783 c-9699)
               \pa^3 T T \Bigg)     
  \Bigg](w)
+\cdots.
\label{44}
\eea
}
All the nonlinear terms in Appendix (\ref{44}) vanish
under the large $c$ limit.

By introducing the ${\cal N}=1$ stress energy
tensor \cite{FMS}, 
\bea
{\bf T} = \frac{1}{2} G(z) + \theta \, T(z),  
\label{boldT}
\eea
and representing the ${\cal N}=1$ lowest higher spin multiplet
\bea
{\bf W}^{(\frac{7}{2})} (Z) = W^{(\frac{7}{2})}(z) + \theta \, W^{(4)}(z),
\label{boldW}
\eea
the above three OPEs,  Appendix (\ref{7half7half}),
Appendix (\ref{7half4}) and Appendix
(\ref{44}),
can be written as
{\small
\bea
&&    {\bf W}^{(\frac{7}{2})}(Z_1) \, {\bf W}^{(\frac{7}{2})}(Z_2)
     =  \frac{1}{z_{12}^7} \, \frac{2c}{7}
     +  \frac{\theta_{12}}{z_{12}^6} \, 6 {\bf T}(Z_2)
       +  \frac{1}{z_{12}^5} \, 2 D {\bf T}(Z_2)
    +  \frac{\theta_{12}}{z_{12}^5} \, 4 \pa {\bf T}(Z_2)
    \nonu \\
    & + &  \frac{1}{z_{12}^4} \,  \pa D {\bf T}(Z_2)
    +   \frac{\theta_{12}}{z_{12}^4} \Bigg[
 7 C_{\frac{7}{2} \frac{7}{2}}^{4} {\bf W}^{(\frac{7}{2})}
  +  \frac{3(c-6) }{ (4 c+21)}
  2 \pa^2 {\bf T} +\frac{90}{(4 c+21)} 2 D {\bf T} {\bf T}
      \Bigg](Z_2)
    \nonu \\
    & + & \frac{1}{z_{12}^3} \Bigg[
 C_{\frac{7}{2} \frac{7}{2}}^{4}
 D  {\bf W}^{(\frac{7}{2})} +  C_{\frac{7}{2} \frac{7}{2}}^{4'} {\bf W}^{(4')}
  + \frac{1}{(4 c+21) (10 c-7)} \Bigg( 8 (37 c+3) D {\bf T} D {\bf  T} 
  \nonu \\
  &+& 3 (2 c-117)
  4 \pa {\bf T} {\bf T} +
  6 (2 c^2-6 c+9)
\pa^2 D {\bf T} \Bigg)
      \Bigg](Z_2)
    \nonu \\
    & + & \frac{\theta_{12}}{z_{12}^3} \Bigg[
 \frac{1}{2}  C_{\frac{7}{2} \frac{7}{2}}^{4'}
  D {\bf W}^{(4')}
  + 4 C_{\frac{7}{2} \frac{7}{2}}^{4}
  \pa {\bf W}^{(\frac{7}{2})}
  +\frac{1}{(4 c+21) (10 c-7)} \Bigg( 16 (37 c+3) 
   2 D {\bf T} \pa {\bf T} \nonu \\
  & + & 
  6(76c-111) 2 \pa D {\bf T}  {\bf T}
  +  4(2c^2 -43c+6)
8 \pa^3 {\bf T} \Bigg)
      \Bigg](Z_2)
    \nonu \\
    & + & \frac{1}{z_{12}^2} \Bigg[
 \frac{1}{2} C_{\frac{7}{2} \frac{7}{2}}^{4}
 \pa D {\bf W}^{(\frac{7}{2})} +
 \frac{1}{2} C_{\frac{7}{2} \frac{7}{2}}^{4'} \pa
 {\bf W}^{(4')}
 + \frac{1}{(4 c+21) (10 c-7)} \Bigg( 8 (37 c+3) \pa D {\bf T} {\bf  T}
 \nonu \\
 & + &
\frac{3}{2} (2 c-117)  4 \pa^2 {\bf T} {\bf T}
+ \frac{4}{3} (2 c^2-6 c+9) 
\pa^3 D {\bf T} \Bigg)
      \Bigg](Z_2)
    \nonu \\
    & + & \frac{\theta_{12}}{z_{12}^2} \Bigg[
 11 C_{\frac{7}{2} \frac{7}{2}}^{6} {\bf W}^{(\frac{11}{2})} 
 +\frac{1}{ (3 c+20)}  C_{\frac{7}{2} \frac{7}{2}}^{4'}
 \Bigg ( \frac{ 59 }{2}
    2 D {\bf T} {\bf W}^{(4')}
 +\frac{1}{6} (5 c-6) 
    \pa D {\bf W}^{(4')} \Bigg)
    \nonu \\
  & + & 
     \frac{1}{(2 c+53)} C_{\frac{7}{2} \frac{7}{2}}^{4} \Bigg ( 119  
     D {\bf T} {\bf W}^{(\frac{7}{2})}
      + \frac{51  }{2}
   2 {\bf T} D {\bf W}^{(\frac{7}{2})}
    +  \frac{1 }{4}(10 c+163)
    \pa^2 {\bf W}^{(\frac{7}{2})} \Bigg)
    \nonu \\
    & +&
    \frac{1}{(c+11) (4 c+21) (10 c-7)} \Bigg( 36 (99 c+31)
    2 D {\bf T} D {\bf T} {\bf T}
   \nonu \\
   &+ &
   6 (50 c^2 + 501 c+519) 
   2 \pa D {\bf T} \pa {\bf T}
   +6 (23 c^2 +34 c-822)
   2 \pa^2 D {\bf T}  {\bf T}
   \nonu \\
   &+& \frac{1}{3} (5 c^3 -135c^2-1139 c+939)
   2 \pa^4 {\bf T}
 +
   4 (55 c^2-19 c-516)  
   2 D {\bf T} \pa^2 {\bf T}
   \Bigg)
      \Bigg](Z_2)
    \nonu \\
    & + & \frac{1}{z_{12}} \Bigg[
 C_{\frac{7}{2} \frac{7}{2}}^{6}
  D {\bf W}^{(\frac{11}{2})} +  C_{\frac{7}{2} \frac{7}{2}}^{6'} {\bf W}^{(6')}
+ \frac{1}{ (c-1) (2 c+53)}  C_{\frac{7}{2} \frac{7}{2}}^{4} \Bigg( -
\frac{1}{2} (c-101) 
2 {\bf T} \pa {\bf W}^{(\frac{7}{2})}
\nonu \\
& + &  \frac{50}{3} (c+1)  
  D {\bf T}  D {\bf W}^{(\frac{7}{2})}
+
 \frac{7}{6} (c-101) 
  2 \pa {\bf T}  {\bf W}^{(\frac{7}{2})}
+ \frac{1}{36} (10 c^2+157 c-567)
  \pa^2 D {\bf W}^{(\frac{7}{2})} \Bigg)
  \nonu \\
  & + & \frac{1}{ (2 c+61) (3 c+20)}
C_{\frac{7}{2} \frac{7}{2}}^{4'}
  \Bigg( \frac{1}{12} (2 c+9) (5 c+112)
  \pa^2  {\bf W}^{(4')}
 \nonu \\
 & + & 
  \frac{1}{2} (4 c-229)  
 2 {\bf T} D {\bf W}^{(4')}
+ (50 c+589) 
 D {\bf T}  {\bf W}^{(4')} \Bigg)
 \nonu \\
 & + & \frac{1}{(c+11) (4 c+21) (10 c-7) (14 c+11)}
 \nonu \\
 &\times & 
 \Bigg( 24 (450 c^2+1199 c+1)
  D {\bf T} D {\bf T} D {\bf T} 
 +  36 (18 c^2-1037 c+169)
       4 D {\bf T} \pa {\bf T} {\bf T}
\nonu \\
&+& 2 (520 c^3+5946 c^2+10067 c+867)
 \pa D {\bf T} \pa D {\bf T} 
+
(10 c^3-303 c^2-14968 c+9111)
     4 \pa^2 {\bf T} \pa {\bf T}
     \nonu \\
     & + &
     2 (620 c^3+2580 c^2-5459 c-891) 
          \pa^2 D {\bf T}  D {\bf T}
+ \frac{3}{2} (8 c^3-482 c^2+619 c-6745)
4 \pa^3 {\bf T} {\bf T}
\nonu \\
&+&
\frac{1}{3} (20 c^4+100 c^3-2477 c^2+2159 c-1302)
\pa^4 D {\bf T} \Bigg)
      \Bigg](Z_2)
    \nonu \\
    & + & \frac{\theta_{12}}{z_{12}} \Bigg[
\frac{1}{2}
C_{\frac{7}{2} \frac{7}{2}}^{6'} D {\bf W}^{(6')}
+ 6
C_{\frac{7}{2} \frac{7}{2}}^{6} \pa {\bf W}^{(\frac{11}{2})}
+ \frac{1}{(c-1) (2 c+53)}  C_{\frac{7}{2} \frac{7}{2}}^{4}
\Bigg( \frac{14}{3} (13c-38) 
  \pa D {\bf T} {\bf W}^{(\frac{7}{2})}
\nonu \\
& + &   (13c-38)
2 {\bf T} \pa D {\bf W}^{(\frac{7}{2})}
 + 
\frac{1}{18} (10 c^2+ c-111) 
\pa^3 {\bf W}^{(\frac{7}{2})}
+
 \frac{2}{3} (101 c-1) 
   D {\bf T} \pa {\bf W}^{(\frac{7}{2})}
   \nonu \\
   & + & \frac{1}{3} (49 c+151) 
    2 \pa {\bf T} D
  {\bf W}^{(\frac{7}{2})} \Bigg)
  \nonu \\
&+& 
  \frac{1}{(2 c+61) (3 c+20)}  C_{\frac{7}{2} \frac{7}{2}}^{4'}
  \Bigg( \frac{3}{2} (19 c+638) 
 2 {\bf T} \pa {\bf W}^{(4')}
 +9 (3c+20)
 D {\bf T} D {\bf W}^{(4')}
 \nonu \\
  &+&
  3(14 c+349) 
    2 \pa {\bf T} {\bf W}^{(4')}
  +  \frac{1}{4} (2c^2+31 c-564)  
\pa^2 D {\bf W}^{(4')} \Bigg)
\nonu \\
& + &
\frac{1}{(c+11) (4 c+21) (10 c-7) (14 c+11)}
\nonu \\
&\times & \Bigg( 72 (450 c^2+1199 c+1)
2 D {\bf T} D {\bf T} \pa {\bf T}
+ 24 (34 c^3-505 c^2-2197 c-332)
2 D {\bf T} \pa^3 {\bf T}
\nonu \\
&+&
216 (234 c^2+81 c+85) 
2 \pa D {\bf T} D {\bf  T} {\bf T} 
+
12 (130 c^3-93 c^2+1970 c-357) 
2 \pa D {\bf T} \pa^2 {\bf T}
\nonu \\
&+&
6(212 c^3+540 c^2-677 c+3075)
2 \pa^2 D {\bf T} \pa {\bf T}
+  9 (48 c^3+102 c^2-2399 c-2351)
2 \pa^3 D {\bf T} {\bf T}
\nonu \\
&+&
\frac{2}{5} (10 c^4-437 c^3+107 c^2+11387 c+3183)
 2 \pa^5 {\bf T} \Bigg)
      \Bigg](Z_2) + \cdots.
    \label{superspaceope}
\eea
}
Here in addition to Appendix (\ref{boldT}) and Appendix
(\ref{boldW}), we also introduce
the following three ${\cal N}=1$ higher spin multiplets
as follows \cite{Ahn1211,Ahn1305}:
\bea
    {\bf W}^{(4')}(Z) & = & W^{(4')}(z) + \theta \, W^{(\frac{9}{2})}(z),
    \nonu \\
          {\bf W}^{(\frac{11}{2})}(Z) & = & W^{(\frac{11}{2})}(z) + \theta \,
          W^{(6)}(z),
    \nonu \\
    {\bf W}^{(6')}(Z) & = & W^{(6')}(z) + \theta \,
          W^{(\frac{13}{2})}(z).
          \label{n1notation}
          \eea
          In the relation (\ref{fusion}),
          the definitions in Appendix (\ref{n1notation}) is used.
          Note that the $\theta_{12}$ independent terms in
          the right hand side of Appendix (\ref{superspaceope})
          are related to
          the OPE Appendix (\ref{7half7half}) while
          the $\theta_{12}$ dependent terms in the right hand side
          of Appendix (\ref{superspaceope}) are related to
          the OPE between $W^{(4)}(z)\, W^{(\frac{7}{2})}(w)$.         
More explicitly, by the projection of $\theta_1=0=\theta_2$ on the
equation Appendix (\ref{superspaceope}),
one obtains the OPE in Appendix (\ref{7half7half}).
By acting $D_1$ on the equation Appendix (\ref{superspaceope})
and putting the condition $\theta_1=0=\theta_2$, one obtains
the OPE $W^{(4)}(z)\, W^{(\frac{7}{2})}(w)$ which can be obtained from
Appendix (\ref{7half4}). The remaining component OPEs
$W^{(\frac{7}{2})}(z) \, W^{(4)}(w)$ and
$W^{(4)}(z)\, W^{(4)}(w)$ can be read off from
Appendix (\ref{superspaceope}):
the result of ${\cal N}=1$ supersymmetry.

The factor $(4c+21)(10c-7)$ occurs in the OPE
between the ${\cal N}=1$ higher spin $\frac{5}{2}$ current in the
unitary case \cite{Ahn1305,Ahn1211,ASS1991}.
For the  ${\cal N}=1$ higher spin $\frac{7}{2}$ current
in the orthogonal case, the extra factor $(c+11)(14c+11)$
occurs in Appendix (\ref{superspaceope}). It would be interesting to
see whether there are any null states at the particular values
of the central charge, $c=-\frac{21}{4}, \frac{7}{10}, -11$ or
$-\frac{11}{14}$.

\section{ 
  The OPE between the two lowest higher spin multiplets
  in ${\cal N}=2$ superspace
}

Let us present the OPE in (\ref{n2ope})
here
{\small
\bea
&& {\bf W}_{\frac{4}{3}}^{(3)}(Z_1) \, 
{\bf W}_{-\frac{4}{3}}^{(3)}(Z_2)  =  
\frac{\theta_{12} \, \bar{\theta}_{12}}{z_{12}^7} \, g_{1} +
\frac{1}{z_{12}^6} \, \frac{c}{3}+
\frac{\theta_{12} \, \bar{\theta}_{12}}{z_{12}^6} \,
g_{2} \, {\bf T}(Z_2)
+ \frac{1}{z_{12}^5} \, g_{3} \, {\bf T}(Z_2)
\nonu \\
&& +\frac{\theta_{12}}{z_{12}^5} \, g_{4} \, D {\bf T}(Z_2)
+ \frac{\bar{\theta}_{12}}{z_{12}^5} \, g_{5} \, \overline{D} {\bf T}(Z_2)
+ \frac{\theta_{12} \, \bar{\theta}_{12}}{z_{12}^5} \, 
\Bigg[ g_{6} \, [ D, \overline{D} ] {\bf T}
+ g_{7} \, {\bf T} {\bf T} + g_{8} \, \pa {\bf T} \Bigg]
(Z_2)
\nonu \\
&&+ \frac{1}{z_{12}^4} \, \Bigg[ g_{9} \,
[D, \overline{D} ] {\bf T} + g_{10} \, {\bf T} 
{\bf T} + g_{21} \, \pa {\bf T} \Bigg](Z_2) 
 + \frac{\theta_{12}}{z_{12}^4} \, \Bigg[ 
g_{11} \, \pa D {\bf T} + g_{12} \, {\bf T} D {\bf T}
\Bigg](Z_2)
\nonu \\
&&+ \frac{\bar{\theta}_{12}}{z_{12}^4} \, \Bigg[ 
g_{13}\, \pa \overline{D} {\bf T} 
+ g_{14} \, {\bf T} \overline{D} {\bf T}
\Bigg](Z_2)
 +  \frac{\theta_{12} \, \bar{\theta}_{12}}{z_{12}^4}
\frac{1}{  (c-1)  (c+6)  (2 c-3) }
\nonu \\
& & \times
 \Bigg[  
g_{15} \, \pa [ D,  \overline{D} ] {\bf T} + 
g_{16} \, {\bf T} [D, \overline{D}] {\bf T}
+ g_{17} \, {\bf T} {\bf T} {\bf T}
+ g_{18} \, \overline{D} {\bf T} D {\bf T}
+   g_{19} \, \pa {\bf T} {\bf T} 
  + 
g_{20} \, \pa^2 {\bf T}
\Bigg](Z_2)
\nonu \\
&&+ \frac{1}{z_{12}^3}
\frac{1}{  (c-1)  (c+6)  (2 c-3) }
\nonu \\
& & \times
 \Bigg[
  g_{22} \, \pa [D, \overline{D}] {\bf T}
+ g_{23} \, {\bf T} [D, \overline{D}] {\bf T} + 
g_{24} \, {\bf T} {\bf T} {\bf T}
+ g_{25} \, \overline{D} {\bf T} D {\bf T} 
 +  g_{26} \, \pa {\bf T} {\bf T}
+ g_{50} \, \pa^2 {\bf T} \Bigg](Z_2)
\nonu \\
&&+ \frac{\theta_{12}}{z_{12}^3}
\frac{1}{  (c-1)  (c+6)  (2 c-3) }
\nonu \\
& & \times
 \Bigg[ 
  g_{27} \, \pa^2 D {\bf T} +
  g_{28} \, {\bf T} {\bf T} D {\bf T}
+ g_{29} \, [D, \overline{D}] {\bf T} D {\bf T}
+ g_{30} \, \pa D {\bf T} {\bf T}
+   g_{31} \, \pa {\bf T} D {\bf T}
\Bigg](Z_2)
\nonu \\
&&+ 
\frac{\bar{\theta}_{12}}{z_{12}^3}
\frac{1}{  (c-1)  (c+6)  (2 c-3) }
\nonu \\
& & \times
 \Bigg[ 
g_{32} \, \pa^2 \overline{D} {\bf T} + 
g_{33} \, {\bf T} {\bf T} \overline{D} {\bf T}
+ g_{34} \, \overline{D} {\bf T} [D, \overline{D}] {\bf T} 
+ g_{35} \, \pa \overline{D} {\bf T} {\bf T}
+   g_{36} \, \pa {\bf T} \overline{D} {\bf T}
\Bigg](Z_2)
\nonu \\
&&+ \frac{\theta_{12} \, \bar{\theta}_{12}}{z_{12}^3}
\frac{1}{  (c-1) (c+1) (c+6)  (2 c-3) (5 c-9)}
\nonu \\
& & \times
\Bigg[ g_{37} \, \pa^2 [ D, \overline{D} ] {\bf T}
+ g_{38} \, {\bf T} {\bf T} [ D, \overline{D} ] {\bf T}
+ g_{39} {\bf T} {\bf T} {\bf T} {\bf T} 
+ g_{40} \, {\bf T} \overline{D} {\bf T}
D {\bf T} + g_{41} \,
\pa \overline{D} {\bf T} D {\bf T}
 \nonu \\
& & +  g_{42} \, [ D, \overline{D} ] {\bf T}  [ D, \overline{D} ] {\bf T}
+ g_{43}\, \pa  [ D, \overline{D} ] {\bf T} {\bf T}
+  g_{44} \, \pa D {\bf T}  \overline{D} {\bf T}
\nonu \\
&&+   g_{45} \, \pa {\bf T} [ D, \overline{D} ] {\bf T}
+ g_{46} \, \pa {\bf T} {\bf T} {\bf T} + 
g_{47} \, \pa {\bf T} \pa {\bf T}
+ g_{48} \, \pa^2 {\bf T} {\bf T} + 
g_{49} \, \pa^3 {\bf T}
\Bigg](Z_2)
\nonu \\
&&+  \frac{1}{z_{12}^2}
\frac{1}{  (c-1) (c+1) (c+6)  (2 c-3) (5 c-9)}
\nonu \\
& & \times
\Bigg[ g_{220} \, \pa^2 [ D, \overline{D} ] {\bf T}
+ g_{51} \, {\bf T} {\bf T} [ D, \overline{D} ] {\bf T}
+ g_{52} \, {\bf T} {\bf T} {\bf T} {\bf T} +
g_{53} \, {\bf T} \overline{D} {\bf T}
D {\bf T} + g_{54} \,
\pa \overline{D} {\bf T} D {\bf T}
 \nonu \\
&& +  g_{55} \, [ D, \overline{D} ] {\bf T}  [ D, \overline{D} ] {\bf T}
+ g_{56} \, \pa  [ D, \overline{D} ] {\bf T} {\bf T}
+  g_{57} \, \pa D {\bf T}  \overline{D} {\bf T}
\nonu \\
&&+   g_{58} \, \pa {\bf T} [ D, \overline{D} ] {\bf T}
+ g_{59} \, \pa {\bf T} {\bf T} {\bf T} 
+ g_{60} \, \pa {\bf T} \pa {\bf T}
+ g_{61} \, \pa^2 {\bf T} {\bf T} 
+ g_{106} \, \pa^3 {\bf T}
\Bigg](Z_2)
\nonu \\
&&+ \frac{\theta_{12}}{z_{12}^2}
\frac{1}{  (c-1) (c+1) (c+6)  (2 c-3) (5 c-9)}
\nonu \\
& & \times
 \Bigg[ 
g_{62} \, \pa^3 D {\bf T} + 
g_{63} \, {\bf T} {\bf T} {\bf T} D {\bf T} 
+ g_{64} \, {\bf T}  [ D, \overline{D} ] {\bf T} D {\bf T}
+ g_{65} \, \pa [ D, \overline{D} ] {\bf T} D {\bf T}
+   g_{66} \, \pa D {\bf T}  [ D, \overline{D} ] {\bf T}
 \nonu \\
&& + 
g_{67} \, \pa D {\bf T} {\bf T} {\bf T}
+ g_{68} \, \pa D {\bf T}  \pa {\bf T}
+   g_{69} \, \pa^2 D {\bf T} {\bf T}
+ g_{70}\,  \pa {\bf T} {\bf T} D {\bf T}
+g_{71} \, \pa^2 {\bf T} D {\bf T}
\Bigg](Z_2)
\nonu \\
&&+ \frac{\bar{\theta}_{12}}{z_{12}^2}
\frac{1}{ (c-1) (c+1) (c+6)  (2 c-3) (5 c-9)}
\nonu \\
& & \times
 \Bigg[ g_{72} \,
\pa^3 \overline{D} {\bf T} + 
g_{73} \, {\bf T} {\bf T} {\bf T} \overline{D} {\bf T} 
+ g_{74} \, {\bf T}  \overline{D} {\bf T} [ D, \overline{D} ] {\bf T} 
+ g_{75} \, \pa \overline{D} {\bf T}  [ D, \overline{D} ] {\bf T}
 +   g_{76} \, \pa \overline{D} {\bf T}  {\bf T} {\bf T}
 \nonu \\
&& +   g_{77}\,  \pa \overline{D} {\bf T} \pa {\bf T}
+ g_{78} \, \pa^2 \overline{D} {\bf T}   {\bf T}
 +   g_{79} \, [ D, \overline{D} ] {\bf T} \overline{D} {\bf T}
+  g_{80} \,\pa {\bf T} {\bf T} \overline{D} {\bf T}
+ g_{81} \, \pa^2 {\bf T} \overline{D} {\bf T}
\Bigg](Z_2)
\nonu \\
&&+ \frac{\theta_{12} \, \bar{\theta}_{12}}{z_{12}^2}
\frac{1}{ (c-2) (c-1) (c+1) (c+6) (c+12) (2 c-3) (5 c-9)}
\nonu \\
& & \times
\Bigg[ g_{82} \, \pa^3 [ D, \overline{D} ] {\bf T} + 
g_{83} \, {\bf T}  {\bf T} {\bf T} 
[ D, \overline{D} ] {\bf T} +
g_{84} \, {\bf T}  {\bf T} {\bf T}   {\bf T}  {\bf T} 
+  g_{85} \, {\bf T}  {\bf T}   \overline{D} {\bf T} D {\bf T}
 \nonu \\
&& +  
g_{86} \, {\bf T} [ D, \overline{D} ] {\bf T} [ D, \overline{D} ] {\bf T}
+  
g_{87} \, \overline{D} {\bf T} [ D, \overline{D} ] {\bf T} D {\bf T}
+  g_{88} \, \pa \overline{D} {\bf T} \pa  D {\bf T}
+  g_{89} \,  \pa \overline{D} {\bf T} {\bf T}  D {\bf T}
\nonu \\
&& +  g_{90} \, \pa^2 \overline{D} {\bf T} D {\bf T} +
 g_{91} \, \pa  [ D, \overline{D} ] {\bf T}  [ D, \overline{D} ] {\bf T}
+ g_{92} \, \pa [ D, \overline{D} ] {\bf T} {\bf T} {\bf T} 
 + g_{93} \, \pa [ D, \overline{D} ] {\bf T} \pa {\bf T} 
\nonu \\
&& + g_{94} \,
\pa^2 [ D, \overline{D} ] {\bf T}  {\bf T} 
+  g_{95} \, \pa D {\bf T} {\bf T} \overline{D} {\bf T}
+  g_{96} \, \pa^2 D {\bf T}  \overline{D} {\bf T}
+  g_{97} \, \pa {\bf T}  {\bf T} [ D, \overline{D} ] {\bf T}
+ 
 g_{98} \, \pa {\bf T} {\bf T} {\bf T} {\bf T} 
\nonu \\
&&
+  g_{99} \, \pa {\bf T} \overline{D} {\bf T} D {\bf T}
+  g_{100} \, \pa {\bf T} \pa {\bf T} {\bf T}
\nonu \\
&& +    g_{101} \, \pa^2 {\bf T} [ D, \overline{D} ] {\bf T}
+   g_{102} \, \pa^2 {\bf T} {\bf T} {\bf T}
+  g_{103} \, \pa^2 {\bf T} \pa {\bf T}
+  g_{104} \, \pa^3 {\bf T}  {\bf T}
+  g_{105} \, \pa^4 {\bf T}
\Bigg](Z_2)
\nonu \\
&&+  \frac{1}{z_{12}}
\frac{1}{ (c-2) (c-1) (c+1) (c+6) (c+12) (2 c-3) (5 c-9)}
\nonu \\
& & \times
\Bigg[ g_{107}\, \pa^3 [ D, \overline{D} ] {\bf T} 
+ g_{108} \, {\bf T}  {\bf T} {\bf T} 
[ D, \overline{D} ] {\bf T} 
+ g_{109} \, {\bf T}  {\bf T} {\bf T}   {\bf T}  {\bf T} 
+  g_{110} \, {\bf T}  {\bf T}   \overline{D} {\bf T} D {\bf T} 
\nonu \\
&&  +  g_{111} \,
{\bf T} [ D, \overline{D} ] {\bf T} [ D, \overline{D} ] {\bf T}
 \nonu \\
&& +  g_{112} \, \overline{D} {\bf T} [ D, \overline{D} ] {\bf T} D {\bf T}
+  g_{113} \, \pa \overline{D} {\bf T} \pa  D {\bf T}
+   g_{114} \, \pa \overline{D} {\bf T} {\bf T}  D {\bf T}
+  g_{115} \, \pa^2 \overline{D} {\bf T} D {\bf T} 
\nonu \\
&& +
 g_{116} \, \pa [ D, \overline{D} ] {\bf T} [ D, \overline{D}] {\bf T} 
+  g_{117} \, \pa [ D, \overline{D} ] {\bf T} {\bf T} {\bf T} 
+  g_{118} \, \pa [ D, \overline{D} ] {\bf T} \pa {\bf T}
+ g_{119} \, \pa^2 [ D, \overline{D} ] {\bf T}  {\bf T} 
\nonu \\
&& +  g_{120} \, \pa D {\bf T} {\bf T} \overline{D} {\bf T}
+  g_{121} \, \pa^2 D {\bf T}  \overline{D} {\bf T}
+   g_{122} \, \pa {\bf T}  {\bf T} [ D, \overline{D} ] {\bf T}
+  g_{123} \,
\pa {\bf T} {\bf T} {\bf T} {\bf T} 
+  g_{124} \, \pa {\bf T} \overline{D} {\bf T} D {\bf T}
\nonu \\
&& +
    g_{125} \, \pa {\bf T} \pa {\bf T} {\bf T}
+    g_{126} \, \pa^2 {\bf T} [ D, \overline{D} ] {\bf T}
+   g_{127} \, \pa^2 {\bf T} {\bf T} {\bf T}
+  g_{128} \, \pa^2 {\bf T} \pa {\bf T}
+  g_{129} \, \pa^3 {\bf T}  {\bf T}
+  g_{219} \, \pa^4 {\bf T}
\Bigg](Z_2)
\nonu \\
&&+ \frac{\theta_{12}}{z_{12}}
\frac{1}{ (c-2) (c-1) (c+1) (c+6) (c+12) (2 c-3) (5 c-9)}
\nonu \\
& & \times
\Bigg[ g_{130} \,
\pa^4 D {\bf T}
+ g_{131} \, {\bf T} {\bf T} {\bf T} {\bf T} D {\bf T}
+ g_{132} \, {\bf T} {\bf T}  [ D, \overline{D} ] {\bf T} D {\bf T}
 +  g_{133} \, [ D, \overline{D} ] {\bf T} [ D, \overline{D} ] {\bf T}
D {\bf T} 
\nonu \\
&&+ 
 g_{134} \, \pa [ D, \overline{D} ] {\bf T} 
\pa D {\bf T}
+ g_{135} \, \pa [ D, \overline{D} ] {\bf T} 
{\bf T} D {\bf T}
+ g_{136} \, \pa^2 [ D, \overline{D} ] {\bf T} 
 D {\bf T}
+ 
g_{137} \, \pa D {\bf T} {\bf T} [ D, \overline{D} ] {\bf T} 
\nonu \\
&& 
+ g_{138} \, \pa D {\bf T} {\bf T} {\bf T} {\bf T}
+ 
  g_{139} \, \pa D {\bf T} \overline{D} {\bf T} D {\bf T}
+ g_{140} \, \pa D {\bf T} \pa {\bf T} {\bf T}
+ g_{141} \, \pa^2 D {\bf T}  [ D, \overline{D} ] {\bf T} 
+  g_{142} \, \pa^2 D {\bf T} {\bf T} {\bf T}
\nonu \\
&&+ 
g_{143} \, \pa^2 D {\bf T} \pa {\bf T} + 
g_{144} \, \pa^3 D {\bf T} {\bf T} 
+  g_{145} \, \pa {\bf T} {\bf T} {\bf T} D {\bf T} 
+ g_{146}\, \pa {\bf T} [ D, \overline{D} ] {\bf T} D {\bf T}
\nonu \\
&&+  
g_{147} \, \pa {\bf T} \pa {\bf T} D {\bf T}
+  g_{148}\, \pa^2 {\bf T} \pa D {\bf T}
+ g_{149}\,  \pa^2 {\bf T} {\bf T} D {\bf T}
+ g_{150} \, \pa^3 {\bf T} D {\bf T}
\Bigg](Z_2)
\nonu \\
&&+ \frac{\bar{\theta}_{12}}{z_{12}}
\frac{1}{ (c-2) (c-1) (c+1) (c+6) (c+12) (2 c-3) (5 c-9)}
\nonu \\
& & \times
\Bigg[ 
g_{151} \, \pa^4 \overline{D} {\bf T}
+ g_{152} \, {\bf T} {\bf T} {\bf T} {\bf T} \overline{D} {\bf T}
+ g_{153} \, {\bf T} {\bf T}  \overline{D} {\bf T} [ D, \overline{D} ] {\bf T}
+  g_{154} \, \overline{D} {\bf T}
[ D, \overline{D} ] {\bf T} [ D, \overline{D} ] {\bf T}
 \nonu \\
&& + g_{155}
 \, \pa \overline{D} {\bf T} \pa [ D, \overline{D} ] {\bf T} 
+ g_{156} \, \pa  \overline{D} {\bf T} {\bf T} [ D, \overline{D} ] {\bf T} 
+   g_{157} \, \pa  \overline{D}  {\bf T} 
  {\bf T} {\bf T} {\bf T}
+ g_{158} \,
\pa \overline{D} {\bf T}  \overline{D}  {\bf T} D {\bf T} 
\nonu \\
&& + g_{159} \, \pa  \overline{D} {\bf T} \pa {\bf T}  {\bf T} 
+ g_{160} \, \pa^2 \overline{D} {\bf T} [ D, \overline{D} ] {\bf T}
+ g_{161} \,
 \pa^2 \overline{D} {\bf T}  {\bf T}  {\bf T}
+  g_{162} \, \pa^2 \overline{D} {\bf T}  \pa  {\bf T}
\nonu \\
&& +  g_{163} \, \pa^3 \overline{D} {\bf T} {\bf T} +
g_{164} \, \pa [ D, \overline{D} ] {\bf T} {\bf T} \overline{D} {\bf T} 
 +  g_{165} \,
\pa^2 [ D, \overline{D} ] {\bf T}  \overline{D} {\bf T} 
+ g_{166} \, \pa {\bf T} {\bf T} {\bf T} \overline{D} {\bf T} 
\nonu \\
&&  
+ g_{167} \, \pa {\bf T}  \overline{D} {\bf T} [ D, \overline{D} ] {\bf T}
+  g_{168} \,
\pa {\bf T} \pa {\bf T} \overline{D} {\bf T}
+ g_{169}\,  \pa^2 {\bf T} \pa \overline{D} {\bf T}
+ g_{170}\, \pa^2 {\bf T} {\bf T} \overline{D} {\bf T}
+ g_{171} \, \pa^3 {\bf T} \overline{D} {\bf T}
\Bigg](Z_2)
\nonu \\
&&+ \frac{\theta_{12} \, \bar{\theta}_{12}}{z_{12}}
\frac{1}{ (c-2) (c-1) (c+1) (c+6) (c+12) (2 c-3) (5 c-9) (7 c-15)}
\nonu \\
& & \times
\Bigg[ g_{172} \, \pa^4  [ D, \overline{D} ] {\bf T}
+ g_{173}\, {\bf T} {\bf T} {\bf T} {\bf T} [ D, \overline{D} ]  {\bf T}
+ g_{174} \, {\bf T} {\bf T} {\bf T} {\bf T} {\bf T}  {\bf T}
+  g_{175} \, {\bf T} {\bf T} {\bf T} \overline{D} {\bf T} D {\bf T}
 \nonu \\
&& + g_{176} \,
 {\bf T} {\bf T}  [ D, \overline{D} ] {\bf T}  [ D, \overline{D} ] {\bf T}
+ g_{177} \, {\bf T} \overline{D} {\bf T}  [ D, \overline{D} ] {\bf T} 
D {\bf T}
+  g_{178} \, \pa \overline{D} {\bf T} {\bf T} {\bf T} D {\bf T}
\nonu \\
&& 
+ g_{179}\, \pa \overline{D} {\bf T}  [ D, \overline{D} ] {\bf T}  D {\bf T}
\nonu \\
&& 
+ g_{180}\, \pa \overline{D} {\bf T}  \pa  D {\bf T} {\bf T}
+  g_{181}\, \pa \overline{D} {\bf T} \pa {\bf T} D {\bf T}
+  g_{182} \, \pa^2 \overline{D} {\bf T} \pa  D {\bf T}
+  g_{183} \, \pa^2 \overline{D} {\bf T} {\bf T}  D {\bf T}
\nonu \\
&& + g_{184} \, \pa^3 \overline{D} {\bf T} D {\bf T}
+ g_{185} \, [ D, \overline{D} ] {\bf T}  [ D, \overline{D} ] {\bf T}  
[ D, \overline{D} ] {\bf T}
+g_{186} \, \pa [ D, \overline{D} ] {\bf T} \pa [ D, \overline{D} ] {\bf T}
\nonu \\
&& +  g_{187} \, 
\pa [ D, \overline{D} ] {\bf T} {\bf T}  [ D, \overline{D} ] {\bf T}
+ g_{188} \, \pa [ D, \overline{D} ] {\bf T} {\bf T} {\bf T} {\bf T}
+ g_{189} \, \pa [ D, \overline{D} ] {\bf T} \overline{D} {\bf T}   D {\bf T}
\nonu \\
&& 
+g_{190}\,  \pa [ D, \overline{D} ] {\bf T} \pa {\bf T} {\bf T}
\nonu \\
&& + g_{191} \,
\pa^2 [ D, \overline{D} ] {\bf T}  [ D, \overline{D} ] {\bf T} 
+ g_{192} \, \pa^2 [ D, \overline{D} ] {\bf T}   {\bf T} {\bf T}
+ g_{193} \, \pa^2 [ D, \overline{D} ] {\bf T}   \pa {\bf T} 
+  g_{194}\, \pa^3 [ D, \overline{D} ] {\bf T}   {\bf T} 
\nonu \\
& &
+   g_{195} \, \pa D {\bf T} {\bf T} {\bf T} \overline{D} {\bf T}
+  g_{196} \, \pa D {\bf T}  \overline{D} {\bf T} [ D, \overline{D} ] {\bf T}
+  g_{197} \, \pa D {\bf T} \pa {\bf T}  \overline{D} {\bf T}
+   g_{198} \, \pa^2 D {\bf T}  \pa \overline{D} {\bf T}
\nonu \\
&& +  g_{199} \, \pa^2 D {\bf T}  {\bf T} \overline{D} {\bf T}
+ g_{200} \, \pa^3 D {\bf T}   \overline{D} {\bf T}
+ g_{201} \, \pa {\bf T} {\bf T} {\bf T}  [ D, \overline{D} ] {\bf T}
+ g_{202} \, \pa {\bf T} {\bf T}   {\bf T} {\bf T} {\bf T}
\nonu \\
&& 
+ g_{203}\, \pa {\bf T} {\bf T}   \overline{D} {\bf T} D {\bf T}
\nonu \\
&&+  g_{204} \,
\pa {\bf T}  [ D, \overline{D} ] {\bf T}   [ D, \overline{D} ] {\bf T} 
+ g_{205} \, \pa {\bf T}  \pa {\bf T}   [ D, \overline{D} ] {\bf T} 
+ g_{206} \, \pa {\bf T}  \pa {\bf T}   {\bf T}   {\bf T} 
\nonu \\
&&+  g_{207} \, \pa {\bf T}  \pa {\bf T}  \pa {\bf T}
+ g_{208}\, \pa^2 {\bf T} \pa  [ D, \overline{D} ] {\bf T} 
+ g_{209} \, \pa^2 {\bf T} {\bf T}  [ D, \overline{D} ] {\bf T} 
+ g_{210} \, \pa^2 {\bf T} {\bf T}  {\bf T}   {\bf T} 
\nonu \\
&&+  g_{211} \, \pa^2 {\bf T} \overline{D} {\bf T} D {\bf T}
+  g_{212} \, \pa^2 {\bf T}  \pa {\bf T} {\bf T}
+ g_{213} \, \pa^2 {\bf T} \pa^2 {\bf T}
+ g_{214} \, \pa^3 {\bf T}   [ D, \overline{D} ] {\bf T} 
\nonu \\
&&+   g_{215} \, \pa^3 {\bf T}  {\bf T} {\bf T}
+  g_{216}\, \pa^3 {\bf T}  \pa {\bf T}
+ g_{217} \, \pa^4 {\bf T} {\bf T} + 
g_{218} \, \pa^5 {\bf T}
\Bigg](Z_2) +\cdots.
\label{aboveexpression}
\eea
}
The structure constants in Appendix (\ref{aboveexpression})
are given by
{\small
\bea
g_1 & = &  -\frac{2c}{9}, \qquad g_2 = \frac{19}{9},
\qquad
g_3 = \frac{4}{3},\qquad
g_4 = -\frac{7}{3},
\qquad
g_5 = \frac{11}{3},
\nonu \\
g_6 & = &  \frac{4 (3 c+23)}{27 (c-1)},
\qquad g_7 = \frac{104}{9 (c-1)},
\qquad g_8 = \frac{23}{9},
\qquad g_9 = -\frac{(9 c-8)}{9 (c-1)},
\qquad g_{10} = -\frac{1}{3 (c-1)},
\nonu \\
g_{11} & = &  -\frac{7 (2 c-7)}{9 (c-1)},\qquad
g_{12}= -\frac{35}{3 (c-1)},\qquad
g_{13} = \frac{11 (2 c+1)}{9 (c-1)},\qquad
g_{14} = \frac{11}{(c-1)},
\nonu \\
g_{15} & = & \frac{1}{108} (36 c^3-859 c^2+5814 c-12264),
\qquad
g_{16} = -\frac{1}{54}(780 c^2-3641 c-7002),
\nonu \\
g_{17} & = &  -\frac{2}{27} (c-3571),
\qquad
g_{18} = \frac{1}{9}(535 c^2-1170 c+2336),
\nonu \\
g_{19} & = &
\frac{155}{9}(c+6)(2c-3),
\qquad
g_{20} = \frac{1}{54}(146 c^3-24 c^2-801 c-1022),
\qquad
g_{21} = \frac{2}{3},
\nonu \\
g_{22} & = &  -\frac{1}{18}(18 c^3+71 c^2-558 c+336),
\qquad
g_{23} = -\frac{2}{9} (36 c^2+121 c+18),
\nonu \\
g_{24} & = & -\frac{8}{9} (19 c+65),
\qquad g_{25} =
\frac{4}{3} (c^2-54 c+32),
\nonu \\
g_{26} & = & -\frac{1}{3} (c+6)(2c-3),
\qquad
g_{27} = -\frac{7}{6} (c^3-4 c^2+24 c-84),
\nonu \\
g_{28} & = & -\frac{14}{3} (3 c+67),
\qquad
g_{29} = \frac{7}{18} (39 c^2-74 c-84),
\nonu \\
g_{30} & = & -\frac{7}{9} (19 c^2+121 c-546),
\qquad
g_{31} = -\frac{7}{6} (11 c^2+2 c+8),
\nonu \\
g_{32} & = &  \frac{11}{36} (6 c^3-9 c^2+10 c-84),
\qquad
g_{33} = -\frac{22}{3} (5 c-19),
\nonu \\
g_{34} & = & -\frac{11}{18} (39 c^2-62 c-12),
\qquad
g_{35} = \frac{11}{9} (13 c^2-13 c+42),
\nonu \\
g_{36} & = & \frac{11}{6} (5 c^2+74 c-128),
\qquad
g_{37} = \frac{1}{27} (9 c^5+273 c^4-2463 c^3+5677 c^2-4392 c-3150),
\nonu \\
g_{38} & = & -\frac{4}{27}
(2334 c^3-3862 c^2-6821 c-5448),
\qquad
g_{39} = -\frac{2}{27} (9572 c^2-22722 c-17825),
\nonu \\
g_{40} &=& \frac{4}{9} (2819 c^3-10177 c^2+12112 c+5816),
\nonu \\
g_{41} & = & \frac{2}{27} (2694 c^4-3327 c^3+2937 c^2-3134 c+7200),
\nonu \\
g_{42} & = & -\frac{1}{162}
(576 c^4+18801 c^3-72025 c^2+57786 c-6300),
\nonu \\
g_{43} & = &  -\frac{1}{54}
(2070 c^4+463 c^3-69242 c^2+112335 c+64218),
\nonu \\
g_{44} & = &
\frac{2}{27} (2676 c^4-13749 c^3+21621 c^2+6938 c-50400),
     \nonu \\
     g_{45} & = & -\frac{2}{27} (c+1)(5c-9)(219 c^2-220 c-1854),
\qquad
g_{46} = -\frac{4}{9} (c+1)(5c-9)(13 c-1147),
     \nonu \\
g_{47} & = & \frac{1}{18}(1036 c^4+3801 c^3-22741 c^2+26866 c-5504),
     \nonu \\
     g_{48} & = & \frac{1}{27}
     (2054 c^4+1994 c^3-1401 c^2-53871 c+43930),
     \nonu \\
g_{49} & = &
\frac{1}{81}
(375 c^5-1365 c^4-2750 c^3+33544 c^2-50706 c-27384),
          \nonu \\
g_{50} & = & \frac{2}{9} (2 c^3+6 c^2+27 c-14),
          \qquad
g_{51} =
          \frac{2}{9} (72 c^3-1576 c^2-2753 c-1560),
               \nonu \\
g_{52}  & = &  -\frac{1}{9} (544 c^2+6720 c+4811),
                      \qquad
g_{53}  = 
-\frac{2}{3}
(13 c^3-731 c^2-2764 c-200),
                      \nonu \\
g_{54} &=&
                      -\frac{1}{9}
                      (3 c^4-183 c^3-3498 c^2-916 c+576),
                      \nonu \\
g_{55} & = &
                      \frac{1}{108}
                      (1728 c^4-2733 c^3-10177 c^2+9348 c+504),
                      \nonu \\
g_{56} & = &  -\frac{1}{18} (360 c^4+961 c^3-3629 c^2-10614 c-924),
                      \nonu \\
g_{57} & = &  \frac{1}{9}(57 c^4-3105 c^3+7902 c^2+5212 c-4032),
                      \nonu \\
g_{58} &= & -\frac{1}{9} (c+1)(5c-9)(36 c^2+121 c+18),
                      \qquad
g_{59}= -\frac{4}{3} (c+1)(5c-9)(19 c+65),
                      \nonu \\
g_{60} & = &
                      -\frac{1}{12}
                      (28 c^4-609 c^3+923 c^2+3136 c-3884),
                      \nonu \\
g_{61} & = &
                      -\frac{1}{18}
                      (2 c^4+746 c^3-2817 c^2-8691 c+3970),
                      \nonu \\
g_{62} & = &
                      -\frac{14}{27} (3 c^5+3 c^4-95 c^3+235 c^2-342 c-126),
                      \nonu \\
g_{63} & = & \frac{14}{9}
                      (404 c^2-1722 c-1307),
                      \qquad
g_{64} = \frac{7}{18} (981 c^3-2009 c^2-554 c-1932), \nonu \\
g_{65} & = & \frac{7}{18}
(99 c^4-353 c^3+720 c^2-1096 c-84),
\nonu \\
g_{66} & = & \frac{7}{54}
(384 c^4-1673 c^3-2771 c^2+8946 c+5292),
                      \nonu \\
g_{67} &=&
                      -\frac{7}{9}
                      (37 c^3+3087 c^2-6190 c-6510),
                      \qquad
g_{68} = -\frac{7}{18} (c+6)(104 c^3-461 c^2+471 c-56), \nonu \\
g_{69} & = & -\frac{7}{18} (69 c^4-192 c^3+3326 c^2-8467 c-3318),
g_{70} = -\frac{7}{6}
(83 c^3+305 c^2+386 c-928), \nonu
                      \\
g_{71} & = & -\frac{7}{18}
                      (58 c^4-191 c^3+861 c^2-1678 c+488),
\nonu \\
g_{72} & = & \frac{11}{108}
                      (c+1) (24 c^4-192 c^3+437 c^2-1642 c+1212),
                      \nonu \\
g_{73} & = &  -\frac{22}{9} (356 c^2+126 c+43),
g_{74} = -\frac{11}{18} (579 c^3-531 c^2+142 c-204),\nonu \\
g_{75} &=& -\frac{11}{54}
(384 c^4-65 c^3+305 c^2-810 c-108),
g_{76} = -\frac{11}{9}
(83 c^3+1013 c^2-118 c-138),\nonu \\
g_{77} &=& \frac{11}{18}
(c+6)(56 c^3-75 c^2+69 c-164),
g_{78} = \frac{11}{36}
(2 c-1) (51 c^3-177 c^2+1804 c+576),
                      \nonu \\
g_{79} & = & -\frac{11}{18}
(99 c^4-305 c^3-372 c^2+844 c+84),
g_{80} = -\frac{11}{6}
(117 c^3-1221 c^2+618 c+1592),
                      \nonu \\
g_{81} &=& \frac{11}{18}
(22 c^4+511 c^3-1161 c^2-742 c+2000),
                      \nonu \\
g_{82} &=& \frac{1}{972}
(36 c^7-2082 c^6-114909 c^5+190147 c^4+979562 c^3
\nonu \\
& - & 2223812 c^2-576408 c+1792224), \nonu \\
g_{83} &=& \frac{1}{81}
(8592 c^4-690638 c^3+492653 c^2+301732 c+152436),
                      \nonu \\
g_{84} & = & -\frac{1}{27}
(14816 c^3+458716 c^2-638341 c-144486), \nonu
                      \\
g_{85} & =& \frac{2}{9} (2320 c^4-101889 c^3+606844 c^2-674264 c+142664),\nonu \\
g_{86} & = & \frac{1}{162}
(21024 c^5-155943 c^4-477323 c^3+2798086 c^2-3446922 c+677124), \nonu \\
g_{87} &=& \frac{1}{54}
(56463 c^5-221675 c^4+287658 c^3-263908 c^2+480360 c-170688),\nonu \\
g_{88} &=& -\frac{1}{81} (10764 c^6+67305 c^5-243325 c^4-549414 c^3+1592164 c^2+1045896 c-1536192), \nonu \\
g_{89} & = & -\frac{1}{54} (46011 c^5+233631 c^4-1403284 c^3+2121364 c^2-3022480 c+322848),\nonu \\
g_{90} &= & -\frac{1}{108} (8091 c^6+10017 c^5-481476 c^4+2033078 c^3-3183264 c^2+3023176 c-347232), \nonu \\
g_{91} &=& -\frac{1}{648} (3 c-10) (384 c^5-19097 c^4+354613 c^3-952818 c^2+808704 c-128016),
                      \nonu \\
g_{92} &=& -\frac{1}{54} (9288 c^5+56174 c^4-230811 c^3-1268420 c^2+2402016 c-410424),\nonu \\
g_{93} &=& -\frac{1}{216} (9000 c^6+92627 c^5-178145 c^4-3381908 c^3+11292180 c^2-6585920 c-5335680),\nonu \\
g_{94} &=& -\frac{1}{108} (1314 c^6-7662 c^5-147177 c^4+893811 c^3-1813360 c^2+1082048 c-284424),\nonu \\
g_{95} &=& \frac{1}{54} (43257 c^5+373377 c^4-3279584 c^3+8020988 c^2-3334496 c-3284064),\nonu \\
g_{96} &=& \frac{1}{108} (8055 c^6-12570 c^5-97429 c^4-511992 c^3+4979588 c^2-8095952 c+3884832),\nonu \\
g_{97} &=& -\frac{2}{27} (c-2)(c+12) (6954 c^3-2861 c^2-23299 c-25314), \nonu \\
g_{98} & = & -\frac{2}{27} (c-2)(c+12) (24202 c^2-53445 c-42157),
                      \nonu \\
g_{99} &=& -\frac{1}{9}
(c-2)(c+12)(5899 c^3-31814 c^2+36521 c+26914),
                      \nonu \\
g_{100} &=& -\frac{1}{18}(734 c^5-32469 c^4-211047 c^3+1109960 c^2-984776 c-211976),\nonu \\
g_{101} &=& -\frac{1}{81}(3405 c^6+18707 c^5-141714 c^4-114585 c^3+778432 c^2+348368 c-1590744),\nonu \\
g_{102} &=& -\frac{1}{27}(c+12)(296 c^4-19977 c^3+2186 c^2+152690 c-196004),
\nonu \\
g_{103} &=& \frac{2}{27} (c-2)(c+12)(644 c^4+977 c^3-6375 c^2+7269 c-9683),
                      \nonu \\
g_{104} &=& \frac{1}{81} (1923 c^6+11271 c^5-83708 c^4+179760 c^3+479801 c^2-27546 c-1116048),\nonu \\
g_{105} &=& \frac{1}{1296}(1524 c^7+192 c^6-228911 c^5+758549 c^4+1510616 c^3 \nonu \\
& - & 7264492 c^2+3218632 c+5989872),\nonu \\
g_{106} &=& \frac{1}{54}
(30 c^5-294 c^4-682 c^3+2933 c^2+10053 c+1302),
                      \nonu \\
g_{107} &=& -\frac{1}{81}(27 c^7+171 c^6-1578 c^5-6019 c^4-101 c^3+55478 c^2+36624 c-14112),\nonu \\
g_{108} &=& \frac{4}{27} (1872 c^4+11665 c^3-57505 c^2-84080 c-57732), \nonu \\
g_{109} & = & \frac{4}{9} (640 c^3+2234 c^2-34793 c-10998),\nonu \\
g_{110} &=& \frac{8}{3} (31 c^4-1488 c^3-1757 c^2+36922 c+32),
                      \nonu \\
g_{111} &=& \frac{1}{27} (1728 c^5+13656 c^4-30415 c^3-117361 c^2+173622 c-74592),\nonu \\
g_{112} &=& \frac{2}{9} (9 c-7) (11 c^4-591 c^3+658 c^2+2544 c+192),\nonu \\
g_{113} &=& -\frac{4}{27} (12 c^6-525 c^5-4828 c^4-8937 c^3+65674 c^2+49488 c-12096),\nonu \\
g_{114} &=& \frac{1}{9} (63 c^5-1713 c^4-44560 c^3+248500 c^2+96752 c+7488),
                      \nonu \\
g_{115} &=& \frac{1}{18} (9 c^6-336 c^5-3111 c^4+77186 c^3-239904 c^2-91088 c-7104),\nonu \\
g_{116} &=& \frac{1}{108} (3 c-10) (576 c^5+6967 c^4-13895 c^3-35142 c^2+32184 c+2016),\nonu \\
g_{117} &=& \frac{1}{9} (72 c^5-670 c^4-29169 c^3-1808 c^2+272004 c+37632),
                      \nonu \\
g_{118} &=& -\frac{1}{36} (360 c^6+4549 c^5-1243 c^4-113704 c^3+79500 c^2+344432 c-172032),
                      \nonu \\
g_{119} &=& -\frac{1}{9} (54 c^6+318 c^5-2403 c^4+13806 c^3-12505 c^2-75262 c+22512),\nonu \\
g_{120} &=& \frac{1}{9}(141 c^5-5319 c^4-106876 c^3+186724 c^2+482768 c+36288),
                      \nonu \\
g_{121} &=& \frac{1}{18} (45 c^6-2166 c^5-8819 c^4+30636 c^3-75380 c^2+316112 c+149184),\nonu \\
g_{122} &=& \frac{2}{9}
(c-2)(c+12)(72 c^3-1576 c^2-2753 c-1560), \nonu \\
g_{123} & = & -\frac{2}{9} (c-2)(c+12)(544 c^2+6720 c+4811),\nonu \\
g_{124} &=& -\frac{1}{3} (c-2)(c+12)(13 c^3-731 c^2-2764 c-200), \nonu \\
g_{125} & = & -\frac{1}{3} (208 c^5+1926 c^4-5559 c^3-14291 c^2+60518 c-38392), \nonu \\
g_{126} &=& -\frac{4}{27} (45 c^6+542 c^5+1065 c^4-1899 c^3-15641 c^2-1786 c+25584),\nonu \\
g_{127} &=& -\frac{2}{9} (172 c^4+744 c^3-1277 c^2-10355 c+7874),
                      \nonu \\
g_{128} &=& -\frac{2}{9} (c-2)(c+12)
                          (7 c^4-149 c^3+48 c^2+93 c-1021),
                      \nonu \\
g_{129} &=& \frac{1}{54} (12 c^6-636 c^5-10495 c^4-53613 c^3+5614 c^2+814848 c+294816), \nonu \\
g_{130} &=& -\frac{7}{648} (30 c^7+80 c^6-5979 c^5+16513 c^4+55058 c^3-248048 c^2+179896 c+82992),
                      \nonu \\
g_{131} &=& \frac{7}{9} (1376 c^3+24184 c^2-66847 c-18402),
                      \nonu \\
g_{132} &=& \frac{7}{9}
(246 c^4+11499 c^3-27490 c^2+10670 c-29484),\nonu \\
g_{133} &=& -\frac{7}{108}
(3 c-10) (675 c^4+295 c^3-7346 c^2+9078 c-3780),\nonu \\
g_{134} &=& \frac{7}{54} (3 c-10) (65 c^5+499 c^4-1428 c^3-9172 c^2+20508 c-504),\nonu \\
g_{135} &=& \frac{7}{9} (249 c^4-1212 c^3+2929 c^2-3716 c+812),
                      \nonu \\
g_{136} &=& \frac{7}{36} (60 c^6-49 c^5-2109 c^4+16386 c^3-38894 c^2+31648 c-12264),\nonu \\
g_{137} &=& \frac{7}{54} (1833 c^5+20341 c^4-149380 c^3+41140 c^2+280008 c+248976), \nonu \\
g_{138} &=& \frac{14}{27} (814 c^4+2355 c^3-146337 c^2+266620 c+131124),\nonu \\
g_{139} &=& -\frac{14}{9} (10 c^5-633 c^4+8734 c^3-43992 c^2+60892 c+1512),\nonu \\
g_{140} &=& -\frac{7}{18} (113 c^5+6013 c^4+23676 c^3-166376 c^2+131240 c+94080),\nonu \\
g_{141} & = &  \frac{7}{108} (285 c^6+169 c^5-670 c^4-60784 c^3+60116 c^2+184872 c-38304), \nonu \\
g_{142} &=& -\frac{7}{9} (9 c^5+614 c^4+1520 c^3+62106 c^2-172210 c+9660),\nonu \\
g_{143} &=& -\frac{7}{36} (75 c^6+331 c^5-1806 c^4+14276 c^3-31932 c^2-47136 c+106512),\nonu \\
g_{144} &=& -\frac{7}{54} (54 c^6+649 c^5+2382 c^4-11713 c^3+34356 c^2-98724 c+24864),\nonu \\
g_{145} &=& \frac{7}{3} (3 c-10) (134 c^3+1167 c^2-904 c-2084),\nonu \\
g_{146} &=& \frac{7}{9}
(270 c^5+868 c^4-6159 c^3+5271 c^2-642 c+5712),\nonu \\
g_{147} &=& -\frac{7}{12} (35 c^5+705 c^4-756 c^3-11834 c^2+37100 c-30752),\nonu \\
g_{148} &=& -\frac{7}{54} (110 c^6+915 c^5-1745 c^4+1154 c^3-99200 c^2+242168 c-82992),\nonu \\
g_{149} &=& -\frac{7}{9} (58 c^5-35 c^4-1575 c^3+27644 c^2-71752 c+40536),
                      \nonu \\
g_{150} &=& -\frac{7}{108} (90 c^6-539 c^5-12299 c^4+76248 c^3-196162 c^2+238668 c-67296),\nonu \\
g_{151} &=& \frac{11}{2592} (120 c^7+460 c^6-8851 c^5+52395 c^4-231412 c^3+528598 c^2+4096 c-52008),\nonu \\
g_{152} &=& \frac{11}{3} (96 c^3-3776 c^2-6907 c-2762),
                      \nonu \\
g_{153} & = & \frac{11}{9}
(558 c^4-4699 c^3-5334 c^2+2914 c-2196),
                      \nonu \\
g_{154} &=& \frac{11}{108} (2025 c^5-4653 c^4-15988 c^3+41354 c^2-19392 c-4536),
                      \nonu \\
g_{155} &=& -\frac{11}{54} (c+12) (195 c^5-561 c^4+84 c^3-1896 c^2+2164 c+168),
                      \nonu \\
g_{156} &=& -\frac{11}{54} (1239 c^5+4231 c^4+8424 c^3+57092 c^2-24552 c-13968),\nonu \\
g_{157} &=& -\frac{22}{27} (742 c^4+4873 c^3+32681 c^2+27908 c-5556),
                      \nonu \\
g_{158} & = & -\frac{22}{9} (20 c^5-1087 c^4+4654 c^3-14524 c^2+10180 c+1800),\nonu \\
g_{159} &=& -\frac{11}{18} (199 c^5+2455 c^4+12816 c^3-63256 c^2+29608 c+4416),\nonu \\
g_{160} &=& -\frac{11}{108} (285 c^6+1058 c^5-9411 c^4+21276 c^3-105890 c^2+55464 c+19368),\nonu \\
g_{161} &=& -\frac{11}{18} (54 c^5+298 c^4+1861 c^3-62578 c^2-48906 c+12468), \nonu \\
g_{162} &=& \frac{11}{36} (c+12) (45 c^5-47 c^4+1182 c^3-4490 c^2+2434 c+372),
                      \nonu \\
g_{163} &=& \frac{11}{54} (42 c^6+151 c^5-3066 c^4+28255 c^3-94028 c^2-26656 c+13080),\nonu \\
g_{164} &=& -\frac{77}{9} (3 c-10)(c+12)
(7 c^3-2 c^2+c-4),
                      \nonu \\
g_{165}&=& -\frac{11}{36} (60 c^6-17 c^5-3045 c^4+5330 c^3+14010 c^2-21392 c-168), \nonu \\
g_{166} &=& -\frac{11}{3} (346 c^4+4327 c^3-15410 c^2+188 c+5656),
                      \nonu \\
g_{167} &=& -\frac{11}{9}
(120 c^5+2494 c^4-8791 c^3+4955 c^2-450 c+5088),\nonu \\
g_{168} & = & -\frac{11}{12} (85 c^5-821 c^4-14700 c^3+52430 c^2-18852 c-38848),\nonu \\
g_{169} &=& \frac{11}{54} (50 c^6+1085 c^5+1845 c^4-16450 c^3+53796 c^2-73808 c+42960), \nonu \\
g_{170} &=& -\frac{11}{9} (54 c^5+79 c^4-9527 c^3+20534 c^2+3572 c-23784),\nonu \\
g_{171} &=& \frac{11}{108} (30 c^6+643 c^5+2623 c^4-18896 c^3+1566 c^2+134124 c-131136),\nonu \\
g_{172} &=& \frac{77}{23328} (1440 c^7-14997 c^6-62076 c^5+1287093 c^4-4637476 c^3 \nonu \\
& + & 5140856 c^2+3881352 c-551040),
                      \nonu \\
g_{173} &=& \frac{308}{81} (4368 c^4-30343 c^3+14437 c^2-28765 c-14850),
                      \nonu \\
g_{174} &=& \frac{2156}{81} (c-10) (20 c-13) (32 c-7),
                      \nonu \\
g_{175} &=& \frac{616}{27} (2912 c^4-22307 c^3+78585 c^2-42856 c+33520),\nonu \\
g_{176} &=& \frac{77}{81} (4032 c^5-16497 c^4-39946 c^3+188221 c^2-233572 c+9240),\nonu \\
g_{177} &=& \frac{308}{27} (2688 c^5-17111 c^4+44127 c^3-52462 c^2+38522 c+4760), \nonu \\
g_{178} &=& \frac{154}{27} (294 c^5-2187 c^4+34562 c^3-82675 c^2+308522 c-52240),\nonu \\
g_{179} &=& \frac{77}{162} (10143 c^6-33504 c^5-9135 c^4+114096 c^3-78904 c^2+157688 c-20160),\nonu \\
g_{180} &=& -\frac{308}{81} (1008 c^6+4230 c^5-13346 c^4-48561 c^3+82163 c^2+298018 c-38360), \nonu \\
g_{181} &=& -\frac{77}{162} (6363 c^6+14892 c^5-323451 c^4+764148 c^3-451800 c^2-5176 c-90880),\nonu \\
g_{182} &=& -\frac{77}{972} (4410 c^7-15369 c^6-120012 c^5+227883 c^4+712944 c^3\nonu \\
& - & 275864 c^2-6730616 c+981120),\nonu \\
g_{183} &=& -\frac{154}{81} (1197 c^6-285 c^5-46635 c^4+276678 c^3-574875 c^2+999112 c-164240), \nonu \\
g_{184} &=& -\frac{77}{486} (882 c^7-5664 c^6-38073 c^5+435645 c^4-1390596 c^3\nonu \\
& + & 1886018 c^2-1948908 c+367360),\nonu \\
g_{185} &=& \frac{77}{1458} (3 c-10) (9168 c^4-41233 c^3+63926 c^2-32925 c+4284),\nonu \\
g_{186} &=& \frac{77}{972} (3 c-10) (2619 c^5-26718 c^4+93996 c^3-148037 c^2+121372 c-29064), \nonu \\
g_{187} &=& \frac{77}{162} (3 c-10) (672 c^5-194 c^4-32876 c^3+96045 c^2-103071 c+7728),\nonu \\
g_{188} &=& \frac{154}{81} (252 c^5-8779 c^4+2830 c^3+137012 c^2-102195 c+169050),\nonu \\
g_{189} &=& \frac{77}{27} (7 c-15)(189 c^5-1045 c^4+3357 c^3-6704 c^2+6324 c-672),\nonu \\
g_{190} &=& -\frac{77}{162} (3780 c^6+20748 c^5-65190 c^4-857481 c^3+2953173 c^2-2006758 c-516040), \nonu \\
g_{191} &=& -\frac{77}{486} (2412 c^6-14484 c^5-43683 c^4+380967 c^3-687016 c^2+234404 c+73920),\nonu \\
g_{192} &=& -\frac{77}{81} (378 c^6-990 c^5-42774 c^4+232857 c^3-400095 c^2+32236 c-8120),\nonu \\
g_{193} &=& -\frac{77}{108} (7 c-15) (18 c^6-27 c^5-594 c^4+4678 c^3-14908 c^2+12436 c+1176), \nonu \\
g_{194} &=& -\frac{77}{972} (252 c^7-309 c^6+7686 c^5+62992 c^4-425287 c^3+561346 c^2+593504 c+146720),\nonu \\
g_{195} &=& -\frac{154}{27} (546 c^5-35523 c^4+210850 c^3-439679 c^2+129322 c-145040),\nonu \\
g_{196} &=& -\frac{77}{162} (10143 c^6-76284 c^5+127473 c^4+181620 c^3-549440 c^2+8344 c+141120), \nonu \\
g_{197} &=& \frac{77}{162} (6237 c^6+28896 c^5-743565 c^4+3456888 c^3-5422224 c^2+843128 c+2656640),\nonu \\
g_{198} &=& \frac{77}{486} (2205 c^7-4947 c^6+32040 c^5-199680 c^4+429834 c^3 \nonu \\
& - & 290084 c^2+1394512 c+809760),\nonu \\
g_{199} &=& \frac{154}{81} (1071 c^6-4551 c^5+28176 c^4-227049 c^3+785763 c^2-674102 c+633640), \nonu \\
g_{200} &=& \frac{77}{486} (882 c^7+1632 c^6-76929 c^5+207717 c^4+408444 c^3\nonu \\
& - & 1785350 c^2+1833900 c+240800),\nonu \\
g_{201} &=& \frac{77}{27} (7 c-15)(96 c^4-5534 c^3-955 c^2+7648 c+7068),
\nonu \\
g_{202} &=& -\frac{77}{27}
(7 c-15)(544 c^3+17756 c^2-23063 c-5058),
                      \nonu \\
g_{203} &=& \frac{154}{9}  (7 c-15)(32 c^4-1482 c^3+11519 c^2-12826 c-1016),
                      \nonu \\
g_{204} &=& \frac{77}{162}  (7 c-15)(288 c^5-261 c^4-9395 c^3+26770 c^2-26670 c+10332), \nonu \\
g_{205} &=& -\frac{77}{162} (2520 c^6+21138 c^5-150681 c^4+206202 c^3+1479 c^2+208354 c-474680),\nonu \\
g_{206} &=& -\frac{77}{9} (1372 c^5+6039 c^4-77008 c^3+184847 c^2-95304 c-55520),\nonu \\
g_{207} &=& -\frac{77}{54} (7 c-15)(14 c^5-481 c^4-1799 c^3+17014 c^2-27804 c+11208),\nonu \\
g_{208} &=& -\frac{77}{324} (630 c^7+2843 c^6-15447 c^5-210354 c^4+1239562 c^3\nonu \\
& - & 2137342 c^2+838484 c+642320),
                      \nonu \\
g_{209} &=& -\frac{77}{81} (1680 c^6+8416 c^5-55197 c^4-36918 c^3+290553 c^2-46874 c-420520),\nonu \\
g_{210} &=& -\frac{154}{81} (2534 c^5+14344 c^4-147247 c^3+177799 c^2+155864 c-469280),\nonu \\
g_{211} &=& -\frac{154}{81} (840 c^6+387 c^5-64812 c^4+317754 c^3-515511 c^2+145294 c+244120),\nonu \\
g_{212} &=& -\frac{77}{27} (7 c-15)(10 c^5-399 c^4-1802 c^3+1411 c^2+8330 c-9216),
                      \nonu \\
g_{213} &=& \frac{77}{243} (210 c^7+1647 c^6-9126 c^5+21660 c^4-106350 c^3+240397 c^2-79586 c-182280),\nonu \\
g_{214} &=& -\frac{77}{972} (1260 c^7+1353 c^6-68409 c^5+374204 c^4-1287730 c^3 \nonu \\
& + & 2948886 c^2-3909156 c+1819920),\nonu \\
g_{215} &=& -\frac{77}{162} (42 c^6-2274 c^5-13142 c^4+131741 c^3-254657 c^2-936966 c+523320),\nonu \\
g_{216} &=& \frac{77}{486} (630 c^7+2016 c^6-47841 c^5-5964 c^4+1391691 c^3\nonu \\
& - & 4114472 c^2+3324268 c-149760),\nonu \\
g_{217} &=& \frac{77}{972} (504 c^7-1728 c^6-60234 c^5+409212 c^4-751281 c^3\nonu \\
& + & 584605 c^2-1013070 c+1718360), \nonu \\
g_{218} &=&  \frac{77}{38880} (840 c^8+720 c^7-76821 c^6+437576 c^5-1951055 c^4 \nonu \\
& + & 6780860 c^3-11809808 c^2+6517336 c-510720),
                      \nonu \\
g_{219} &=& \frac{1}{216} (24 c^7-288 c^6-5713 c^5+28303 c^4+22516 c^3-97784 c^2-4360 c+83328),\nonu \\
g_{220} &=& -\frac{1}{18}
(27 c^5-105 c^4+234 c^3-371 c^2-1395 c+252).
\label{coeff}   
\eea
}
Note that in the large $c \rightarrow \infty$, all the nonlinear
terms in Appendix (\ref{aboveexpression}) together with
Appendix (\ref{coeff})
disappear.
Let us emphasize that the field contents in
the right-hand side of 
Appendix
(\ref{aboveexpression}) can be read off from the ${\cal N}=2$
higher spin $\frac{7}{2}$ current with $U(1)$ charge $\frac{1}{3}$
and the higher spin $\frac{7}{2}$ current with $U(1)$ charge $-\frac{1}{3}$
in the unitary case \cite{Ahn1604} where  the factor
$(c-2) (c-1) (c+1) (c+6) (c+12) (2 c-3) (5 c-9)$ occurs also.
One can analyze the various null states at $c=2,1,-1,-6,-12,\frac{3}{2}$
or $\frac{9}{5}$. See also \cite{Blumenhagen}.


\end{document}